\documentclass[12pt]{article}
\usepackage{newtxtext,newtxmath}
\usepackage{hyperref}
\usepackage{subcaption}
\usepackage{booktabs}%

\usepackage[x11names]{xcolor}

\newcommand{\Tabref}[1]{Table~\ref{#1}}
\newcommand{\Figref}[1]{Fig.~\ref{#1}}

\newcommand{\Appref}[1]{(Appendix~\ref{#1})}

\newcommand{\xhdr}[1]{\vspace{1.7mm}\noindent{{\textbf{#1.}}}}

\usepackage{graphicx}

\usepackage[letterpaper,margin=1in]{geometry}

\renewenvironment{abstract}
	{\quotation}
	{\endquotation}

\date{}

\makeatletter
\renewcommand{\fnum@figure}{\textbf{Figure \thefigure}}
\renewcommand{\fnum@table}{\textbf{Table \thetable}}
\makeatother

\usepackage{scicite}

\usepackage{url}

\def\scititle{
Reranking partisan animosity in algorithmic social media feeds alters affective polarization
}
\title{\bfseries \boldmath \scititle}

\author{
	Tiziano~Piccardi$^{1\ast\dagger\ddagger}$,
	Martin~Saveski$^{2\ast\dagger}$,
	Chenyan~Jia$^{3\dagger}$,\and
        Jeffrey~Hancock$^{4}$,
        Jeanne~Tsai$^{4}$,
        Michael~Bernstein$^{4}$\and
	\small$^{1}$Johns Hopkins University, Baltimore, MD, USA.\and
	\small$^{2}$University of Washington, Seattle, WA, USA.\and
    \small$^{3}$Northeastern University, Boston, MA, USA.\and
     \small$^{4}$Stanford University, Stanford, CA, USA.\and
	\small$^\ast$Corresponding authors. Emails: piccardi@jhu.edu, msaveski@uw.edu\and
	\small$^\dagger$These authors contributed equally to this work. \\
    \small$^\ddagger$Work done while at Stanford University.
}

\begin{document} 

\maketitle

\vspace{-1cm}
\begin{abstract} \bfseries \boldmath
Today, social media platforms hold sole power to study the effects of feed ranking algorithms. We developed a platform-independent method that reranks participants' feeds in real-time and used this method to conduct a preregistered 10-day field experiment with 1,256 participants on X during the 2024 U.S. presidential campaign. Our experiment used a large language model to rerank posts that expressed antidemocratic attitudes and partisan animosity (AAPA). Decreasing or increasing AAPA exposure shifted out-party partisan animosity by two points on a 100-point feeling thermometer, with no detectable differences across party lines, providing causal evidence that exposure to AAPA content alters affective polarization. This work establishes a method to study feed algorithms without requiring platform cooperation, enabling independent evaluation of ranking interventions in naturalistic settings.
\end{abstract}

\noindent
Social media algorithms profoundly impact our lives: they curate what we see~\cite{bakshy2015exposure} in ways that can shape our opinions~\cite{brady2023overperception,lee2018does,feezell2021exploring}, our moods~\cite{kramer2014experimental,brady2021social,allcott2020welfare}, and our actions~\cite{bond201261,tufekci2017twitter,jost2018social,smith2023digital,zuckerberg2021us}. Due to the power that these ranking algorithms have to direct our attention, the research literature has articulated many theories and results detailing the impact that ranking algorithms have on us~\cite{rose2023outrage,lorenz2023systematic,tucker2018social,oldemburgo2024twitter,beam2020facebook}. However, validating these theories and results has remained extremely difficult because the ranking algorithm behavior is determined by the social media platforms, and only the platforms themselves can test alternative feed designs and causally assess their impact. Platforms, however, face political and financial pressures that constrain the kinds of experiments they can launch and share~\cite{haidt_bail_social_media}. Concerns about lawsuits and the need to preserve engagement-driven revenue further limit what platforms are willing to test, leaving massive gaps in the design space of ranking algorithms that have been explored in naturalistic settings and at scale.  

To address this gap, in this work, we present an approach that enables researchers to rerank participants' social media feeds in real-time as they browse, without requiring platform permission or cooperation. We built a browser extension---a small add-on to a web browser that modifies how web pages appear or behave, similar to an ad blocker.  Our extension intercepts and modifies X's web-based feed in real-time and reranks the feed using Large Language Model (LLM)-based rescoring, with only a negligible increase in page load time. This web extension allows us to rerank content according to experimentally controlled conditions. The design opens a new paradigm for algorithmic experimentation: it provides external researchers with a tool for conducting independent field experiments and evaluating the causal effects of algorithmic content curation on user attitudes and behaviors while preserving ecological validity. 

This capability allowed us to investigate a pressing question: can feed algorithms cause affective polarization---hostility toward opposing political parties~\cite{iyengar2019origins,boxell2024cross,iyengar2018strengthening,saveski2022perspective}? This concern has grown since the 2016 U.S. presidential election~\cite{nyhan2023like}, and the debate remains ongoing after the 2020 and 2024 elections. If social media algorithms are causing affective polarization, they might not only bear responsibility for rising political incivility online~\cite{frimer2023incivility} but also pose a risk to trust in democratic institutions~\cite{druckman2023does}. In this case, isolating the algorithmic design choices that cause polarization could offer alternative algorithmic approaches~\cite{stray2021designing}.

A major hypothesized mechanism for how feed algorithms cause polarization is a self-reinforcing engagement loop: users engage with content aligning with their political views, the feed algorithm interprets this engagement as a positive signal, and the algorithm exposes even more politically aligned content to users, leading to a polarizing cycle. Some studies support this hypothesis, finding that online interactions exacerbate polarization~\cite{barnidge2017exposure}, potentially because of the increased visibility of hostile political discussions~\cite{bor2022psychology}, divisive language~\cite{banks2021polarizedfeeds,rathje2021out,van2021social,suhay2018polarizing,kim2021distorting}, and content that reinforces existing beliefs~\cite{cho2020search}. However, large-scale field experiments aimed at reducing polarization by intervening on the feed algorithm---for example, by increasing exposure to out-party content—have found both a decrease~\cite{levy2021social} and an increase~\cite{bail2018exposure} in polarization. Similarly, recent large-scale experiments on Facebook and Instagram found no evidence that reduced exposure to in-party sources or a simpler reverse-chronological algorithm affected polarization and political attitudes~\cite{guess2023social,nyhan2023like} during the 2020 U.S. election. These mixed results reveal the difficulty in identifying what, if any, algorithmic intervention might help reduce polarization, especially during politically charged times.

We distill the goals of these prior interventions to a direct hypothesis that we can operationalize via real-time LLM reranking: that feed algorithms cause affective polarization by exposing users specifically to content that polarizes. An algorithm that upranks content reflecting genuine political dialogue is less likely to polarize than one that upranks demagoguery. This content-focused hypothesis has been difficult to operationalize into interventions, making studies that intervene on cross-partisan exposure and reverse-chronological ranking attractive but more diffuse in their impact and thus more likely to observe mixed results. However, by connecting our real-time reranking infrastructure with recent advances in LLMs, we can create a ranking intervention that more directly targets the focal hypothesis~\cite{bernstein2023embedding} without needing platform collaboration. We draw, in particular, on a recent large-scale field experiment that articulated eight categories of antidemocratic attitudes and partisan animosity as bipartisan threats to the healthy functioning of democracy~\cite{voelkel2024megastudy}. We operationalize these eight categories into an AI classifier that labels expressions of these constructs in social media posts, does so with accuracy comparable to trained annotators, and produces depolarization effects in a lab setting on a fixed feed~\cite{jia2024embedding}. This real-time classification enables us to perform a scalable, content-based, reranking experiment on participants' own feeds in the field~\cite{piccardi2024reranking}.

We conducted a preregistered field experiment on X, the most used social media platform for political discourse in the U.S.~\cite{mcclain2024americans}, using our extension to dynamically rerank participants' social media content by either increasing or decreasing exposure to content that expresses these eight factors of antidemocratic attitudes and partisan animosity (AAPA) over the course of a week. The experiment was conducted during a pivotal moment in the 2024 United States election cycle, from July to August 2024---an important period for understanding how social media feeds impact affective polarization. Major political events during the study period included the attempted assassination of Donald Trump, the withdrawal of Joe Biden from the 2024 presidential race, and the nomination of Kamala Harris as the Democratic Party’s candidate. These events allow us to examine the impact of heterogeneous AAPA content on partisan polarization and hostility. We measured the intervention’s effect on affective polarization~\cite{iyengar2012affect} and emotional experience~\cite{russell1980circumplex}. Compared to control conditions that did not rerank the feed, decreased  AAPA exposure led to warmer feelings toward the political outgroup, whereas increased AAPA exposure led to colder feelings. These changes also affected participants’ levels of negative emotions (anger and sadness) as measured through in-feed surveys. 

\section*{Exposure to Antidemocratic Attitudes and Partisan Animosity}

\subsection*{Antidemocratic Attitudes and Partisan Animosity (AAPA)}
There are many viable theories as to which content, if amplified, could impact affective polarization. In this experiment, we draw specifically on prior work that articulates eight factors as problematic outcomes for healthy democratic functioning, referred to as ``antidemocratic attitudes and partisan animosity''~\cite{voelkel2024megastudy}: (1)~\textit{partisan animosity}, (2)~\textit{support for undemocratic practices}, (3)~\textit{support for partisan violence}, (4)~\textit{support for undemocratic candidates}, (5)~\textit{opposition to bipartisanship}, (6)~\textit{social distrust}, (7)~\textit{social distance}, and (8)~\textit{biased evaluation of politicized facts} \Appref{sup:poda}. 
We refer to these eight factors collectively as \textit{AAPA} (for \textit{Antidemocratic Attitudes and Partisan Animosity}), and hypothesize that increased or decreased exposure to AAPA content causes corresponding changes in affective polarization.
Our feed ranking intervention first identified only political content in the feed, using a broad definition from the Pew Research Center~\cite{bestvater2022politics} that includes mentions of officials and activists, social issues, news, and current events \Appref{sup:political_classifier}. 
We then scored each political post from zero to eight based on the number of the eight AAPA factors expressed in its text, and we classified a post as AAPA if it reflected at least four of these eight factors. 
This scoring was performed by a large language model~\Appref{sup:scoring}, which showed a high correlation with human judgment on identifying these eight factors in social media posts~\cite{jia2024embedding}.

\subsection*{Experimental Design}

\begin{figure}[htbp]
\centering
\includegraphics[width=\textwidth]{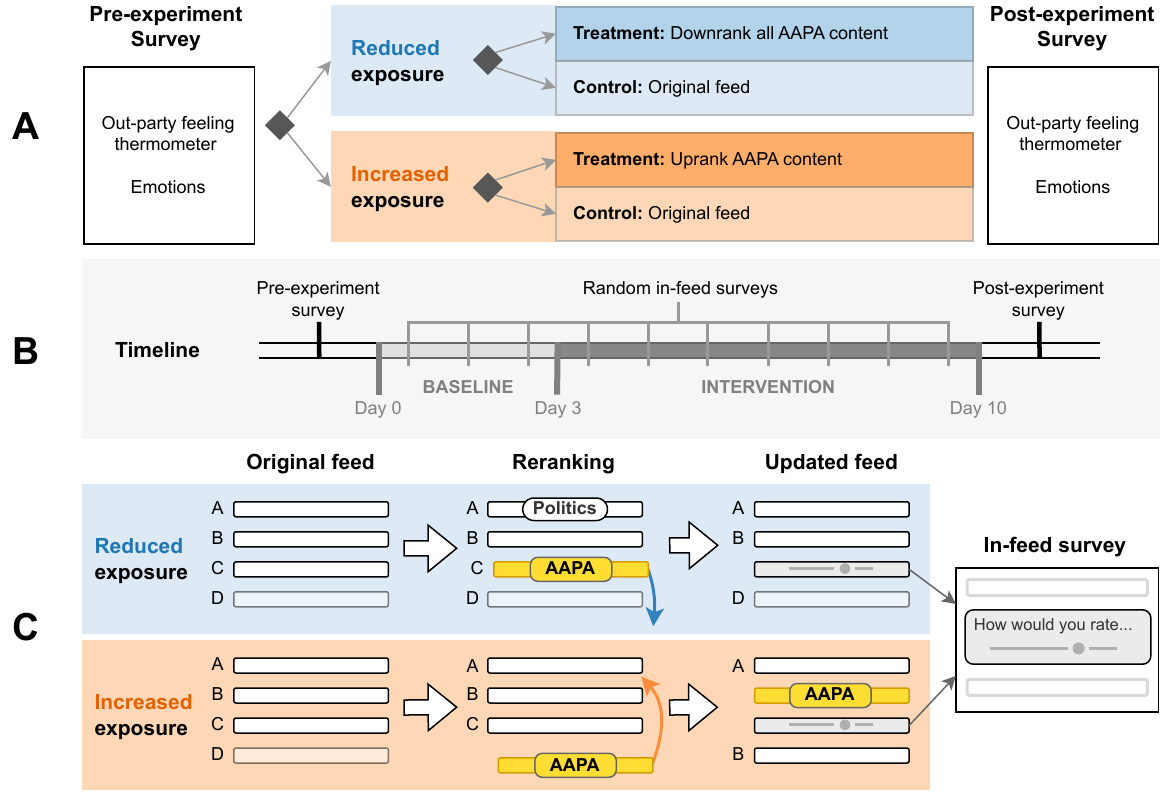}
\caption{Overview of the field experiment on X.
(A) Participants completed a pre-experiment survey to rate their feelings toward the opposing party and report their emotions. Then, they were randomly assigned to one of the two parallel experiments, reduced or increased AAPA content exposure, and further randomized into treatment and control groups. After ten days, they completed a post-experiment survey. (B) Timeline of the experiment. During the first three days, no intervention was applied to measure the participants' baseline responses. For the next seven days, participants in the treatment groups received increased or decreased AAPA exposure. In-feed surveys were periodically shown throughout the baseline and the intervention period.
(C) The interventions downranked or upranked AAPA posts in the participants' feeds, depending on their treatment assignment.}\label{fig:experiment}
\end{figure}

We designed two parallel experiments to measure the impact of reducing and increasing exposure to AAPA posts (\Figref{fig:experiment}). The first, a ``Reduced Exposure'' experiment, downranked all AAPA content, pushing it lower in their feeds. In contrast, the second, an ``Increased Exposure'' experiment upranked AAPA posts into participants' feeds. These posts came from the inventory of posts recommended by X to all participants with the same political leaning. Both experiments were randomized controlled trials. After providing informed consent and installing our web extension, participants were randomly assigned to one of the two experiments and then randomly divided into treatment or control conditions for each experiment. 
The overall experiment spanned ten days, with the first three days serving as a baseline period, during which no intervention was applied. During the subsequent seven days, participants in the treatment groups received the modified feed (downranked or upranked AAPA posts), while those in the control groups continued to receive the original, unmodified feed. 

We assessed outcomes at two different time points: post-experiment and in-feed. To measure post-experiment effects, we administered a survey asking participants how warmly they feel towards members of the opposing party (0--100 degrees on a feeling thermometer) and about the various emotions (e.g., anger, sadness, excitement, calm) they felt when reading social media over the previous week (5-point scale between ``never'' and ``all the time''). We surveyed the participants both before and after the experiment, enabling us to quantify the impact of the intervention at the end of the week-long treatment \Appref{method:survey}. 

To measure in-feed effects, we added posts containing a small survey directly in the participant's feed. 
These surveys, modeled on ecological momentary assessments~\cite{shiffman2008ecological}, resembled typical social media posts but were colored to stand out and contained a slider survey instrument. Each survey randomly selected one of two outcomes to assess: (1)~affective polarization, where participants rated their feelings toward the opposing party, or (2)~emotional experience, where participants reported the extent to which they felt two emotions: one randomly selected positive emotion (excited or calm), and one randomly selected negative emotion (angry or sad).
For both experiments, the in-feed surveys were shown at an equivalent rate and position in the control conditions \Appref{method:survey}. To reduce fatigue effects, each time participants completed the in-feed survey, they were less likely to receive another one on the same day. 

\subsection*{Feed Reranking}
To administer the interventions, we developed a browser extension for Google Chrome and Microsoft Edge that reranked participants' feeds in real-time~\cite{piccardi2024reranking}. The extension activated when participants accessed the ``For you'' feed on X. It intercepted the participants' original feed and sent it to a remote backend, which scored and reranked the posts~\Appref{sup:platform}. This process added a latency of about three seconds when opening the feed for the first time, which was consistent across all experimental conditions.
To rerank the original feeds, we first identified political posts~\Appref{sup:political_classifier}, then classified whether and how much AAPA the political posts contained. 

In the Reduced Exposure treatment condition, we downranked all AAPA posts in the participants' feeds. If participants continued to scroll down far enough, e.g., across several incremental loads of more content, these demoted posts would appear (\Figref{fig:seen_by_day_reduced}). Before downranking all the AAPA posts, we randomly selected the position of one AAPA post and, if the in-feed survey was sampled to be included, inserted the post with the survey in the next available position \Appref{sup:reduced}. This survey position was consistent across treatment and control conditions. Thus, participants in the control condition saw AAPA posts and the in-feed survey, while participants in the treatment condition saw the survey in the position where the first AAPA post used to be, but all AAPA posts were placed further down. The survey was limited to only a few times per day to avoid participant fatigue \Appref{method:survey}.
In the Increased Exposure experiment, since we lacked access to the platform's full post inventory, we sourced AAPA posts from the inventory of all posts in the ``For you'' feeds of all other study participants. We selected a random position in the original feed and upranked a random AAPA post to that position. The post containing the survey, if included, was inserted in the next available position. In the control condition, the survey was still added in the randomly sampled position, but no post was upranked. As in the Reduced Exposure experiment, the survey was placed consistently in the treatment and control conditions. During continuous scrolling sessions, this process repeated approximately every 30 posts \Appref{sup:increased}.

The Reduced Exposure treatment downranked 85.4 AAPA posts on average (median = 22) per user per day, and the Increased Exposure upranked 10.9 AAPA posts on average (median~=~5.43) per user per day~\Appref{sup:intervention_impact}. To avoid excessive exposure to polarizing content, we upranked fewer but higher-scoring AAPA posts. Despite the difference in the number of posts affected by the two interventions, the average AAPA score of the feed consumed during treatment days exhibited symmetric changes across the two experiments (Fig. \ref{fig:feed_scores}).

\section*{Results}\label{result}
We recruited participants from two online survey platforms, CloudResearch and Bovitz~\cite{stagnaro2024representativeness}. We targeted participants who used X, were 18 or older, resided in the United States, and self-identified as either Democrats or Republicans \Appref{sup:participants}. After giving informed consent, participants installed our web extension. 
To ensure the participants had enough political content, we used a pre-screening task to filter for users whose ``For you'' feeds contained at least 5\% of posts related to politics or social issues, retaining 73.4\% of the screened participants. The participants were recruited during the U.S. election season, between July 7th and August 14th, 2024, and instructed to use X daily only from the web interface. 

We preregistered a target of at least 1,100 participants to complete the full study (600 in the Reduced Exposure experiment; 500 in the Increased Exposure experiment) based on power analyses performed on pilot studies~\cite{piccardi2024feedrankings}. Toward this goal, a total of 1,662 participants completed the initial survey and qualified for inclusion in the study. Participants were randomly assigned to either the Reduced or Increased Exposure experiment until the preregistered experiment-specific quotas were reached. Of the qualified participants, N=1,256 completed the full study by responding to a follow-up survey at the end of the experiment: 727 in the Reduced Exposure experiment and 529 in the Increased Exposure experiment (\Figref{fig:study_flow} shows the participant recruitment funnel). Both treatment and control groups were well balanced across covariates, and there was no evidence of differential attrition between conditions \Appref{sec:experimental-design}. To maintain a consistent sample across the post-experiment and in-feed analyses, the reported treatment effects are based on participants who completed both the post-experiment survey and at least one in-feed survey for the corresponding outcome (N = 1,090). The results are similar when the full population is considered (Appendix \ref{sup:regression_tables} and \ref{sup:emotions_regression_tables}).

Participants received \$20 upon completing the study. The sample included 48\% women, 50\% men, and 2\% who identified as another gender. Participants ranged in age from 18 to 81 years (mean = 41.6, SD = 14.0). Party affiliation was 66\% Democrats and 34\% Republicans. Educational attainment was as follows: 17 (1\%) had less than a high school degree, 158 (13\%) were high school graduates, 267 (21\%) had some college but no degree, 145 (12\%) held an associate degree, 419 (33\%) held a bachelor’s degree, 189 (15\%) held a master’s degree, 30 (2\%) held a professional degree (e.g., JD, MD), and 31 (2\%) held a doctoral degree. Participants reported their race/ethnicity, and were allowed to choose multiple responses. Overall, 945 (75\%) identified as White, 183 (15\%) as Black or African American, 153 (12\%) as Hispanic or Latino, 66 (5\%) as Asian, 34 (3\%) as South Asian, 28 (2\%) as American Indian or Alaska Native, 5 ($<$1\%) as Middle Eastern, and 9 ($<$1\%) as Other.

On average, participants viewed 154 posts per day (median = 92) and spent 24.1 minutes (median = 13.1 minutes) per day with a browser tab in the foreground on their X feed. During the baseline period, an average of 31.6\% of the posts participants viewed were related to politics or social issues. Of these posts, 32\% were classified as AAPA, which comprised an average of 10.1\% of the total feed~content. 

The interventions had the intended effects (\Figref{fig:si_daily_aapa}). During the baseline period, participants across all experimental conditions were exposed to a similar fraction of AAPA posts. However, after the intervention began, participants in the treatment condition of the Reduced Exposure experiment experienced a noticeable decrease in exposure to such posts (treatment: 1.04\%, control: 10\% AAPA posts of all feed exposures; 89.6\% relative reduction), while those in the treatment condition of the Increased Exposure experiment were exposed to more AAPA posts (treatment: 13.4\%, control: 10.6\% AAPA posts; 26.4\% relative increase).

Over the full ten days of the experiment, participants answered an average of 13 (median~=~9) in-feed affective polarization questions in the Reduced Exposure experiment and 14.7 (median~=~11) in the Increased Exposure experiment. Similarly, they answered an average of 13.2 (median = 9) in-feed emotion surveys in the Reduced Exposure experiment and 15.0 (median~=~12) in the Increased Exposure experiment, with each survey asking them to rate two emotions.

In the post-survey, we asked participants whether the browser extension had impacted their experience on X.
Across conditions, most participants reported that they did not notice any impact on their online experience (74.2\%), while others guessed changes that were not part of the intervention~\Appref{sup:impact}. These results indicate a limited awareness of the experimental feed ranking intervention, reducing the likelihood that the observed results were due to demand effects.

\subsection*{Affective Polarization}
\label{sec:polarization}
\begin{figure}[tbp]
    \centering
    \includegraphics[width=\textwidth]{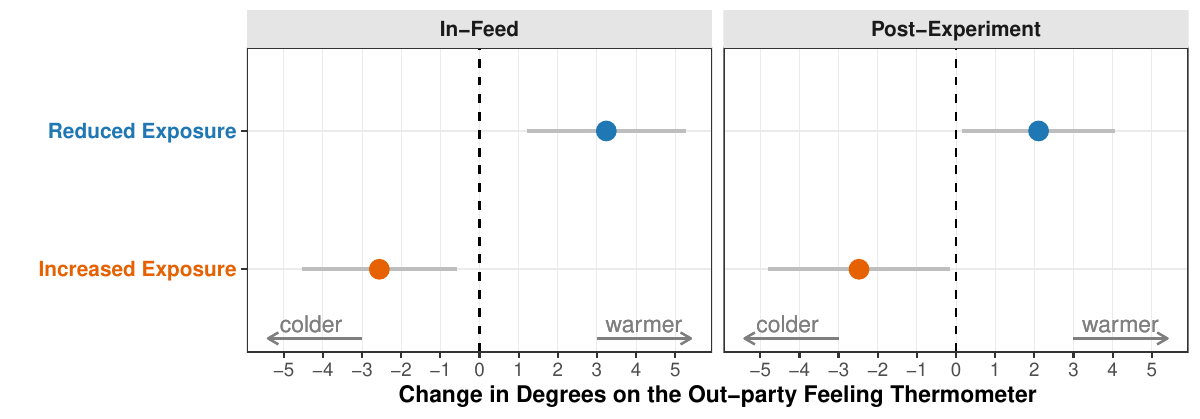}
    \caption{Effects of reducing and increasing exposure to AAPA content in participants' feeds on their feeling towards the out-party (feeling thermometer scale between 0: cold, and 100: warm) relative to the corresponding control group. Participants were surveyed within the feed during the intervention and off-platform after the experiment. The error bars represent 95\% confidence intervals.}
    \label{fig:results_polarization}
\end{figure}

Exposure to AAPA significantly influenced affective polarization (\Figref{fig:results_polarization}). 
Participants in the reduced AAPA exposure experiment reported a significant post-experiment effect, corresponding to an outparty warmth increase of 2.11 degrees (95\% CIs: [0.15, 4.06]; \textit{P}=0.035). Conversely, participants in the Increased Exposure experiment exhibited a symmetric reduction in affective warmth toward the opposing party of -2.48 degrees (95\% CIs: [-4.79, -0.17]; \textit{P}=0.036). We did not observe any significant heterogeneity in treatment effects across the preregistered moderators~\Appref{sup:heterogeneous}, including party identification, indicating that the effect was consistent and bipartisan. Like other studies~\cite{santoro2022promise}, we contextualize these effects against the rising levels of partisan animosity in American society over time: our effects correspond to a decrease and increase in affective polarization equivalent to 3 and 3.6 years, respectively. We observed even stronger effects with the in-feed assessment, with Reduced Exposure causing an increase of 3.24 degrees (95\% CIs: [$1.21$,~$5.27$]; \textit{P}=0.002) and Increased Exposure causing a cooling of -2.56 degrees (95\% CIs: [-4.53, -0.59];~\textit{P}=0.011). Supplementary exploratory analysis reweighting our sample to match the population of X users reduced our statistical power but replicated our results, indicating substantial robustness~\Appref{sec:reweighting}. The only exception was the post-treatment effects in the Reduced Exposure experiment, which had a similar point estimate but were not statistically significant.

In an exploratory analysis, we investigated which types of AAPA content were predictive of changes in affective polarization. Specifically, a higher proportion of posts supporting \textit{Partisan Animosity} (95\% CIs: [-13.49, -2.61]; \textit{P}=0.003), \textit{Opposition to Bipartisan Cooperation} (95\% CIs: [-26.55, -4.40]; \textit{P}=0.006), and \textit{Biased Evaluation of Politicized Facts} (95\% CIs: [-12.83, -2.21]; \textit{P}=0.005;) predicted a colder feeling toward the opposing party by the end of the study (\Figref{fig:factors_contribution}).

In addition to affective polarization (or partisan animosity), we also surveyed participants about their attitudes toward the remaining seven AAPA factors. We found that, although reduced and increased exposure to AAPA content impacted affective polarization, it had no significant effects on these other outcomes \Appref{sec:political-attitude-shifts}. These results are consistent with recent studies finding that changes in affective polarization are not necessarily linked to changes in antidemocratic attitudes~\cite{voelkel2024megastudy, landry2024partisan}.

\begin{figure}[htbp]
    \centering
    \includegraphics[width=\textwidth]{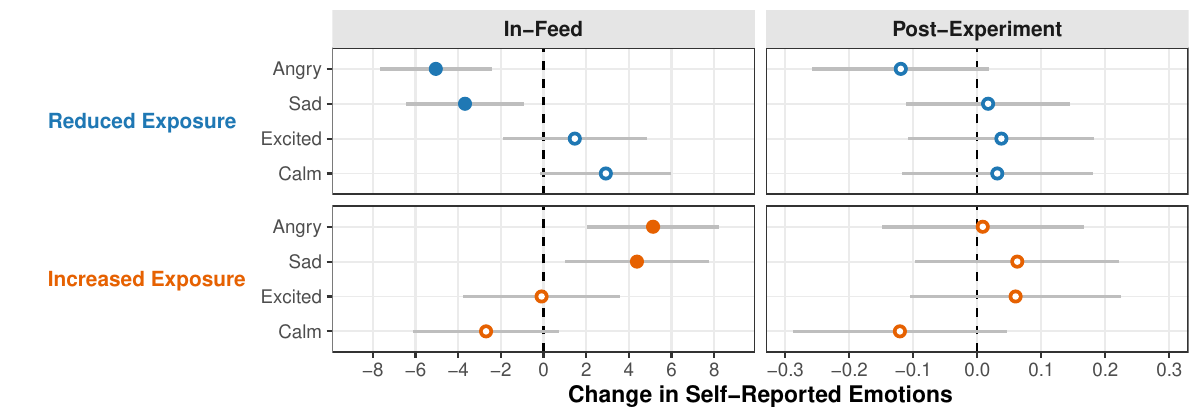}
    \caption{
    Effects of reducing and increasing exposure to AAPA content in participants' feeds on their experiences of emotion, relative to the corresponding control group. Participants were surveyed within the feed during the intervention (scale ranged from 0: ``none at all'' to 100: ``extremely'') and off-platform after the experiment (scale ranged from 1: ``never'' to 5: ``all the time'').  The filled circles represent statistical significance ($P<0.05$, adjusted for multiple hypothesis testing), and the error bars represent 95\% confidence intervals.
    }
    \label{fig:results_emotions}
\end{figure}

\subsection*{Emotions}
Changes to the feed algorithm caused in-feed, but not post-experiment, changes to participants' negative emotions (\Figref{fig:results_emotions}).  Reduced Exposure caused a change of -5.05 points (95\% CIs: [$-7.68$,~$-2.42$]; $P_{adj}$=0.002) for ``angry'' and -3.68 points (95\% CIs: [-6.42, -0.94]; $P_{adj}$=0.024) for ``sad.''  Likewise, in the Increased Exposure, participants reported a symmetric increase in negative emotions: 5.13 points (95\% CIs: [2.05, 8.22]; $P_{adj}$=0.013) for ``angry'' and 4.38 points (95\% CIs: [$1.00$,~$7.76$]; $P_{adj}$=0.036) for ``sad.'' Neither intervention had an effect on the positive emotions ``calm'' and ``excited.'' Post-experiment emotion effects were not significant, suggesting that the effects were likely short-lived.

\subsection*{Engagement}
We tested whether our interventions impacted platform engagement \Appref{sup:engagement}. Neither experiment observed differential attrition rates across conditions.

In the Reduced Exposure experiment, participants in the treatment group had a similar number of daily sessions (estimate: -0.06; 95\% CI: [-0.16, 0.04]), but spent $-4.71$ fewer minutes per day (95\% CI: [-8.22, -1.20]) compared to the control group. Over the 7-day intervention period, they viewed -169.52 fewer posts (95\% CI: [-325.48, -13.56]), liked -24.06 fewer posts (95\% CI: [$-39.53$,~$-8.59$]), but reposted a similar number of posts (estimate: -4.18; 95\% CI: [-9.89, 1.53]). On the other hand, they engaged with the posts they viewed at higher rates, with a 0.61\% higher like rate (95\% CI: [0.11\%, 1.12\%]) and a 0.058\% higher repost rate (95\% CI: [0.009\%, 0.11\%]). While the per-view engagement rates increased, these findings suggest that downranking AAPA content comes at the cost of slightly reduced engagement overall.

In the Increased Exposure experiment, we did not find any significant differences in engagement between the treatment and control groups. Participants in the treatment group had a similar number of daily sessions (estimate: -0.04; 95\% CI: [-0.16, 0.08]) and spent a comparable amount of time on the platform (estimate: -1.76 minutes per day; 95\% CI: [-6.17, 2.66]). They also viewed (estimate:~$-132.3$; 95\% CI: [-292.5, 27.94]), liked (estimate: 10.25; 95\% CI: [-5.23, 25.74]), and reposted (estimate: 3.38; 95\% CI: [-0.06, 6.82]) a similar number of posts, and liked and reposted the posts they viewed at similar rates (effect on like rate: 0.22\%, 95\% CI: [-0.57\%, 1.01\%]); effect on repost rate: -0.052\%; 95\% CI: [-0.17\%, 0.066\%]).

\section*{Discussion}\label{discussion}
This article demonstrates a deployable framework for conducting reranking field experiments on social media feeds without requiring direct involvement from the platforms. We developed a browser extension infrastructure that enables real-time manipulation of what consenting users see in their feed, using LLM-based classifiers to dynamically assess and rerank posts based on a custom ranking function. This approach opens new avenues for auditing, evaluating, and prototyping alternative feed-ranking strategies in real-world settings. 

Our specific experimental findings demonstrate that repeated exposure to AAPA posts, particularly during heightened political tensions such as the months leading up to the 2024 U.S. election, significantly influences users' affective polarization and emotional responses. A feed-ranking algorithm that controls exposure to such content produces measurable changes, equivalent to approximately three years of affective polarization, over the period it was deployed. Importantly, our results indicate that the intervention does not disproportionately impact any specific political leaning or demographic group, suggesting that the approach is bipartisan. The success of this method shows that it can be integrated into social media AIs to mitigate harmful personal and societal consequences. At the same time, our engagement analyses indicate a practical trade-off: interventions that downrank AAPA may reduce short-term engagement volume, posing challenges for engagement-driven business models and supporting the hypothesis that content that provokes strong reactions generates more engagement.

\subsection*{Methodological advances and policy implications}
Our work opens the door to conducting large-scale causal inference on social media feed algorithms without requiring ``independence by permission''~\cite{wagner2023independence} from platforms. While prior work made important advances in randomized controlled trials on the Facebook and Instagram algorithms\cite{guess2023social}, these experiments could only proceed with negotiated support from Meta's executives---an arrangement that may limit the scope and independence of scientific work. Our approach enables scientists to test a much broader set of interventions beyond those that platforms might agree to test, fostering a more robust ecosystem for independent social media research.

\subsection*{Limitations and future directions}
Our study focused on a sample of partisan U.S. users restricted to the web version of X during the months leading up to a U.S. presidential election. This context, marked by a high volume of political content and emotional engagement, enabled us to test interventions effectively, but may limit the generalizability of the findings~\cite{budak2024misunderstanding,eckles2022algorithmic} beyond a period of heightened political activity. Our population is limited to X users with at least 5\% of political content, but we intentionally selected this group to ensure the interventions were viable. Additionally, the experiment examines short-term effects---immediate in-feed responses and post-survey responses measured soon after the experiment---but does not address whether these effects persist over time. In additional analysis examining the variation of the treatment effects by experiment day and the time gap between the end of the experiment and when the participants started the post-survey, we did not find any evidence for immediate decay of the observed effects~\Appref{sec:time_gap}. We also note that in scenarios where this ranking approach is integrated into the feed algorithm, the intervention would be continuously applied, and any effects may dissipate once it is turned off. Other technical limitations of our approach include the difficulty in performing cluster-randomized experiments testing for social influence~\cite{eckles2017design}, and the inability to collect the full feed inventory for a given user. 

Our reliance on in-feed surveys to assess immediate responses to stimuli follows established practices of ecological momentary assessment~\cite{shiffman2008ecological}. To mitigate the risk of demand characteristics associated with the close proximity of the assessment, we followed best practices in designing the experiment~\cite{zizzo2010experimenter}, providing only essential information about the study's purpose. Post-survey responses to an open-ended question about participants'  experiences during the study suggest that these precautions were effective, as they did not report any impact on their experience attributable to their assigned condition~\Appref{sup:impact}. 

Describing the causal relation between social media and polarization as a unified phenomenon remains challenging: platforms are complex socio-technical systems, and small variations in design or exogenous factors can lead to different outcomes~\cite{yarchi2021political}. Future longitudinal research should investigate the impact of sustaining the intervention over longer periods of time---especially on long-term metrics such as user retention~\cite{KGI2025betterfeeds} and news knowledge---and evaluate the efficacy of feed reranking across different platforms and cultural settings. For example, open questions remain about how our approach can be extended to the broader online information ecosystem and if our findings hold true in different political systems with unique cultural contexts or party structures. In addition, future research can move beyond the initial eight factors related to antidemocratic attitudes and operationalize new factors grounded in other social science theories~\cite{bernstein2023embedding}, and measure how this content can impact other experiences online, such as mental health, trust in institutions, civic engagement, or the quality of democratic discourse. Using a similar design as our study, these future investigations can also examine different time horizons, from immediate reactions captured through in-feed surveys to long-term impacts assessed through follow-up surveys conducted days or even months later.

As political polarization and societal division become increasingly linked to social media activity, our findings provide a potential pathway for platforms to address these challenges through algorithmic interventions. Together, these interventions may result in algorithms that not only reduce partisan animosity but also promote greater social trust and healthier democratic discourse across party lines~\cite{bernstein2023embedding,jia2024embedding}.

\section*{Materials and Methods Summary}

\paragraph{Participant recruitment.}
We recruited participants through two online platforms, CloudResearch and Bovitz, targeting U.S. residents over 18 years old who self-identified as either Republican or Democrat and were active users of X (Section~\ref{methods:participants}). Qualified individuals were invited to complete a screening task, which included installing a browser extension that analyzed their X feed. To ensure the interventions could have a meaningful impact on participants’ feeds, only those with at least 5\% of posts related to politics or social issues were invited to participate. Figure~\ref{fig:study_flow} summarizes the recruitment funnel, including the number of individuals at each stage of the process. Participants were not instructed to use X in any particular way, but they received daily reminders if they had not used the platform that day.

\paragraph{Surveys.}
Participants were required to complete a pre-experiment survey that included questions about their demographics, personal values, affective experiences on social media, and political attitudes (Section~\ref{method:survey}). At the end of the study, they were asked to complete a post-experiment survey with a similar structure, but without questions on demographics or personal values. The surveys took about 20 and 10 minutes to complete, respectively. The section of the survey measuring political attitudes included a question related to our primary outcome, assessing participants’ feelings toward out-partisans, while the section on affective experiences on social media included questions related to our secondary outcomes, measuring how often participants felt angry, sad, excited, and calm.

Throughout the experiment, we also inserted short surveys into participants’ feeds to measure their feelings in the moment. Each in-feed survey included either one question about participants’ feelings toward members of the out-party or two questions asking how much they felt one positive emotion (excited or calm) and one negative emotion (angry or sad), randomly selected. To minimize survey fatigue, once a participant answered a question, they were not shown the same type of question for the next ten minutes and were subsequently less likely to receive that question again.

\paragraph{Feed reranking implementation.}
To administer the interventions, we developed a browser extension that intercepted the participants’ original X feeds, sent the posts to a remote backend for real-time analysis, and displayed the reranked feeds (Section~\ref{subsec:feed-reranking-implementation}). The backend used a fast, custom-built classifier to detect political posts (Section~\ref{sup:political_classifier}), and OpenAI's GPT-3.5 Turbo model to score the political posts on the eight AAPA dimensions (Section~\ref{sup:scoring}). The reranking process took approximately three seconds and was consistent across all experimental conditions. The extension also administered in-feed surveys and logged participants’ engagement with the feed, including which posts they viewed, liked, and reposted.

\paragraph{Experimental design.}
We used Bernoulli randomization to assign participants to either the Reduced or Increased Exposure experiment, and subsequently to treatment and control groups with equal probability (Section~\ref{sec:experimental-design}). Due to differences in how the two interventions were operationalized, each experiment had its own control group. To determine the sample size, we conducted a power analysis based on data from a pilot study with 70 participants that followed the same experimental design and analysis procedure. We chose target quotas that enabled us to detect effects of at least two degrees on the feeling thermometer scale for affective polarization with 95\% power. We used randomization inference to test for imbalances in participants’ characteristics, differential attrition rates, and differential attrition patterns between the treatment and control groups in each experiment. We found no evidence of covariate imbalance or differential attrition.

\paragraph{Analysis.}
As preregistered, to estimate the effects of the interventions on the post-experiment outcomes, we used ordinary least squares, regressing the outcome variable on the treatment assignment indicator, the participant’s response to the same question on the pre-survey, and a recruitment platform indicator (Section~\ref{method:analysis}). To estimate the effects on the outcomes measured with in-feed surveys, we used linear mixed-effects models to account for the multiple observations per participant. We regressed the outcome on the treatment assignment, the participant’s average response on the same question during the three-day baseline period, and the recruitment platform indicator. To estimate heterogeneous treatment effects, we used the same model specifications but included an interaction between the treatment assignment and the moderator, running one regression per moderator (Section~\ref{sup:heterogeneous}). As preregistered, we considered party identification, party strength, and the proportion of political content in the participant’s feed during the baseline period as primary moderators, and gender, race, age, income, and socioeconomic ladder as secondary moderators. To account for testing multiple hypotheses, we applied False Discovery Rate adjustment (Section~\ref{sec:p-values-corrections}). Finally, to appropriately model the nature of the outcomes, we used quasi-Poisson models to estimate the effects on engagement volume and Beta models to estimate the effects on engagement rates (Section~\ref{sup:engagement}).

\clearpage %

\bibliographystyle{sciencemag}
\bibliography{science_template}

\begin{thebibliography}{10}
\providecommand{\url}[1]{\texttt{#1}}
\expandafter\ifx\csname urlstyle\endcsname\relax
  \providecommand{\doi}[1]{doi:\discretionary{}{}{}#1}\else
  \providecommand{\doi}{doi:\discretionary{}{}{}\begingroup \urlstyle{rm}\Url}\fi

\bibitem{bakshy2015exposure}
E.~Bakshy, S.~Messing, L.~A. Adamic, Exposure to ideologically diverse news and opinion on Facebook. \emph{Science} \textbf{348}~(6239), 1130--1132 (2015).

\bibitem{brady2023overperception}
W.~J. Brady, \emph{et~al.}, Overperception of moral outrage in online social networks inflates beliefs about intergroup hostility. \emph{Nature human behaviour} \textbf{7}~(6), 917--927 (2023).

\bibitem{lee2018does}
C.~Lee, J.~Shin, A.~Hong, Does social media use really make people politically polarized? Direct and indirect effects of social media use on political polarization in South Korea. \emph{Telematics and Informatics} \textbf{35}~(1), 245--254 (2018).

\bibitem{feezell2021exploring}
J.~T. Feezell, J.~K. Wagner, M.~Conroy, Exploring the effects of algorithm-driven news sources on political behavior and polarization. \emph{Computers in human behavior} \textbf{116}, 106626 (2021).

\bibitem{kramer2014experimental}
A.~D. Kramer, J.~E. Guillory, J.~T. Hancock, Experimental evidence of massive-scale emotional contagion through social networks. \emph{Proceedings of the National academy of Sciences of the United States of America} \textbf{111}~(24), 8788 (2014).

\bibitem{brady2021social}
W.~J. Brady, K.~McLoughlin, T.~N. Doan, M.~J. Crockett, How social learning amplifies moral outrage expression in online social networks. \emph{Science Advances} \textbf{7}~(33), eabe5641 (2021).

\bibitem{allcott2020welfare}
H.~Allcott, L.~Braghieri, S.~Eichmeyer, M.~Gentzkow, The welfare effects of social media. \emph{American economic review} \textbf{110}~(3), 629--676 (2020).

\bibitem{bond201261}
R.~M. Bond, \emph{et~al.}, A 61-million-person experiment in social influence and political mobilization. \emph{Nature} \textbf{489}~(7415), 295--298 (2012).

\bibitem{tufekci2017twitter}
Z.~Tufekci, \emph{Twitter and tear gas: The power and fragility of networked protest} (Yale University Press, USA) (2017).

\bibitem{jost2018social}
J.~T. Jost, \emph{et~al.}, How social media facilitates political protest: Information, motivation, and social networks. \emph{Political psychology} \textbf{39}, 85--118 (2018).

\bibitem{smith2023digital}
L.~G. Smith, L.~Piwek, J.~Hinds, O.~Brown, A.~Joinson, Digital traces of offline mobilization. \emph{Journal of Personality and Social Psychology} \textbf{125}~(3), 496 (2023).

\bibitem{zuckerberg2021us}
M.~Zuckerberg, J.~Dorsey, S.~Pichai, US House Hearing on" Disinformation Nation: Social Media's Role in Promoting Extremism and Misinformation"  (2021).

\bibitem{rose2023outrage}
T.~Rose-Stockwell, \emph{Outrage machine: How tech amplifies discontent, disrupts democracy—And what we can do about it} (Hachette UK, UK) (2023).

\bibitem{lorenz2023systematic}
P.~Lorenz-Spreen, L.~Oswald, S.~Lewandowsky, R.~Hertwig, A systematic review of worldwide causal and correlational evidence on digital media and democracy. \emph{Nature human behaviour} \textbf{7}~(1), 74--101 (2023).

\bibitem{tucker2018social}
J.~A. Tucker, \emph{et~al.}, Social media, political polarization, and political disinformation: A review of the scientific literature. \emph{Political polarization, and political disinformation: a review of the scientific literature (March 19, 2018)}  (2018).

\bibitem{oldemburgo2024twitter}
V.~Oldemburgo~de Mello, F.~Cheung, M.~Inzlicht, Twitter (X) use predicts substantial changes in well-being, polarization, sense of belonging, and outrage. \emph{Communications Psychology} \textbf{2}~(1), 15 (2024).

\bibitem{beam2020facebook}
M.~A. Beam, M.~J. Hutchens, J.~D. Hmielowski, Facebook news and (de) polarization: Reinforcing spirals in the 2016 US election, in \emph{Digital media, political polarization and challenges to democracy} (Routledge, US), pp. 26--44 (2020).

\bibitem{haidt_bail_social_media}
J.~Haidt, C.~Bail, Social media and political dysfunction: A collaborative review (ongoing), \url{https://heystacks.com/doc/1185/social-media-and-political-dysfunction}, unpublished manuscript.

\bibitem{iyengar2019origins}
S.~Iyengar, Y.~Lelkes, M.~Levendusky, N.~Malhotra, S.~J. Westwood, The origins and consequences of affective polarization in the United States. \emph{Annual review of political science} \textbf{22}~(1), 129--146 (2019).

\bibitem{boxell2024cross}
L.~Boxell, M.~Gentzkow, J.~M. Shapiro, Cross-country trends in affective polarization. \emph{Review of Economics and Statistics} \textbf{106}~(2), 557--565 (2024).

\bibitem{iyengar2018strengthening}
S.~Iyengar, M.~Krupenkin, The strengthening of partisan affect. \emph{Political Psychology} \textbf{39}, 201--218 (2018).

\bibitem{saveski2022perspective}
M.~Saveski, N.~Gillani, A.~Yuan, P.~Vijayaraghavan, D.~Roy, Perspective-taking to reduce affective polarization on social media, in \emph{Proceedings of the International AAAI Conference on Web and Social Media}, vol.~16 (2022), pp. 885--895.

\bibitem{nyhan2023like}
B.~Nyhan, \emph{et~al.}, Like-minded sources on Facebook are prevalent but not polarizing. \emph{Nature} \textbf{620}~(7972), 137--144 (2023).

\bibitem{frimer2023incivility}
J.~A. Frimer, \emph{et~al.}, Incivility is rising among American politicians on Twitter. \emph{Social Psychological and Personality Science} \textbf{14}~(2), 259--269 (2023).

\bibitem{druckman2023does}
J.~N. Druckman, D.~P. Green, S.~Iyengar, Does Affective Polarization Contribute to Democratic Backsliding in America? \emph{The ANNALS of the American Academy of Political and Social Science} \textbf{708}~(1), 137--163 (2023).

\bibitem{stray2021designing}
J.~Stray, Designing recommender systems to depolarize. \emph{arXiv preprint arXiv:2107.04953}  (2021).

\bibitem{barnidge2017exposure}
M.~Barnidge, Exposure to political disagreement in social media versus face-to-face and anonymous online settings. \emph{Political communication} \textbf{34}~(2), 302--321 (2017).

\bibitem{bor2022psychology}
A.~Bor, M.~B. Petersen, The psychology of online political hostility: A comprehensive, cross-national test of the mismatch hypothesis. \emph{American political science review} \textbf{116}~(1), 1--18 (2022).

\bibitem{banks2021polarizedfeeds}
A.~Banks, E.~Calvo, D.~Karol, S.~Telhami, \# polarizedfeeds: Three experiments on polarization, framing, and social media. \emph{The International Journal of Press/Politics} \textbf{26}~(3), 609--634 (2021).

\bibitem{rathje2021out}
S.~Rathje, J.~J. Van~Bavel, S.~Van Der~Linden, Out-group animosity drives engagement on social media. \emph{Proceedings of the National Academy of Sciences} \textbf{118}~(26), e2024292118 (2021).

\bibitem{van2021social}
J.~J. Van~Bavel, S.~Rathje, E.~Harris, C.~Robertson, A.~Sternisko, How social media shapes polarization. \emph{Trends in Cognitive Sciences} \textbf{25}~(11), 913--916 (2021).

\bibitem{suhay2018polarizing}
E.~Suhay, E.~Bello-Pardo, B.~Maurer, The polarizing effects of online partisan criticism: Evidence from two experiments. \emph{The International Journal of Press/Politics} \textbf{23}~(1), 95--115 (2018).

\bibitem{kim2021distorting}
J.~W. Kim, A.~Guess, B.~Nyhan, J.~Reifler, The distorting prism of social media: How self-selection and exposure to incivility fuel online comment toxicity. \emph{Journal of Communication} \textbf{71}~(6), 922--946 (2021).

\bibitem{cho2020search}
J.~Cho, S.~Ahmed, M.~Hilbert, B.~Liu, J.~Luu, Do search algorithms endanger democracy? An experimental investigation of algorithm effects on political polarization. \emph{Journal of Broadcasting \& Electronic Media} \textbf{64}~(2), 150--172 (2020).

\bibitem{levy2021social}
R.~Levy, Social media, news consumption, and polarization: Evidence from a field experiment. \emph{American economic review} \textbf{111}~(3), 831--870 (2021).

\bibitem{bail2018exposure}
C.~A. Bail, \emph{et~al.}, Exposure to opposing views on social media can increase political polarization. \emph{Proceedings of the National Academy of Sciences} \textbf{115}~(37), 9216--9221 (2018).

\bibitem{guess2023social}
A.~M. Guess, \emph{et~al.}, How do social media feed algorithms affect attitudes and behavior in an election campaign? \emph{Science} \textbf{381}~(6656), 398--404 (2023).

\bibitem{bernstein2023embedding}
M.~Bernstein, \emph{et~al.}, Embedding Societal Values into Social Media Algorithms. \emph{Journal of Online Trust and Safety} \textbf{2}~(1) (2023).

\bibitem{voelkel2024megastudy}
J.~G. Voelkel, \emph{et~al.}, Megastudy testing 25 treatments to reduce antidemocratic attitudes and partisan animosity. \emph{Science} \textbf{386}~(6719), eadh4764 (2024).

\bibitem{jia2024embedding}
C.~Jia, M.~S. Lam, M.~C. Mai, J.~T. Hancock, M.~S. Bernstein, Embedding democratic values into social media AIs via societal objective functions. \emph{Proceedings of the ACM on Human-Computer Interaction} \textbf{8}~(CSCW1), 1--36 (2024).

\bibitem{piccardi2024reranking}
T.~Piccardi, \emph{et~al.}, Reranking Social Media Feeds: A Practical Guide for Field Experiments. \emph{arXiv preprint arXiv:2406.19571}  (2024).

\bibitem{mcclain2024americans}
C.~McClain, M.~Anderson, R.~Gelles-Watnick, How Americans Navigate Politics on TikTok, X, Facebook and Instagram  (2024).

\bibitem{iyengar2012affect}
S.~Iyengar, G.~Sood, Y.~Lelkes, Affect, not ideology: A social identity perspective on polarization. \emph{Public opinion quarterly} \textbf{76}~(3), 405--431 (2012).

\bibitem{russell1980circumplex}
J.~A. Russell, A circumplex model of affect. \emph{Journal of personality and social psychology} \textbf{39}~(6), 1161 (1980).

\bibitem{bestvater2022politics}
S.~Bestvater, S.~Shah, G.~River, A.~Smith, Politics on twitter: One-third of tweets from us adults are political  (2022).

\bibitem{shiffman2008ecological}
S.~Shiffman, A.~A. Stone, M.~R. Hufford, Ecological momentary assessment. \emph{Annu. Rev. Clin. Psychol.} \textbf{4}, 1--32 (2008).

\bibitem{stagnaro2024representativeness}
M.~N. Stagnaro, \emph{et~al.}, Representativeness versus response quality: Assessing nine opt-in online survey samples. \emph{OSF Preprints} \textbf{2} (2024).

\bibitem{piccardi2024feedrankings}
T.~Piccardi, \emph{et~al.}, Do Feed Rankings Promoting or Demoting Political Distrust and Animosity Impact Affective Experience and Polarization?, OSF (2024), \url{https://doi.org/10.17605/OSF.IO/QY9AX}.

\bibitem{santoro2022promise}
E.~Santoro, D.~E. Broockman, The promise and pitfalls of cross-partisan conversations for reducing affective polarization: Evidence from randomized experiments. \emph{Science advances} \textbf{8}~(25), eabn5515 (2022).

\bibitem{landry2024partisan}
E.~J. Finkel, A.~P. Landry, J.~Druckman~N., J.~J. Van~Bavel, R.~H. Hoyle, Partisan Antipathy and the Erosion of Democratic Norms  (2024).

\bibitem{wagner2023independence}
M.~W. Wagner, Independence by permission. \emph{Science} \textbf{381}~(6656), 388--391 (2023).

\bibitem{budak2024misunderstanding}
C.~Budak, B.~Nyhan, D.~M. Rothschild, E.~Thorson, D.~J. Watts, Misunderstanding the harms of online misinformation. \emph{Nature} \textbf{630}~(8015), 45--53 (2024).

\bibitem{eckles2022algorithmic}
D.~Eckles, Testimony before the Senate Subcommittee on Communications, Media, and Broadband: Algorithmic transparency and assessing effects of algorithmic ranking  (2022), \url{https://www.commerce.senate.gov/services/files/62102355-DC26-4909-BF90-8FB068145F18}.

\bibitem{eckles2017design}
D.~Eckles, B.~Karrer, J.~Ugander, Design and analysis of experiments in networks: Reducing bias from interference. \emph{Journal of Causal Inference} \textbf{5}~(1), 20150021 (2017).

\bibitem{zizzo2010experimenter}
D.~J. Zizzo, Experimenter demand effects in economic experiments. \emph{Experimental Economics} \textbf{13}, 75--98 (2010).

\bibitem{yarchi2021political}
M.~Yarchi, C.~Baden, N.~Kligler-Vilenchik, Political polarization on the digital sphere: A cross-platform, over-time analysis of interactional, positional, and affective polarization on social media. \emph{Political Communication} \textbf{38}~(1-2), 98--139 (2021).

\bibitem{KGI2025betterfeeds}
A.~Moehring, \emph{et~al.}, Better Feeds: Algorithms That Put People First  (2025), \url{https://kgi.georgetown.edu/research-and-commentary/better-feeds/}.

\bibitem{piccardi2025reranking}
T.~Piccardi, \emph{et~al.}, Reranking partisan animosity in algorithmic social media feeds alters affective polarization (2025), \url{https://doi.org/10.5061/dryad.hmgqnk9tj}.

\bibitem{schwartz2012refining}
S.~H. Schwartz, \emph{et~al.}, Refining the theory of basic individual values. \emph{Journal of personality and social psychology} \textbf{103}~(4), 663 (2012).

\bibitem{graham2013moral}
J.~Graham, \emph{et~al.}, Moral foundations theory: The pragmatic validity of moral pluralism, in \emph{Advances in experimental social psychology} (Elsevier, US), vol.~47, pp. 55--130 (2013).

\bibitem{lin2016standard}
W.~Lin, D.~Green, A.~Coppock, Standard operating procedures for Don Green’s lab at Columbia. \emph{Github} pp. 1--31 (2016).

\bibitem{lalmas2014measuring}
M.~Lalmas, H.~O'Brien, E.~Yom-Tov, \emph{Measuring user engagement} (Morgan \& Claypool Publishers, US) (2014).

\bibitem{twitter_algorithm_ml}
Twitter, The Algorithm - ML, \url{https://github.com/twitter/the-algorithm-ml} (2023), accessed: 2024-09-29.

\bibitem{straub2024public}
V.~J. Straub, J.~W. Burton, M.~Geers, P.~Lorenz-Spreen, Public attitudes towards social media field experiments. \emph{Scientific Reports} \textbf{14}~(1), 26110 (2024).

\bibitem{fiske2022twitter}
S.~T. Fiske, Twitter manipulates your feed: Ethical considerations (2022).

\bibitem{levitsky2019democracies}
S.~Levitsky, D.~Ziblatt, \emph{How democracies die} (Crown, US) (2019).

\bibitem{svolik2018polarization}
M.~Svolik, When polarization trumps civic virtue: Partisan conflict and the subversion of democracy by incumbents. \emph{Available at SSRN 3243470}  (2018).

\bibitem{kalmoe2022radical}
N.~P. Kalmoe, L.~Mason, Radical American partisanship: Mapping violent hostility, its causes, and the consequences for democracy, in \emph{Radical American Partisanship} (University of Chicago Press, US) (2022).

\bibitem{graham2020democracy}
M.~H. Graham, M.~W. Svolik, Democracy in America? Partisanship, polarization, and the robustness of support for democracy in the United States. \emph{American Political Science Review} \textbf{114}~(2), 392--409 (2020).

\bibitem{hetherington2020washington}
M.~J. Hetherington, T.~J. Rudolph, \emph{Why Washington won't work: Polarization, political trust, and the governing crisis} (University of Chicago Press, US) (2020).

\bibitem{putnam2000bowling}
R.~D. Putnam, Bowling alone: America’s declining social capital, in \emph{Culture and politics: A reader} (Springer, US), pp. 223--234 (2000).

\bibitem{bishop2009big}
B.~Bishop, R.~G. Cushing, \emph{The big sort: Why the clustering of like-minded America is tearing us apart} (Houghton Mifflin Harcourt, US) (2009).

\bibitem{nyhan2010corrections}
B.~Nyhan, J.~Reifler, When corrections fail: The persistence of political misperceptions. \emph{Political Behavior} \textbf{32}~(2), 303--330 (2010).

\bibitem{barbera2015tweeting}
P.~Barber{\'a}, J.~T. Jost, J.~Nagler, J.~A. Tucker, R.~Bonneau, Tweeting from left to right: Is online political communication more than an echo chamber? \emph{Psychological science} \textbf{26}~(10), 1531--1542 (2015).

\bibitem{barbera2015birds}
P.~Barber{\'a}, Birds of the same feather tweet together: Bayesian ideal point estimation using Twitter data. \emph{Political analysis} \textbf{23}~(1), 76--91 (2015).

\bibitem{yang2022botometer}
K.-C. Yang, E.~Ferrara, F.~Menczer, Botometer 101: Social bot practicum for computational social scientists. \emph{Journal of computational social science} \textbf{5}~(2), 1511--1528 (2022).

\bibitem{anderson2008multiple}
M.~L. Anderson, Multiple inference and gender differences in the effects of early intervention: A reevaluation of the Abecedarian, Perry Preschool, and Early Training Projects. \emph{Journal of the American statistical Association} \textbf{103}~(484), 1481--1495 (2008).

\bibitem{smithson2006better}
M.~Smithson, J.~Verkuilen, A better lemon squeezer? Maximum-likelihood regression with beta-distributed dependent variables. \emph{Psychological methods} \textbf{11}~(1), 54 (2006).

\bibitem{kish1965survey}
L.~Kish, \emph{Survey sampling.} (John Wiley and Sons, USA) (1965).

\bibitem{rsurvey2024}
T.~Lumley, survey: analysis of complex survey samples (2024), r package version 4.4.

\bibitem{lumley2010}
T.~Lumley, \emph{Complex Surveys: A Guide to Analysis Using R: A Guide to Analysis Using R} (John Wiley and Sons, USA) (2010).

\end{thebibliography}
\clearpage

\section*{Acknowledgments}
We thank Pablo Barber\'{a}, Dean Eckles, Matthew Gentzkow, Desmond Ong, Jennifer Pan, Sandy Pentland, Johan Ugander, Robb Willer, and the anonymous reviewers for helpful feedback and discussions.  Michael Bernstein discloses an additional affiliation as a co-founder of Simile AI, Inc., which builds AI human behavioral simulations.

\paragraph*{Funding:}
This work was supported in part by the NSF under awards IIS-2403433, IIS-2403434, IIS-2403435; BCS-2214203; Swiss NSF under award P500PT-206953, and a Hoffman-Yee grant from the Stanford Institute for Human-Centered Artificial Intelligence. 

\paragraph*{Author contributions:} 

Conceptualization: T.P., M.S., C.J., J.T., J.H., M.B.

Methodology: T.P., M.S., C.J., J.T., J.H., M.B.

Software: T.P.

Validation: T.P., M.S.

Formal analysis: T.P., M.S.

Investigation: T.P., M.S.

Resources: T.P., M.S., C.J.

Data curation: T.P., M.S.

Writing – original draft: T.P., M.S., C.J., M.B.

Writing – review \& editing: T.P., M.S., C.J., J.H., J.T., M.B.

Supervision: J.T., J.H., M.B.

Project administration: T.P., M.S., C.J., M.B.

Funding acquisition: M.S., C.J., J.H., M.B

\paragraph*{Competing interests:}
The authors declare that they have no competing interests.

\paragraph*{Data and materials availability:}
The repository associated with the manuscript, containing aggregated data and code required to reproduce the paper results, is available at: \cite{piccardi2025reranking}

The pre-registration of the study is available at: \cite{piccardi2024feedrankings}

\newpage

\renewcommand{\thefigure}{S\arabic{figure}}
\renewcommand{\thetable}{S\arabic{table}}
\renewcommand{\theequation}{S\arabic{equation}}
\renewcommand{\thepage}{S\arabic{page}}
\setcounter{figure}{0}
\setcounter{table}{0}
\setcounter{equation}{0}
\setcounter{page}{1} %

\begin{center}
\section*{Supplementary Materials for\\ \scititle}

\author{
	Tiziano~Piccardi$^{\ast\dagger}$,
	Martin~Saveski$^{\ast\dagger}$,
	Chenyan~Jia$^{\dagger}$, \\
        Jeffrey~Hancock,
        Jeanne~Tsai,
        Michael~Bernstein\and
        
	\small$^\ast$Corresponding author. Email: piccardi@jhu.edu (T.P.); msaveski@uw.edu (M.S.)\and

    \small$^\dagger$These authors contributed equally to this work.
}
\end{center}

\subsubsection*{This PDF file includes:}
Methods and Materials\\
Ethical Considerations\\
Supplementary Text\\
Figs. S1 to S24\\
Tables S1 to S32\\
References (59-81)

\newpage

\clearpage
\section*{Supplementary Material}
\renewcommand{\thesection}{S\arabic{section}}
\setcounter{section}{0} %

\section{Methods}
\label{methods}

\subsection{Participant Recruitment}
\label{methods:participants}
\begin{figure}[htbp!]
\centering
\includegraphics[width=0.90\textwidth]{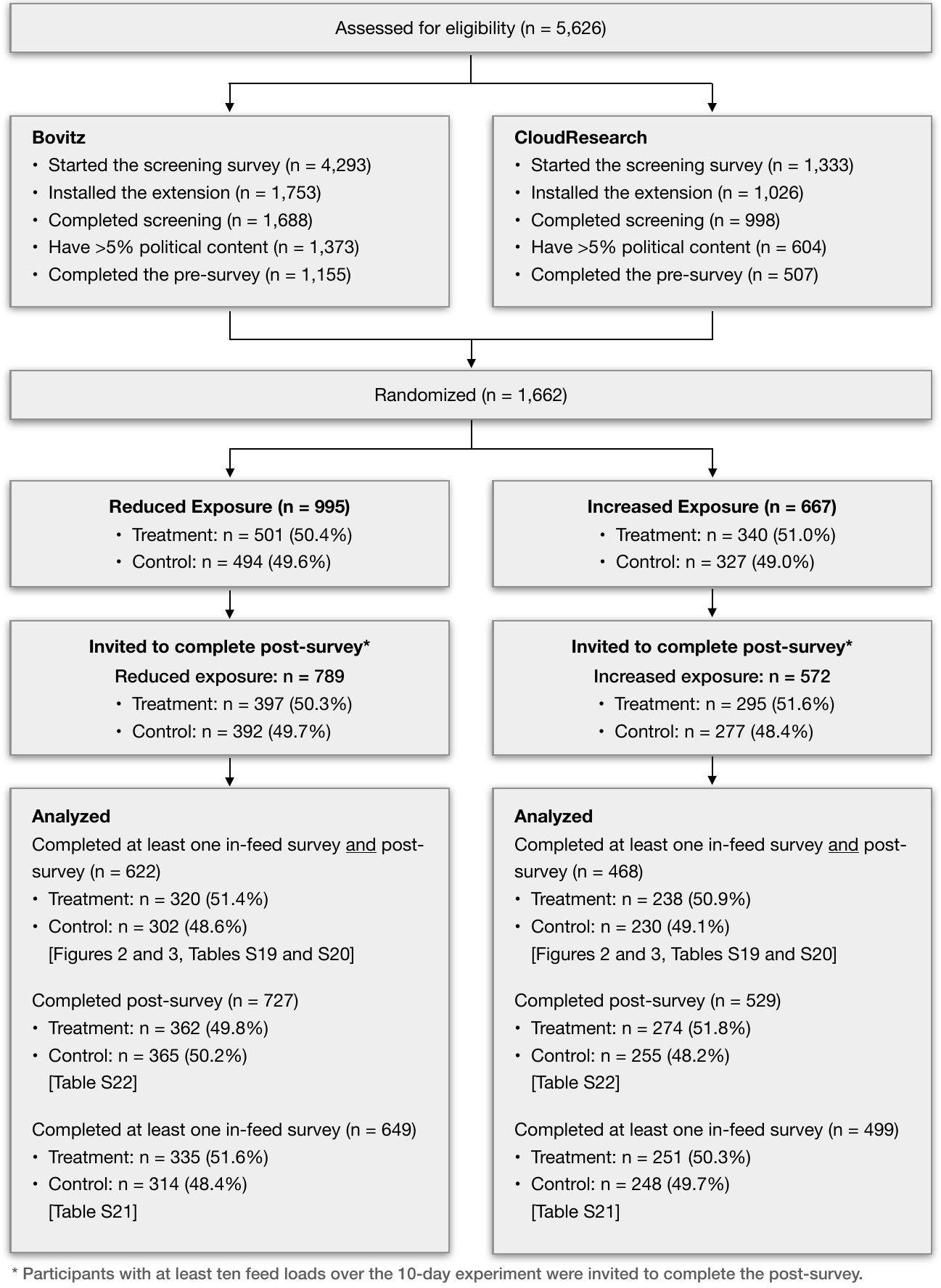}
\caption{Summary of the participants' recruitment funnel. }
\label{fig:study_flow}
\end{figure}

We recruited participants from two online platforms, Bovitz and CloudResearch. Using platform-specific filters, we targeted individuals who met the following criteria: US residents aged 18 or older, self-identifying as either Republican or Democrat, and active users of X. Participants were invited to complete a two-minute screening task to assess their eligibility for the study. The experiment was restricted to desktop users on Google Chrome or Microsoft Edge browsers. Upon accepting the consent form, participants were instructed to install a web extension that launched X in a new browser window, automatically scrolled through their ``For you'' feed to capture approximately 180 posts, and calculated the proportion of posts related to politics or social issues using our classifier \Appref{sup:political_classifier}. Only participants with at least 5\% of posts related to politics or social issues were invited to join the study, ensuring that participants regularly engaged with political content, a necessary condition for the intervention to have an impact on their feed. Participants were compensated \$0.50 for completing the screening task and \$20 upon finishing the study. \Figref{fig:study_flow} provides a visual summary of the full recruitment process and the number of participants at each step.

On Bovitz, 4,293 participants initiated the screening task, and 1,753 successfully installed the extension and accessed X. The observed drop in participation was primarily due to the use of unsupported devices or browsers (e.g., mobile). Participants for whom we collected fewer than 100 posts due to technical issues (e.g., conflicts with other browser extensions) or who deliberately interrupted the screening were excluded from the study, leaving 1,688 eligible participants. After applying the 5\% political content filter, 1,373 participants were invited to continue with the study, of which 1,155 completed the pre-experiment survey and were successfully enrolled. Of these, 720 were randomly assigned to the Reduced Exposure experiment (368 in the treatment and 352 in the control group) and 435 to the Increased Exposure experiment (223 in the treatment and 212 in the control group), as our pilot study revealed that the first group required more participants to detect the effect of the intervention~\cite{piccardi2024feedrankings}.

On CloudResearch, 1,333 participants began the screening, and 1,036 successfully installed the extension and accessed X. The lower dropout rate compared to Bovitz is attributed to the platform's use of more specific filters, such as restricting participation to desktop users and highlighting the need for software installation, as well as advanced filters targeting active X users. After accounting for successful screening completion (998 participants) and applying the 5\% political content criterion, 604 participants were invited to the study. Among these, 507 participants joined, with 275 in the Reduced Exposure experiment (133 in the treatment and 142 in the control group) and 232 in the Increased Exposure experiment (117 in the treatment and 115 in the control group).

\Figref{fig:political_distribution} shows the distribution of political posts among participants before the filtering (mean~=~17.1\%, median~=~12.1\%), illustrating that the 5\% political content filter excluded 26.5\% of the initial participant pool. Participants were invited to engage daily and received reminders via email. After ten days, participants with minimal participation, i.e., those for whom we recorded fewer than ten feed load events, were excluded from the study and did not receive the post-experiment survey. The final completion rates were 69.1\% (798) for Bovitz and 90.3\% (458) for CloudResearch.

Of the participants who completed the study, 602 (47.9\%) identified as female, 631 (50.2\%) as male, and 23 (1.9\%) as other gender identities. In terms of political affiliation, 831 (66.1\%) were Democrats, and 425 (33.9\%) were Republicans. \Figref{fig:participants_demographics} summarizes additional demographics such as race, education level, income, and perceived socioeconomic status.

\subsection{Surveys}
\label{method:survey}
Participants were required to complete a pre-experiment survey, which lasted approximately 20 minutes. This survey collected demographic information, personal values~\cite{schwartz2012refining,graham2013moral}, affective experiences on social media, and political attitudes such as feelings toward members of the opposing political party and the remaining seven measures of antidemocratic attitudes. The political feeling question was asked first, whereas the order of the other questions was randomized. To ensure contextual relevance, survey questions were adapted based on participants' political leanings. We followed the structure of the survey used in the study that introduced the eight AAPA variables~\cite{voelkel2024megastudy}. At the end of the study, participants completed a post-experiment survey, which took approximately 10 minutes. This survey asked about political attitudes and affective experiences, excluding demographics and personal values. Political attitudes were assessed using a feeling thermometer on a scale from ``Very cold or unfavorable feeling" (0) to ``Very warm or favorable feeling" (100). To evaluate emotions, participants rated how frequently they felt 15 different emotions on a scale from ``Never'' (0) to ``All the time" (5).

For the in-feed response, the survey for the political feeling assessment asked ``At the moment, how do you feel about [out-party]?" where out-party is replaced with ``Republicans'' for Democratic participants and vice versa. For emotional responses, the survey asks, ``How much do you feel" for one negative (sad or angry) and one positive (excited or calm) emotion. These in-feed surveys were designed to resemble posts with a yellow background. Both political feelings and emotions were measured on a scale from 0 to 100, and participants clicked on a slider to indicate their response. The numeric value was not displayed to prevent anchoring effects, and the slider was initially shown empty with no default setting. For political feelings, the slider was labeled with ``Very cold or unfavorable feeling" (0), ``No feeling at all" (50), and ``Very warm or favorable feeling" (100). For emotions, the labels were ``None at all" (0), ``A little" (25), ``Moderately" (50), ``A lot" (75), and ``Extremely" (100). 
To avoid survey fatigue, the survey was not shown with every post. Each day, participants had an equal probability (50\%) of receiving either a political feeling or emotion question. Once a survey was answered, the same type of question would not appear for the next 10 minutes, and the likelihood of receiving another survey that day would be halved. When selected, the survey was inserted into the feed according to the logic described in the following sections.

\subsection{Experimental Design} 
\label{sec:experimental-design}

\xhdr{Randomization} We used Bernoulli randomization to allocate participants to the experiments with either Reduced or Increased Exposure to AAPA content until we reached the respective target quotas from our power analyses~\cite{piccardi2024feedrankings} and to a treatment arm (treatment or control) with an equal probability. 
We opted for a Bernoulli randomization since we did not have access to the complete participant pool and their characteristics in advance. 

\xhdr{Control groups}
We ran the Reduced and Increased Exposure experiments as separate, parallel studies, each with its own control group. This was necessary because the interventions were operationalized differently. In the Increased Exposure experiment, we could uprank AAPA content at any point in the feed, allowing for consistent placement of in-feed surveys. In the Reduced Exposure experiment, by contrast, we could only intervene when AAPA content was already present, which limited both the intervention and the placement of surveys.
To ensure internal validity, we designed each control group to mirror the structural features of its corresponding treatment in terms of survey timing and content availability.

\xhdr{Power analyses}
To inform the sample size for our experiments, we ran a power analysis based on data from a pilot study with 70 participants that followed the same experimental design and analysis procedure. Due to the complexity of the mixed-effects models, we used simulations to conduct the power analyses (using the \verb|simR| package) with models that correspond to those described in Section \ref{method:analysis}. We optimized for power in detecting effects on our primary dependent variables, i.e., the in-feed and post-experiment effects on affective polarization. 

Due to the uncertainty in the recruitment process, we preregistered two sample size targets that, based on a pilot study, would enable us to detect effects of 2 and 2.5 degrees on the feeling thermometer scale with 95\% power. We were able to reach and slightly exceed the required sample size to detect effects of at least 2 degrees in both the Reduced and Increased Exposure experiments. 

\xhdr{Covariate balance} We used randomization inference to test whether the imbalance in the background characteristics of the participants assigned to treatment vs. control in each of the two experiments was larger than we would expect from chance~\cite{lin2016standard}. We regressed the treatment indicator on all background characteristics and computed a heteroskedasticity-robust Wald statistic for the hypothesis that all the coefficients are equal to zero. Then, we sampled 10,000 assignments using the same randomization procedure, performed the same regression, and compared the Wald statistics to the observed. 

We considered all questions in the pre-experiment survey as covariates, including demographics (gender, race/ethnicity, age, education, SES ladder, and income), party affiliation (party ID, party ID strength, party importance), affective polarization (in- and out-party feeling thermometer), antidemocratic attitudes (support for undemocratic practices, support for undemocratic candidates, support for partisan violence, biased evaluation of politicized facts, social distance, social distrust, opposition to bipartisan cooperation), emotions (enthusiastic, happy, still, lonely, sad, nervous, satisfied, calm, relaxed, tired, fearful, aroused, excited, bored, and angry), and recruitment platform. We coarsened the education, income, and SES ladder covariates to avoid small categories. For the sample of participants that completed enough of the in-feed surveys, we also included their average answer to the affective polarization in-feed questions during the three-day baseline period. 

We did not find evidence of covariate imbalance in the Reduced Exposure ($p=0.59$) or the Increased Exposure experiment ($p = 0.18$). 
For reference, the distributions of the demographic covariates and univariate comparisons can be found in Tables \ref{table:covariate-balance-reduce} and \ref{table:covariate-balance-increase}.

\xhdr{Attrition} We also used randomization inference to test for differential attrition rates and differential attrition patterns. 

In the Reduced Exposure experiment, the attrition rate was 36.1\% in the treatment group and 38.9\% in the control group. In the Increased Exposure experiment, the attrition rate was 30.0\% in the treatment group and 29.7\% in the control group.

To test whether the difference in attrition rates is larger than we would expect by chance, we computed the t-score of a two-tailed unequal-variances t-test assessing whether the treatment affects the attrition rate, and compared it against the empirical distribution under 10,000 random reassignments of the treatment~\cite{lin2016standard}. We found no evidence of differential attrition rates in either of the two experiments: Reduced Exposure, $p=0.38$; and Increased Exposure, $p=0.92$.

To test for differential attrition patterns, we regressed an attrition indicator on treatment, baseline covariates, and treatment-covariate interactions, and performed a heteroskedasticity-robust F-test of the hypotheses that all interaction coefficients are equal to zero~\cite{lin2016standard}. We considered the same set of covariates as those described in the covariate balance test: demographics (gender, race/ethnicity, age, education, SES ladder, and income), party affiliation (party ID, party ID strength, party importance), affective polarization (in- and out-party feeling thermometer), antidemocratic attitudes (support for undemocratic practices, support for undemocratic candidates, support for partisan violence, biased evaluation of politicized facts, social distance, social distrust, opposition to bipartisan cooperation), emotions (enthusiastic, happy, still, lonely, sad, nervous, satisfied, calm, relaxed, tired, fearful, aroused, excited, bored, and angry), and recruitment platform. Then, we compared the observed F-statistics with the empirical distribution under 10,000 random reassignments of the treatment. We did not find any evidence of differential attrition patterns in either of the two experiments: Reduced Exposure, $p=0.34$; and Increased Exposure, $p=0.72$.

\subsection{Analyses}
\label{method:analysis}
\vspace{-3mm}
All analyses follow our preregistration plan filed at the Open Science Foundation~\cite{piccardi2024feedrankings}. The reported results focus on the participants who completed the post-survey and at least one in-feed survey for the corresponding outcome. We observe similar results when we consider only those who completed the post-survey when analyzing the post-experiment effect and only those who completed at least one in-feed survey when analyzing the in-feed effect (Sections \ref{sup:regression_tables} and \ref{sup:emotions_regression_tables}).

\vspace{-1mm}
\xhdr{Post-experiment effects}
To estimate the average treatment effects of the intervention on the post-experiment versions of outcomes (i.e., measured via pre- and post-experiment survey), we used OLS regressing the outcome variable on the treatment assignment indicator, the participant's response to the same question on the pre-experiment survey, and a variable indicating the platform they were recruited from (``platform indicator''). We also preregistered that we would consider including other control covariates selected via LASSO regression on the outcomes, but found that only the pre-experiment survey answers were selected. The regression tables for the affective polarization and emotions outcomes are reported in Sections \ref{sup:regression_tables} and \ref{sup:emotions_regression_tables}, respectively.

\vspace{-1mm}
\xhdr{In-feed effects}
To estimate the treatment effects on the outcomes measured using in-feed surveys, we used linear mixed-effects models that account for the fact that there are multiple observations for each participant. We regressed the outcome on the treatment indicator, the participants' average answer to the same question during the three-day baseline period, and the platform indicator. In cases where we do not have any responses from the participants in the baseline period, we used a linear model to impute their average based on their response to the equivalent pre-survey question. The model was trained to predict the participant's baseline average given their pre-survey response and fitted on data from all other participants in the corresponding experiment. Participants without answers in the experimental period were excluded from the analysis. The regression tables for the affective polarization and emotions outcomes are reported in Sections~\ref{sup:regression_tables} and \ref{sup:emotions_regression_tables}, respectively.

\vspace{-1mm}
\xhdr{Engagement analysis} We analyzed traditional engagement metrics by focusing on events captured by the web extension. The extension logs each post that enters the browser viewport and remains visible for at least one second, as well as any user interactions such as likes or reposts. Our analysis investigates the return rate and time spent on the platform, key metrics for social media platforms~\cite{lalmas2014measuring}, along with volume and rate of likes and reposts, which play a significant role in how X's feed algorithm ranks posts, as documented in the open-source code released by X~\cite{twitter_algorithm_ml}.

\section{Ethical Considerations}
\label{sec:ethics}
To minimize the risks, we have complied with all relevant ethical regulations. All procedures were approved by our university's Institutional Review Board. All participants were provided with informed consent before installing the browser extension and participating in the study. Participants were given the option to withdraw their consent or discontinue participation at any time without penalty. At the end of the experiment, participants received financial compensation through the recruitment platform and a debriefing statement via email.

For any social media field experiment, we must consider the ethical implications given the high ecological validity of such work~\cite{straub2024public}. Just like other algorithmic feed ranking methods (e.g.,~\cite{guess2023social}), our work may also inherently pose risks to participants by altering the ranking of content on their news feeds. We balance this risk against an acknowledgment that today's social media platforms (e.g., X) already intervene in people's feeds and take control over users' information stream~\cite{fiske2022twitter}. While this work shares limitations with existing algorithmic ranking methods in that participants may cede control over their feeds to a certain extent, our work does not introduce substantial additional risks. Our down-ranking condition reduced exposure to AAPA posts, while the up-ranking condition inserted AAPA posts that already existed on X and up-ranked an order of magnitude fewer posts. Our aim was to mitigate risks while still carefully and causally identifying whether, and how, changes to feed ranking affect polarization.

\section{Interventions}
\label{sup:intervention}
\subsection{Posts Expressing Antidemocratic Attitudes and Partisan Animosity}
\label{sup:poda}

We define AAPA based on eight variables identified as potentially harmful to democratic functioning~\cite{voelkel2024megastudy}. While the previous study tested multiple interventions to measure the impact on these variables, we focus on content expressing support for these outcomes, following findings from an in-lab experiment~\cite{jia2024embedding}.

\noindent
The eight variables are:
\begin{enumerate}
    \item \textit{Partisan animosity (v1)}: defined as dislike for opposing partisans~\cite{iyengar2019origins}.
    \item \textit{Support for undemocratic practices (v2)}: defined as willingness to forgo democratic principles for partisan gain~\cite{levitsky2019democracies,svolik2018polarization}
    \item \textit{Support for partisan violence (v3)}: defined as willingness to use violent tactics against outpartisans~\cite{kalmoe2022radical}
    \item \textit{Support for undemocratic candidates (v4)}: defined as willingness to ignore democratic practices to elect inparty candidates \cite{graham2020democracy}
    \item \textit{Opposition to bipartisan cooperation (v5)}: defined as resistance to cross-partisan collaboration \cite{hetherington2020washington}
    \item \textit{Social distrust (v6)}: defined as distrust of people in general~\cite{putnam2000bowling}
    \item \textit{Social distance (v7)}: defined as resistance to interpersonal contact with outpartisans~\cite{bishop2009big}
    \item \textit{Biased evaluation of politicized facts (v8)}: defined as skepticism of facts that favor the worldview of the other party~\cite{nyhan2010corrections}
\end{enumerate}

\begin{table}[ht]
\small
\centering
\begin{tabular}{lp{12cm}}
\toprule
Score & Post Example \\
\midrule
0 & JUST IN: K. Harris has rolled out plans for a \$6,000 baby tax credit. \\
1 & BREAKING — TRUMP CLAIMS he’ll sit for an interview with E. Musk this Monday night. \\
2 & Kamala mocking Trump after he spirals in paranoia over her crowd numbers [fire emoji]. Just perfect \\
3 & Drop a “MAGA” [US flag emoji] if you believe Trump will destroy Kamala H. in a Presidential debate\\
\midrule
4 & Trump literally boarding Epstein’s jet. \#TrumpPedoFiles deserves to be trending again. \\
5 & OMG!!! M. T. Greene just threw a tantrum at the GOP convention during an interview, yelling “you little liar.” “You’re to blame for President Trump nearly getting assassinated. \\
6 & I don’t give a damn if Trump got hit or it was just broken glass. Who the fuck cares? He’s still a lying traitor, brain-dead scumbag, fraudulent asshole who stole the presidency. The fact that he STILL walks free is a goddamn disgrace and failure. \\

7 & Fuck the NYT, fuck A. Schiff, fuck Rob R., fuck Clooney, fuck every rich, entitled bastard who thinks their whims outweigh 17M Democratic voters. Trump + \#Project2025 will be on you. \\
8 & FRAT HOE KAMALA: Ex-criminal court judge confirms Kamala Harris is a corrupt witch, corrupt her whole career, slept her way up, and isn’t Black. Biden’s a racist dog who once said “Negro kids are like bugs”. If Democrats win, we must respond \\
\bottomrule
\end{tabular}
\caption{Examples of Non-AAPA (Scores 0--3) and AAPA (Scores 4--8) posts. The posts were randomly selected from those containing the keywords Biden, Harris, Trump, Republican, or Democrat. To preserve the anonymity of the post authors, the example posts in this table have been modified from their original text to preserve the original authors' intent as closely as possible while not being directly searchable on the web.}
\label{tab:postexamples}
\end{table}

\begin{figure}[htbp]
\centering
\includegraphics[width=0.72\textwidth]{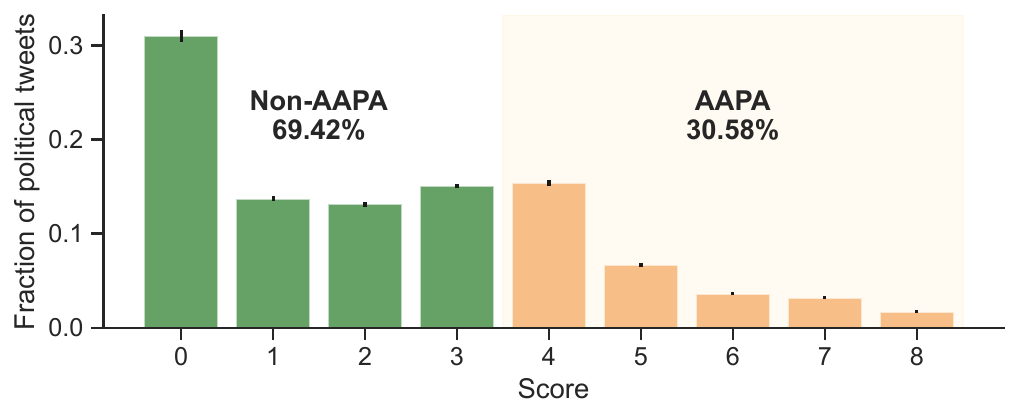}
\caption{Distribution of the AAPA scores assigned to political posts in the ``For you'' feed on X (pre-intervention). The green bars represent the fraction of non-AAPA, while the orange bars describe the AAPA posts recommended by X (not necessarily seen by the users). }\label{fig:scores_distribution}
\end{figure}

\begin{figure}[htbp]
\centering
\includegraphics[width=0.72\textwidth]{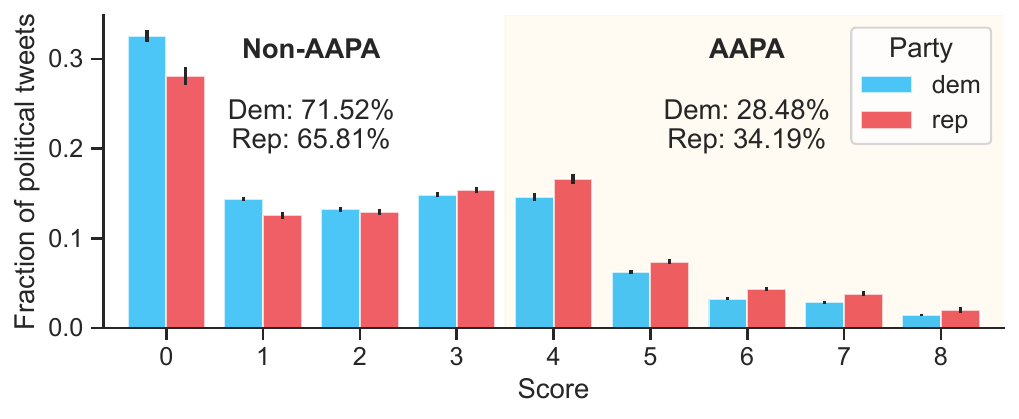}
\caption{AAPA score distribution for political posts, by party.}\label{fig:scores_distribution_party}
\end{figure}

\begin{figure}[htbp]
\centering
\includegraphics[width=0.95\textwidth]{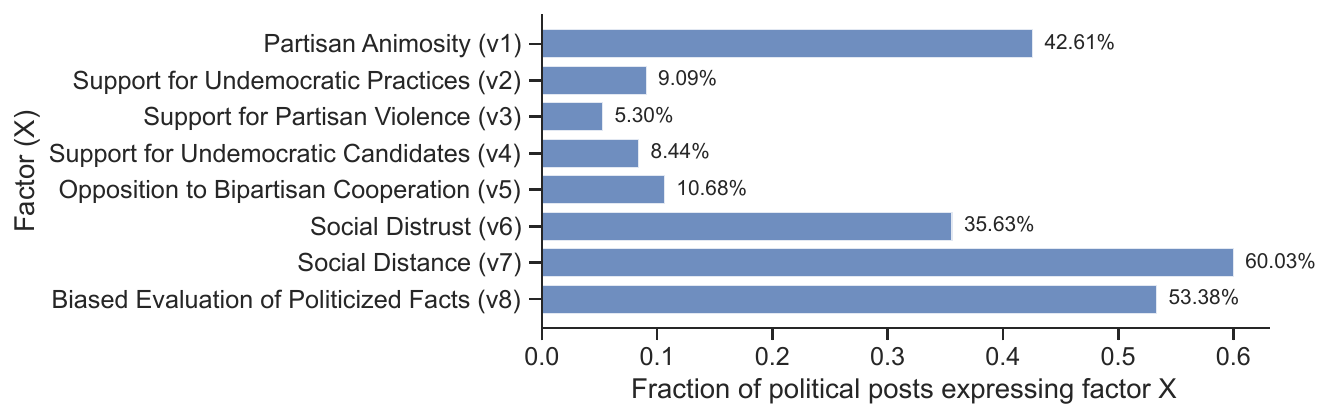}
\caption{Distribution of AAPA factors in political posts seen by participants.}\label{fig:factors_count}
\end{figure}

\begin{figure}[htbp]
\centering
\includegraphics[width=\textwidth]{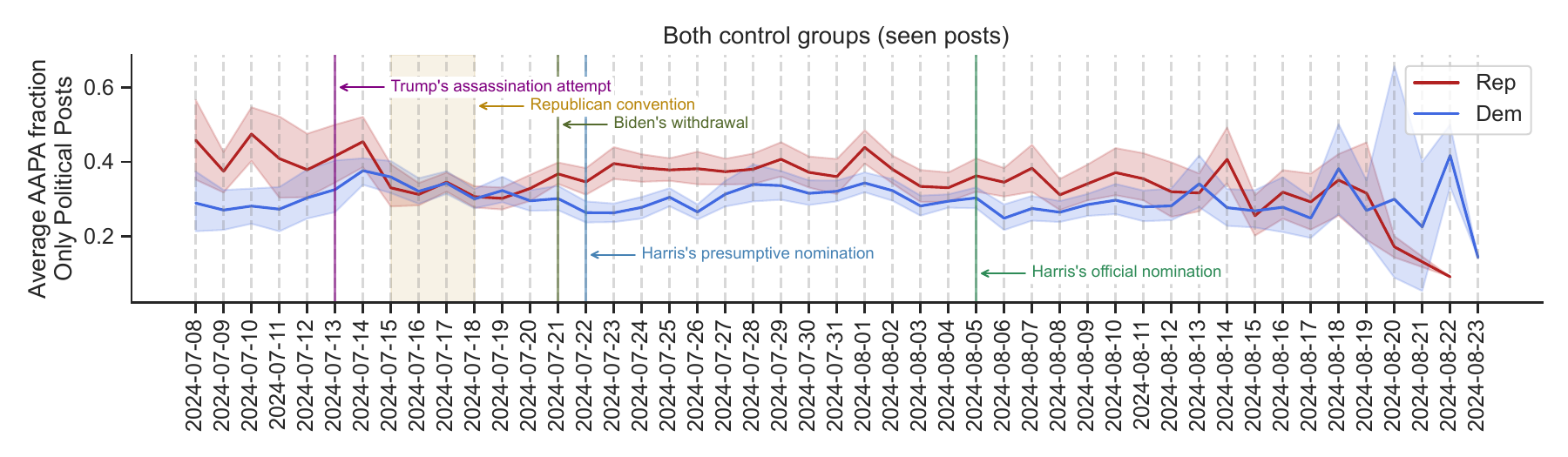}
\caption{Average fraction of AAPA content seen over time among political posts for participants in the two control groups (including during the baseline period), separated by party. Major political events are annotated. Shaded bands represent 95\% bootstrapped confidence intervals.}\label{fig:aapa_fraction_political}
\end{figure}

\noindent
These definitions, evaluated in previous work~\cite{jia2024embedding}, were used to prompt the LLM (\Tabref{tab:prompts}). A post is classified as AAPA when at least four of these factors are present. 
\Tabref{tab:postexamples} shows sample posts with different scores.
Over the course of 10 days, on average, 30.6\% of the political posts in the pre-intervention feed of participants were classified as AAPA (\Figref{fig:scores_distribution}). The most common factors are Social Distance, Biased Evaluation of Politicized Facts, and Partisan Animosity (\Figref{fig:factors_count}).  The distribution of AAPA varied by participants' political orientation, with 34.25\% of political posts in Republican feeds and 28.54\% in Democratic feeds classified as AAPA (\Figref{fig:scores_distribution_party}). These percentages vary over time and can be influenced by major political events~\Figref{fig:aapa_fraction_political}. For instance, during the experiment, external events (such as the presidential nomination of Kamala Harris) were followed by an increase in the proportion of AAPA content, particularly in the feeds of Republican participants.

\begin{figure}[htbp]
\centering
\includegraphics[width=0.50\textwidth]{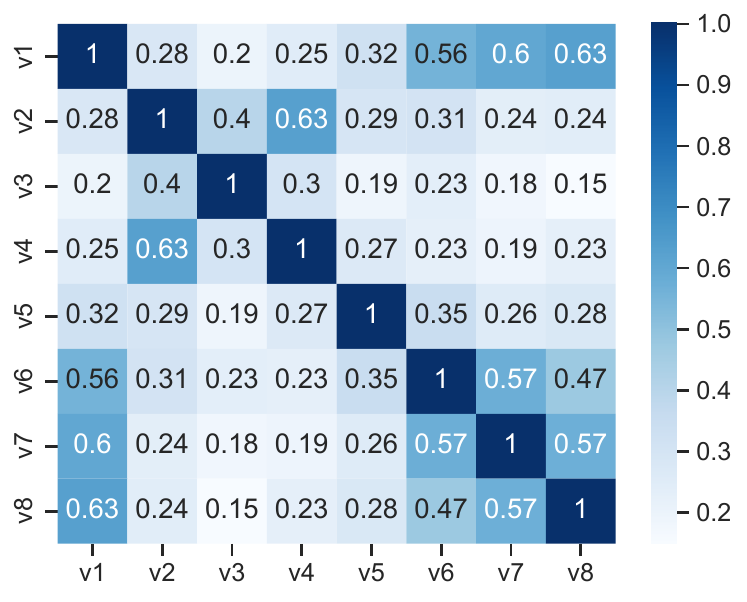}
\caption{Correlation between the eight AAPA variables based on the co-occurrence in all the posts recommended by X to all participants in their ``For you" feed.}\label{fig:factors_correlation}
\end{figure}

\begin{figure}[htbp]
\centering
\includegraphics[width=0.50\textwidth]{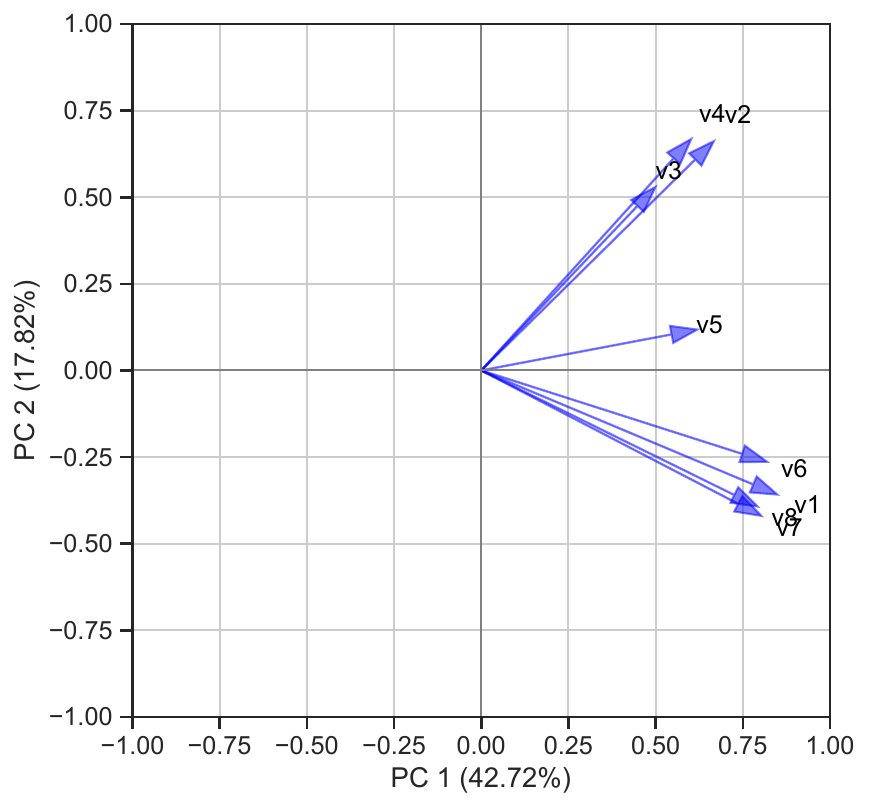}
\caption{PCA analysis of the eight AAPA variables. Reduction of the binary vectors representing the factors present in all the posts recommended by X to all participants in their ``For you" feed. The eight factors cluster into three groups.}\label{fig:factors_analysis}
\end{figure}

\Figref{fig:factors_correlation} shows the correlation between these factors, showing that some of them have a high chance of occurring together. \Figref{fig:factors_analysis} shows that these factors cluster in 3 groups: \{v2, v3, v4\}, \{v5\}, and \{v1, v6, v7, v8\}, suggesting that they tend to appear together.

\subsection{Feed Reranking Implementation}\label{subsec:feed-reranking-implementation}
We developed a browser extension to run the field experiment intervening in real time in the participants' original feeds~\cite{piccardi2024feedrankings}. Our extension was distributed through the Google Chrome Store and supported Chrome and Edge browsers. Once the participants consented to participate, the extension is enabled every time they visit their feed on X. The extension is enabled only when the participants use the ``For you'' feed, which is the feed generated by the recommendation system. Participants can read their chronological feed at any time, but a tooltip regularly reminds them to switch to the ``For you" tab to participate in the study. Our implementation relies on the browser extension to intercept the feed and on a remote backend to rescore and rerank the posts.

Every time the user loads the feed, the extension receives a working set of 35 posts (5 of which are ads) from the server, and for each of these batches, the reranking procedure works as follows. First, the extension intercepts the response returned from the server and extracts the text of all the posts. If the post contains a link, we append the title and description visible in the preview with the prefix ``Attached article's description: '' and discard the URL. If the text quotes another post, we append the quoted text with the prefix ``Quoting: .''
Second, since our reranking logic is applied only to political posts, our reranking service selects all relevant posts in the feed using a RoBERTa-based political classifier \Appref{sup:political_classifier}. Filtering for political posts with a lighter model allows us to reduce the latency by scoring only the relevant posts. Then, the political posts are scored in real-time using GPT-3.5 using prompts based on previous work~\cite{jia2024embedding} and summarized in \Tabref{tab:prompts}. For each post, the resulting data is a vector of 8 binary indicators flagging the presence of each factor. The final score of each post can vary between 0 and 8, and it is obtained by counting how many factors the classifier identified.

Finally, depending on the participant's exposure group, we apply the relevant intervention and return the modified feed to the browser, potentially with an in-feed survey. Each in-feed survey contained a slider for participants to respond using a 0--100 scale on the feeling thermometer. The design required two control conditions since the Reduced Exposure intervention placed the in-feed survey instrument where down-ranked content had previously been, and the Increased Exposure intervention placed the in-feed assessment instrument where up-ranked content would have been inserted.

\subsubsection{Reduced Exposure Feed}
\label{sup:reduced}
Participants in the treatment condition of the Reduced Exposure group receive a feed with less AAPA content. This is achieved by penalizing these posts and moving them to a lower position in the feed. 
Once the relevant posts have been scored, we proceed as follows. First, we select one random AAPA item and flag the next position as the potential location for the in-feed survey. Then, if the participant is currently in the treatment condition, we down-rank all AAPA posts. If the participant is in the control condition, we return the feed with the original ranking. The penalization offset for down-ranked posts depends on their score (position × score × 10) to ensure that posts with higher scores require more effort to be viewed in the feed. Our implementation does not remove the posts, and the participants can read the down-ranked content by scrolling down the feed in a continuous session. This implementation ensures that the selection logic for the survey position is equivalent for the two experimental conditions, differing only in the down-ranking of AAPA posts for the treatment group. Finally, the in-feed survey is injected in the assigned position if the current feed load is selected for inserting the survey.
If the current feed does not contain any AAPA posts, there is no intervention, and we do not add any surveys.

\subsubsection{Increased Exposure Feed}
\label{sup:increased}
The participants in the Increased Exposure group are presented with a treatment opposite to the Reduced Exposure group, as they can find AAPA content in the top positions of their feed. The treatment condition of this group receives a feed where AAPA posts are up-ranked. Since we cannot access the full inventory available on the platform to generate the custom recommendation, we source the AAPA posts from a custom inventory compiled from all the items recommended to the participants enrolled in the study. In particular, to ensure recency and relevance, we limit the candidate set to posts not already seen by the current participant and recommended by the feed algorithm to other participants with the same political leaning in the last 24 hours. This list may include posts recommended to the same participants that were not previously displayed. The content to up-rank is randomly selected from the 100 posts with the highest AAPA score matching these characteristics.
Once the post is selected, we select one random position in the feed, and if the participant is in the treatment condition, the item is inserted there. Alternatively, if the participant is assigned to the control condition, the post is not added. Despite not adding the post, mirroring the same procedure for the control group ensures the similarity of the two conditions in all respects, including potential server latency. Finally, if the event is sampled to include an in-feed survey, we append the in-feed survey in the next available position.

\subsubsection{Manipulation Checks}
\label{sup:intervention_impact}

To assess whether the interventions altered participants' feeds as intended, we analyzed both the number of AAPA posts that were up-ranked or down-ranked and the resulting exposure to political and AAPA content. 

\begin{figure}[htbp]
\centering
\includegraphics[width=0.8\textwidth]{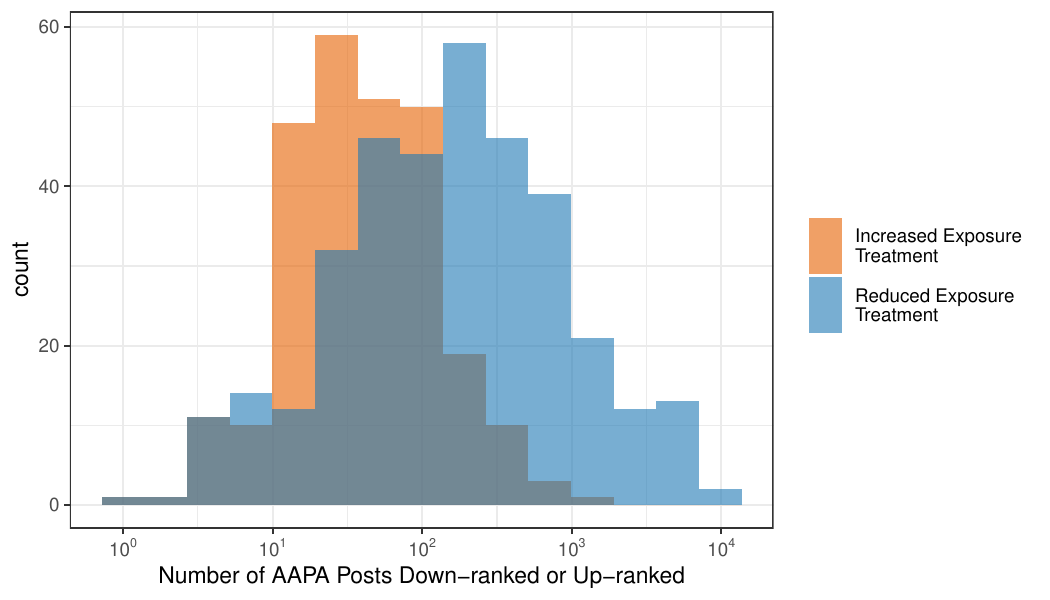}
\caption{Distribution of the number of AAPA posts up-ranked and down-ranked per participant, per day, in the feeds of participants assigned to the corresponding treatment groups. The horizontal axis is on a logarithmic scale. The Reduce Exposure treatment down-ranked approximately an order of magnitude more AAPA posts per participant than the number of posts the Increased Exposure treatment up-ranked.}
\label{fig:dose_paper}
\end{figure}

To avoid excessive exposure to potentially polarizing content, the two experiments were not symmetric in terms of the amount of content reranked: in the Reduced Exposure condition, all AAPA posts detected in a participant's feed were down-ranked, while in the Increased Exposure condition, only one high-scoring AAPA post was up-ranked per batch (typically containing 30 posts). As shown in \Figref{fig:dose_paper}, this design resulted in a substantial difference in the number of reranked posts per user per day, with the Reduced Exposure condition reranking significantly more content. Because AAPA content was already prevalent during the study period, reducing exposure was achieved by down-ranking many posts, whereas increasing exposure required up-ranking fewer but more intense AAPA posts, making this condition a comparatively harder test.

\begin{figure}[htbp]
\centering
\includegraphics[width=\textwidth]{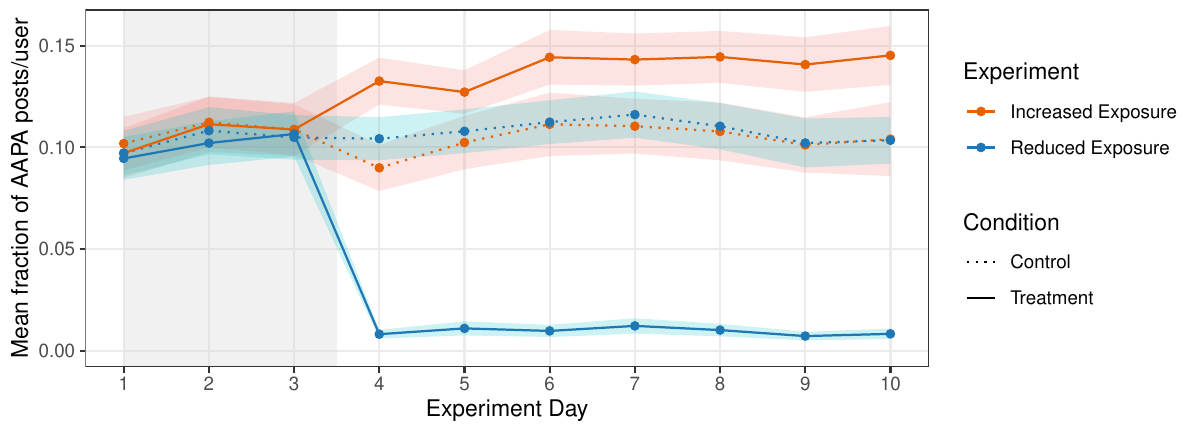}
\caption{Average fraction of AAPA posts seen by participants for each day of the experiment. In the Reduced Exposure condition, the fraction drops at the start of the treatment period but does not reach zero, as some down-ranked posts still appear with sufficient scrolling.}
\label{fig:si_daily_aapa}
\end{figure}

\begin{figure}[htbp]
\centering
\includegraphics[width=\textwidth]{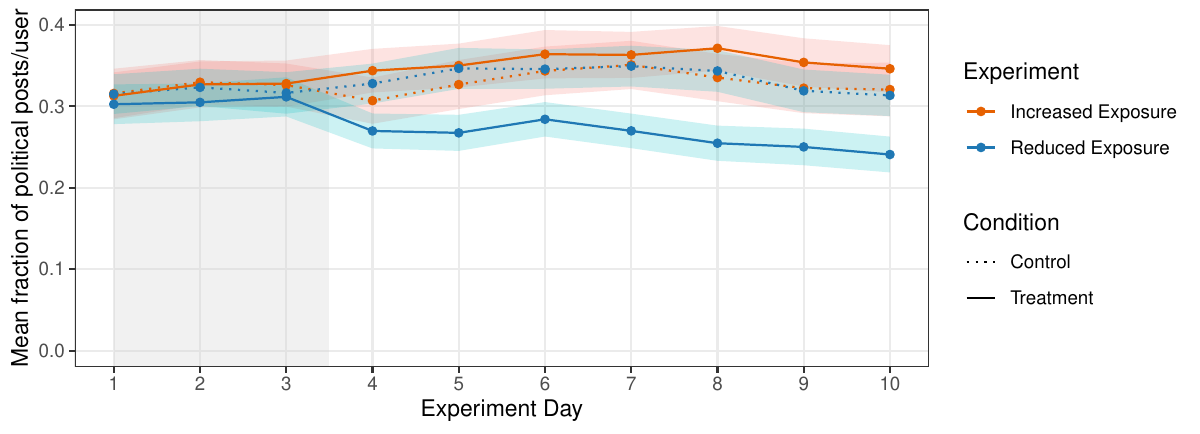}
\caption{Average fraction of political posts seen by participants for each day of the experiment. Despite some variation, participants' feeds in both experiments continued to contain a significant amount of political content.}
\label{fig:si_daily_politics}
\end{figure}

\Figref{fig:si_daily_aapa} shows the effect of the interventions on overall feed composition, summarizing the fraction of AAPA posts in participants' feeds for each day of the experiment. During the three-day baseline period, participants in all four experimental groups were exposed to a similar fraction of AAPA posts per day. However, during the subsequent seven-day intervention period, participants in the two control groups continued to be exposed to a similar fraction of AAPA posts, whereas participants in the treatment groups of the Reduced and Increased Exposure experiments experienced significant decreases and increases in AAPA exposure, respectively. As discussed above, the Increased Exposure treatment up-ranked fewer AAPA posts, and thus the relative increase in AAPA exposure is smaller than the decrease observed in the Reduced Exposure treatment.
We note that the fraction of AAPA posts in the Reduced Exposure treatment does not fall all the way to zero, since posts are down-ranked rather than removed, and participants can still reach them by scrolling much further down their feeds. Importantly, the interventions specifically targeted AAPA content: as shown in~\Figref{fig:si_daily_politics}, participants continued to be exposed to substantial amounts of political content in both experiments. In the Reduced Exposure treatment, the decline in exposure to political content is limited to the subset of AAPA content, whereas in the Increased Exposure condition, the overall volume of political content remains nearly unchanged.

\begin{figure}[htbp]
\centering
\includegraphics[width=\textwidth]{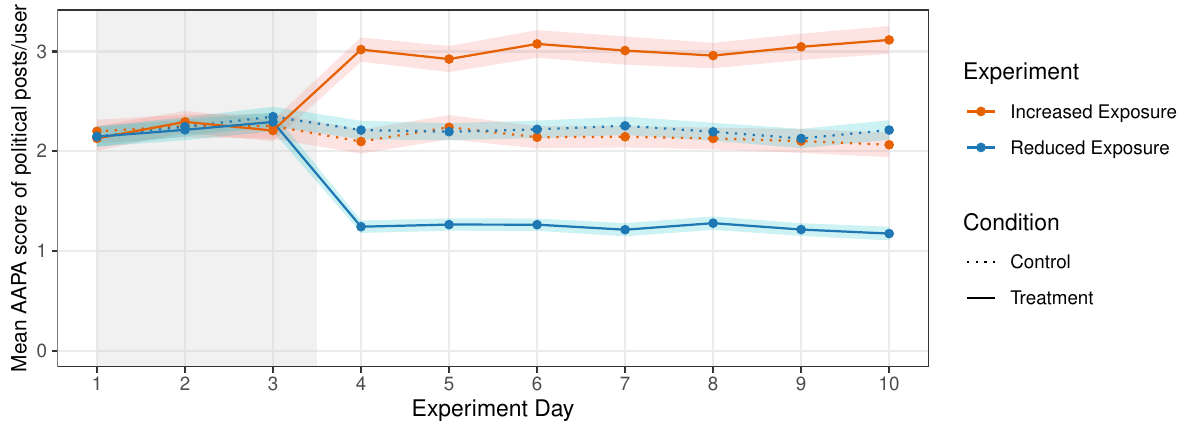}
\caption{Mean AAPA score for political posts by day and condition; first averaged per participant and then averaged across all users in the corresponding group. The results demonstrate that the intervention increased and decreased AAPA exposure, as intended.}
\label{fig:feed_scores}
\end{figure}

Figure~\ref{fig:feed_scores} shows the average AAPA score of all political posts to which participants were exposed to in each of the four experimental groups. Although the Increased Exposure treatment up-ranked fewer AAPA posts, it focused on posts with high AAPA scores. As a result, the Reduced and Increased Exposure treatments induced a similar relative change in average AAPA scores in the respective directions.

\begin{figure}[htbp]
\centering
\includegraphics[width=0.98\textwidth]{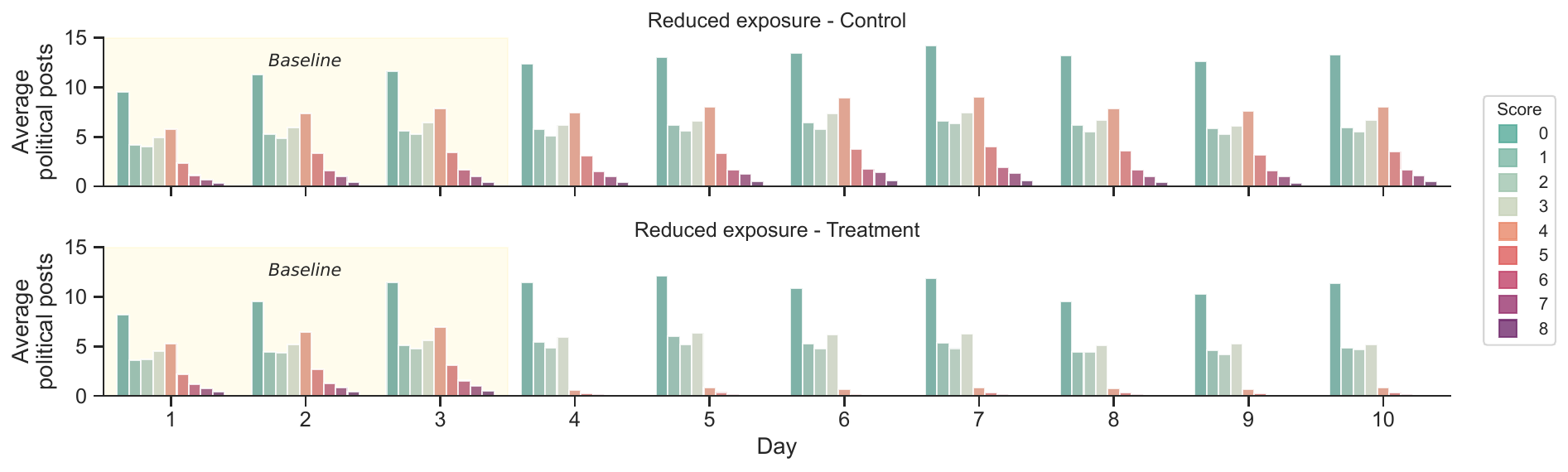}
\caption{Reduced Exposure: Distribution of the average number of political posts that each participant saw by day for each score. During the treatment period, participants experienced a significant reduction (-89.6\%) in their exposure to AAPA content ($\ge$ 4 factors). While AAPA content was down-ranked, participants who scrolled far enough through their feeds still encountered a small amount of AAPA posts.}\label{fig:seen_by_day_reduced}
\end{figure}

\begin{figure}[htbp]
\centering
\includegraphics[width=0.98\textwidth]{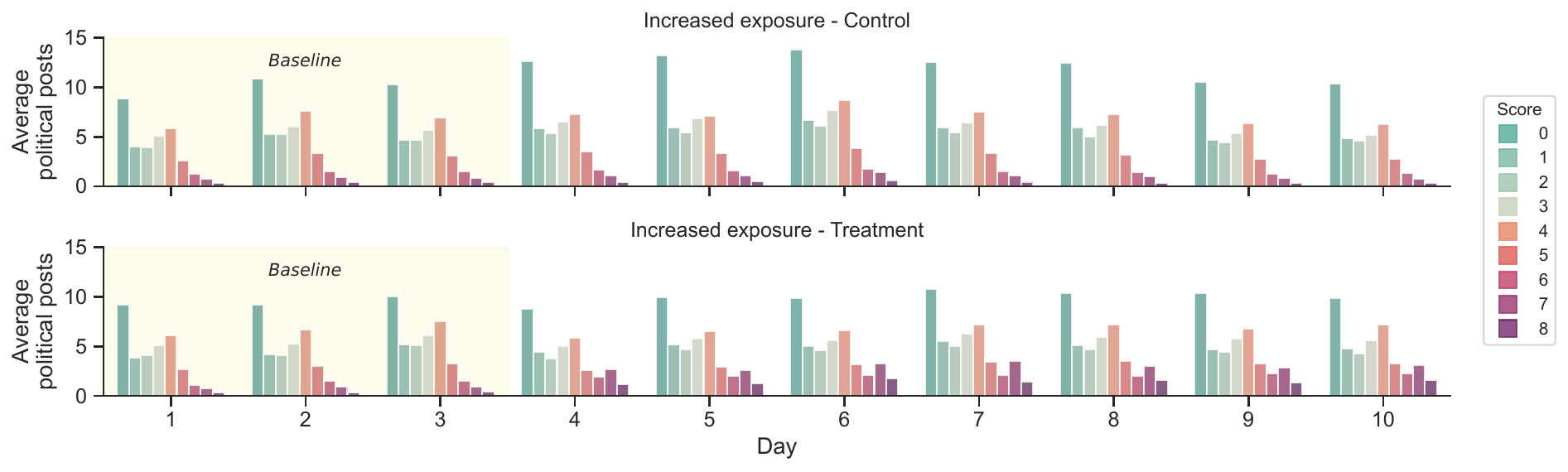}
\caption{Increased Exposure: Distribution of the average number of political posts that each participant saw by day for each score. During the treatment period, participants experienced a significant increase (+26.4\%) in their exposure to AAPA content ($\ge$ 4 factors).}\label{fig:seen_by_day_increased}
\end{figure}

Finally, \Figref{fig:seen_by_day_reduced} and \Figref{fig:seen_by_day_increased} show the daily average exposure to political posts by AAPA score, separately for treatment and control groups in each experiment. In the Reduced Exposure treatment, we observe a clear drop in the number of high-AAPA posts seen by participants after the intervention begins, while the control group remains stable. In contrast, the Increased Exposure treatment shows a rise in exposure to high-AAPA posts at the beginning of the intervention period. 

Overall, these results suggest that the treatments successfully induced the intended variation in AAPA exposure.

\section{Experimental Platform}
\label{sup:platform}

We developed a custom experimental platform focused on performance to ensure the experiment introduces minimal disruption to the user experience on X. Our implementation consists of a browser extension and a REST backend deployed as a cloud service.

\subsection{Browser Extension}
The browser extension handles editing the X feed and managing all interactions with the participants, including the onboarding steps. It is compatible with Google Chrome and Microsoft Edge and requires only \textit{storage} permissions to save the user identifier.

The extension manages the ranking manipulation by modifying the content available on the X feed in real time. Our implementation allows us to inject custom code on the main page to override the methods of the native \textit{XMLHttpRequest} object. This approach allows us to intercept the communication with the server and modify the execution flow of the requests. Our browser extension subscribes to the request endpoints used to retrieve the new content to add to the feed, interrupts the execution flow, sends the server response to our backend for scoring and reranking, and finally restores the execution flow by calling the original methods with the response customized by our server.
This approach requires careful understanding of the communication protocol with the server, but it offers significant advantages: first, the content can be modified by editing structured JSON objects without manipulating the final HTML, and second, by selecting what requests to intercept, it allows logging the interactions with the web page tracked by X without adding custom listeners. 

Additionally, the browser extension interacts with the main page by editing the DOM to add visual elements. These widgets include the in-feed surveys that are added to the feed as regular posts and a top fixed banner used to notify the participant about possible registration errors or that the experiment is completed, with the invitation to take the final survey and remove the extension.

Finally, the extension manages the onboarding flow to ensure that participants can register and join the study only by following links that automatically guide them in installing the extension and setting the identifier from the recruitment platform~\cite{piccardi2024reranking}.

In this work, we have focused on desktop-based usage, but a similar approach can be implemented on mobile phones through web views and emerging support for mobile browser extensions (e.g., Microsoft Edge for Mobile). 

\subsection{Backend}
The backend is responsible for reranking the feed, logging the client events, sending daily participation reminders by email, and orchestrating the participants' registration. The server is written in Python and Java using MongoDB and Memcached for storage and caching. When the browser extension interrupts the execution flow to request the reranking of the feed, X's rotating loading indicator is extended until our reranking is complete. The implementation choices are guided by minimizing response time to ensure minimal disruption to the user experience.

The feed reranking proceeds in four steps: first, the response from X's server, typically containing 35 posts, is sent to our backend and parsed to extract the text from the posts. The implementation focused on individual posts and in-feed conversations, ignoring the other types of content, such as ads or users' recommendations. Second, all the posts are evaluated by a political classifier \Appref{sup:political_classifier} to filter the posts that need to be scored for the presence of the eight AAPA factors. This service runs as a standalone Python web service, accelerated by a GPU, that introduces, on average, a latency of 490 ms (STD = 111 ms). Then, the political posts are dispatched to a Java server that manages an HTTP connection pool with OpenAI APIs. The service batches multiple posts in a single prompt to reduce the number of requests and uses a pool of tokens to minimize latency and manage rate limits. Finally, the scores are parsed and returned to the main backend, which applies the reranking intervention and adds the in-feed surveys. These steps are replicated in the same way in the control condition, but in that case, the backend returns the original feed. In the case of down-ranking, the posts are saved in a cache (Memcached) and added to the feed when the user reaches the new position in a continuous scrolling session.

The backend also stores the logs of interactions with the feed. When a participant interacts with the feed on X, the extension regularly shares with the backend what actions are performed on the main page. This data is limited to the actions on the main feed and does not include any private content. The actions recorded include when a post enters the main viewport, likes, reposts, clicks on external links, and the time spent on the browser tab.

To facilitate communication with the participants in the registration flow, they are invited to share an email address. The platform uses this address to send automatic messages that guide the participant in completing the study successfully. These messages include a copy of the onboarding instructions and the consent form for future reference, a daily reminder that they are enrolled in the study sent at 6 PM (participant's local time), and the end-of-study notification with the link to the post-experiment survey and the instruction to remove the browser extension.

Finally, the backend orchestrates the registration flow of the participants~\cite{piccardi2024feedrankings}. Since the browser extension does not support passing custom installation parameters, the backend is responsible for guiding the onboarding flow to configure the extension for the current participant. After the completion of the pre-experiment survey, the participants are redirected to our web server using a URL that encodes all the parameters relevant for the extension to work, such as the identifier and political position. Our server adds a browser cookie with this data and displays the step-by-step instructions for installing the extension. Participants are then redirected to the Chrome store, where they install the browser extension. As soon as the extension is installed, it opens a browser tab and loads to a welcome page hosted on our coordination server, where it has access to the cookies information previously set. This approach allows the extension to save these parameters in persistent storage without forcing the participants to enter this information manually after the installation of the extension. Lastly, the welcome message contains a link to redirect the participants back to the recruitment platform to set the task as completed.

\subsection{Political Classifier}
\label{sup:political_classifier}
Since our down-ranking intervention applies only to political posts, we developed a classifier to prefilter the relevant posts. Since GPT has a pay-per-use business model and calling an external API service can introduce additional latency, we ensure that GPT is used to score only the content in which the eight variables are relevant. The political classifier is a distilled language model fine-tuning a RoBERTa model to identify political posts, including mentions of officials and activists, social issues, news, and current events, based on an annotated training set. 

The posts used to train the model were collected with the X API in July 2023. To select the posts, we replicated the reverse chronological feeds of users across the political spectrum for two weeks and took a random sample of 50,000 posts. We first selected 300 users and then collected the posts by the users they followed. 

To select the users, we first obtained the followers of all members of Congress. Then, we categorized each user in the superset of followers based on how many Republican and Democrat members of Congress they followed: (a) \textit{far-left}: followed 4-6 Democrats and no Republicans; (b) \textit{left}: followed 1-3 Democrats and no Republicans; (c) \textit{middle}: followed 1-3 Democrats and 1-3 Republicans; (d) \textit{far-middle}: followed 4-6 Democrats and 4-6 Republicans; (e) \textit{right}: followed no Democrats and 1-3 Republicans; (f) \textit{far-right}: followed no Democrats and 4-6 Republicans. We started with members of Congress since their political affiliations are known and are often used to estimate the political leanings of users on X~\cite{barbera2015tweeting, barbera2015birds}. Next, we removed follower accounts that are likely to be bots. We used the Botometer API and filtered out accounts with Complete Automation Probability above 0.6~\cite{yang2022botometer}. Finally, we took a stratified sample of 300 accounts, selecting 50 users from each of the six groups defined above.

Next, for each of the selected users, we collected all of the accounts they followed and collected all the posts by those users as far back as we could. Then, to replicate the reverse chronological feed of each user, we merged all posts by the accounts they followed and sorted from most to least recent. We retain only the English posts that would have appeared in their reverse chronological feeds in the previous two weeks. Finally, we merged the two-week feeds of all users, removed duplicate posts, and took a random sample of 50,000 posts. We later randomly split this dataset into 45,000 posts for training and 5,000 posts for testing.

\begin{table}[h!]
\small
\centering
\begin{tabular}{lllll}
\toprule
\textbf{Ground truth} & \textbf{Test set size} & \textbf{Precision [95\% CI]} & \textbf{Recall [95\% CI]} & \textbf{F1 score} \\
\midrule
GPT-4 & 5,000 & 0.937 [0.931, 0.942] & 0.920 [0.911, 0.928] & 0.936 \\
Human labels & 300 & 0.921 [0.868, 0.966] & 0.905 [0.847, 0.954] & 0.913\\
\bottomrule
\end{tabular}
\caption{Performance of the fine-tuned RoBERTa political classifier. Bootstrapped 95\% confidence intervals.}
\label{tab:political_classifier}
\end{table}

We annotated all 50,000 posts by prompting GPT-4 using a definition proposed by the Pew Research Center~\cite{bestvater2022politics}:

\begin{quote}
\footnotesize
\begin{verbatim}
Political content on Twitter is varied and can be about officials and 
activists, social issues, or news and current events. 
Looking at the following tweet, would you categorize it as POLITICAL 
or NOT POLITICAL content? 
Answer 1 if it is POLITICAL, 0 otherwise.
\end{verbatim}
\end{quote}

This step annotated 34.2\% of the posts as political and 65.8\% as non-political. We used this dataset to fine-tune a pre-trained RoBERTa model (\textit{roberta-base}). 

We evaluate the model on a held-out dataset composed of 5,000 posts annotated by GPT-4 and a random subset of 300 posts annotated by humans. For each post, we collect three labels using Prolific and assign the value based on majority voting. The model achieves an F1 score of 93.6\% on the GPT-4 labels and an F1 score of 91.3\% on the human ground~truth.

\subsection{Scoring Service}
\label{sup:scoring}
The scoring service is a Java backend responsible for evaluating each political post across the eight AAPA dimensions. It is designed to minimize latency. When a new request is received, it proceeds as follows: after the political classifier identifies the posts to score, they are sent to this service in one request as a list of textual items. This service first splits the list into chunks of at most ten items and then compiles eight prompts for each of them. The prompts are in the format:
\begin{quote}
\footnotesize	
\begin{verbatim}
Do the following messages express partisan animosity? 
Partisan animosity is defined as "dislike for opposing partisans".

FORMAT:
The input messages are given as JSON lines in the format 
{"id": <message_id>, "message": <message>}.
The output must be a JSON array of objects in the format 
[{"id": <message_id>, "answer": <YES or NO>}, ... ]. 

INPUT MESSAGES:
...
\end{verbatim}
\end{quote}

The prompt starts with the question and the definition of the factor as described in Table~\Tabref{tab:prompts}, specifies the input and output format, and ends with the list of posts encoded in JSON objects, one for each line. The posts are grouped into at most ten items per prompt, as a trade-off between making too many API calls and keeping the response time low. This implementation ensures that a feed containing ten political posts results in 8 requests to OpenAI instead of 80.

These requests run in parallel using a custom connection pool that shares HTTP connections and iterates round-robin on multiple API tokens. Each request takes a few seconds to receive a response, but each connection has a maximum waiting time of 8 seconds to ensure a slow API call does not disrupt the user experience. When the timeout is reached, the event is logged, and the factor is set as non-present. To ensure a fast response from OpenAI, we use GPT-3.5 Turbo, which, at the time of the experiment, offered a good trade-off between speed and accuracy. We use a fixed seed for reproducibility. Prior work validated that this approach matches expert annotations~\cite{jia2024embedding}.

Since many popular posts appear in multiple feeds, to avoid scoring the same text multiple times, each post score is cached in a backend database and reused when necessary without including the post in the request to OpenAI.

The model we used at the time of our deployment was not multimodal, so we did not process images in the posts. Current models do afford this capability, and we recommend it for future work in this vein.

\begin{table}[t]
\centering
\footnotesize
\begin{tabular}{p{0.18\textwidth} p{0.765\textwidth}}
\toprule
\textbf{Variable} & \textbf{Definition} \\
\midrule
Partisan Animosity & \textit{Do the following messages express partisan animosity? Partisan animosity is defined as "dislike for opposing partisans".} \\
\midrule
Support for Undemocratic Practices & \textit{Do the following messages express support for undemocratic practices? Support for undemocratic practices is defined as "willingness to forgo democratic principles for partisan gain". Undemocratic practices are undemocratic tendencies or actions such as reducing polling stations in areas that support their opponents, attacking the independence of the judiciary, undermining the free press, challenging the legitimacy of election results, or encouraging political violence.} \\
\midrule
Support for Partisan Violence & \textit{Do the following messages express support for partisan violence? Support for partisan violence is defined as a "willingness to use violent tactics against outpartisans". Examples of partisan violence include sending threatening and intimidating messages to the opponent party, harassing the opponent party on the Internet, using violence in advancing their political goals or winning more races in the next election.} \\
\midrule
Support for Undemocratic Candidates & \textit{Do the following messages express support for undemocratic candidates? Support for undemocratic candidates is defined as "willingness to ignore democratic practices to elect inparty candidates." Undemocratic candidates often support undemocratic practices such as reducing polling stations in areas that support their opponents, attacking the independence of the judiciary, undermining the free press, challenging the legitimacy of election results, or encouraging political violence.} \\
\midrule
Opposition to Bipartisan Cooperation & \textit{Do the following messages express opposition to bipartisanship? Opposition to bipartisanship is defined as "resistance to cross-partisan collaboration".} \\
\midrule
Social Distrust & \textit{Do the following messages express social distrust? Social distrust is defined as "distrust of people in general".} \\
\midrule
Social Distance & \textit{Do the following messages express social distance? Social distance is defined as "resistance to interpersonal contact with outpartisans". Messages that increase social distance may contain terms that increase distrust, distance, insecurity, hate, prejudice, or discrimination.} \\
\midrule
Biased Evaluation of Politicized Facts & \textit{Do the following messages express a biased evaluation of politicized facts? Biased evaluation of politicized facts is defined as "skepticism of facts that favor the worldview of the other party". Messages supporting a biased evaluation of politicized facts may partially present political facts or discuss a controversial issue with a certain political stance.}\\
\bottomrule
\end{tabular}
\caption{Definition of the eight factors used in the GPT prompts.}
\label{tab:prompts}
\end{table}

\section{Participant Characteristics}
\label{sup:participants}
\begin{figure}[p]
    \centering
\includegraphics[width=\textwidth]{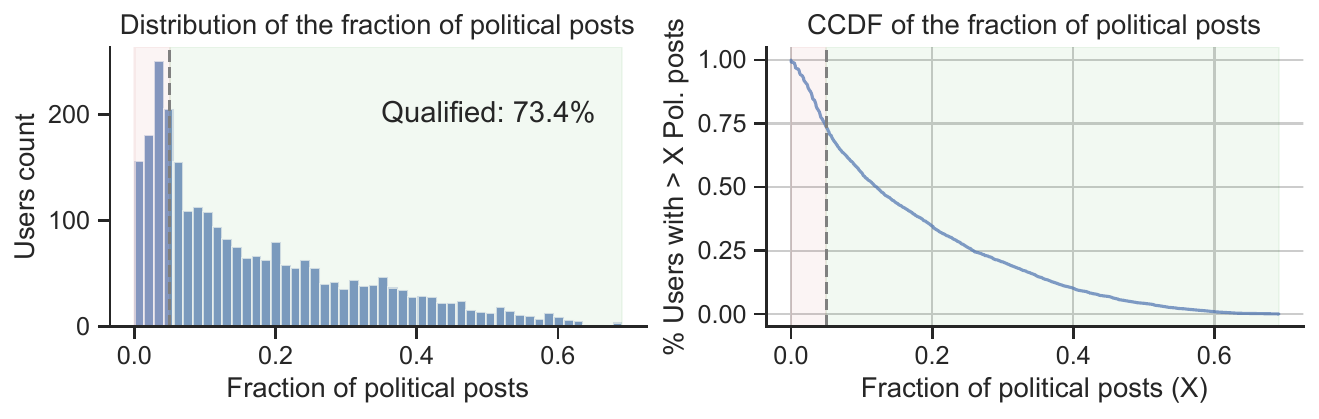}
    \caption{Distribution of the fraction of political content in the feed of the participants screened. The green area represents the participants who were qualified to join the study.}\label{fig:political_distribution}
\end{figure}

To enroll in the experiment, participants had to take a screening task by installing a browser extension. The extension opened X, asked participants to log in if necessary, scrolled through their feed automatically, and verified that at least 5\% of the posts in their ``For you'' feed were about politics or social issues. \Figref{fig:political_distribution} summarizes the distribution of users by the fraction of political content. Around 73\% of the participants qualified and were invited to the full study.

\begin{figure}[p]
\centering
\includegraphics[width=\textwidth]{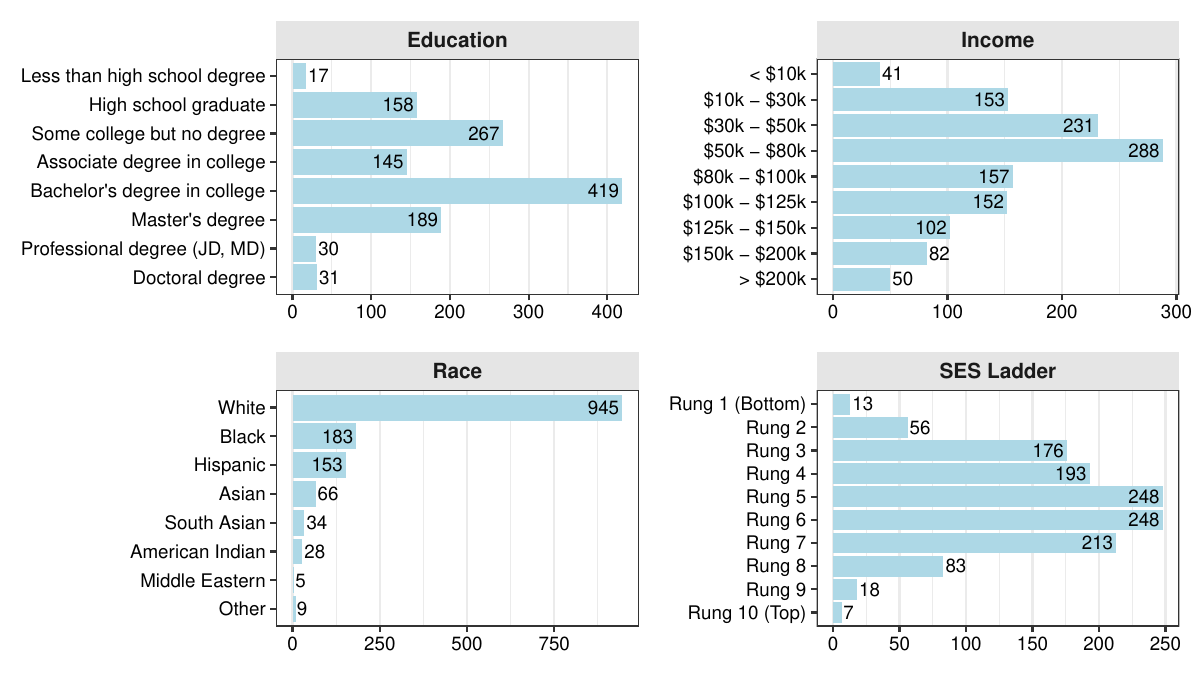}
\caption{Distribution of demographic characteristics among participants who completed the post-experiment survey.}
\label{fig:participants_demographics}
\end{figure}

Figure~\ref{fig:participants_demographics} shows the distributions of some of the participants' demographic information collected in the pre-experiment survey, including race, education level, perceived socioeconomic status, and income. 

Tables \ref{table:covariate-balance-reduce} and \ref{table:covariate-balance-increase} show univariate analyses of covariate balance in the Reduced and Increased Exposure experiments, respectively. Omnibus balance tests using randomization inference are reported in Section~\ref{methods}. We do not observe any evidence of imbalance.

\begin{table}
\small
\resizebox{\textwidth}{!}{
    \begin{tabular}{lccccc}
    \toprule
     & Treatment & Control & Difference & p(diff != 0) & p adjusted\\
    \midrule
    Party: Democrat (\%) & 0.672 & 0.692 & -0.020 & 0.590 & 1.000\\
    Party: Republican (\%) & 0.328 & 0.308 & 0.020 & 0.590 & 1.000\\
    \addlinespace
    Party strength: Weak (\%) & 0.497 & 0.470 & 0.027 & 0.507 & 1.000\\
    Party strength: strong (\%) & 0.503 & 0.530 & -0.027 & 0.507 & 1.000\\
    \addlinespace
    Party importance (0 - 100) & 64.747 & 68.430 & -3.684 & 0.100 & 1.000\\
    \addlinespace
    Gender: Female (\%) & 0.459 & 0.467 & -0.008 & 0.851 & 1.000\\
    Gender: Male (\%) & 0.522 & 0.513 & 0.009 & 0.830 & 1.000\\
    Gender: Other (\%) & 0.019 & 0.020 & -0.001 & 0.920 & 1.000\\
    \addlinespace
    Age & 41.081 & 42.434 & -1.353 & 0.223 & 1.000\\
    \addlinespace
    Race: White (\%) & 0.784 & 0.742 & 0.043 & 0.212 & 1.000\\
    Race: Black (\%) & 0.153 & 0.149 & 0.004 & 0.886 & 1.000\\
    Race: Hispanic (\%) & 0.094 & 0.109 & -0.016 & 0.523 & 1.000\\
    Race: Asian (\%) & 0.078 & 0.083 & -0.005 & 0.831 & 1.000\\
    Race: Other (\%) & 0.038 & 0.030 & 0.008 & 0.595 & 1.000\\
    \addlinespace
    Education: High school or less (\%) & 0.144 & 0.129 & 0.015 & 0.596 & 1.000\\
    Education: Some college (\%) & 0.284 & 0.318 & -0.034 & 0.364 & 1.000\\
    Education: Bachelor's degree (\%) & 0.356 & 0.348 & 0.009 & 0.823 & 1.000\\
    Education: Master's, PhD, JD, MD (\%) & 0.216 & 0.205 & 0.010 & 0.753 & 1.000\\
    \addlinespace
    Income: less than \$30,000 (\%) & 0.125 & 0.166 & -0.041 & 0.153 & 1.000\\
    Income: \$30,000 - \$80,000 (\%) & 0.394 & 0.437 & -0.043 & 0.274 & 1.000\\
    Income: \$80,000 - \$125,000 (\%) & 0.303 & 0.222 & 0.081 & 0.021 & 1.000\\
    Income: more than \$125,000 (\%) & 0.178 & 0.175 & 0.003 & 0.932 & 1.000\\
    \addlinespace
    SES Ladder: Bottom rungs (\%) & 0.166 & 0.182 & -0.016 & 0.588 & 1.000\\
    SES Ladder: Middle rungs (\%) & 0.741 & 0.725 & 0.015 & 0.664 & 1.000\\
    SES Ladder: Top rungs (\%) & 0.094 & 0.093 & 0.001 & 0.965 & 1.000\\
    \addlinespace
    Recruitment platform: Bovitz (\%) & 0.659 & 0.599 & 0.060 & 0.122 & 1.000\\
    Recruitment platform: CloudResearch (\%) & 0.341 & 0.401 & -0.060 & 0.122 & 1.000\\
    \bottomrule
    \end{tabular}
}
\caption{Univariate Covariate Balance: Reduced Exposure Experiment}
\label{table:covariate-balance-reduce}
\end{table}

\begin{table}
\small
\resizebox{\textwidth}{!}{
    \begin{tabular}{lccccc}
    \toprule
     & Treatment & Control & Difference & p(diff != 0) & p adjusted\\
    \midrule
    Party: Democrat (\%) & 0.710 & 0.622 & 0.088 & 0.043 & 0.289\\
    Party: Republican (\%) & 0.290 & 0.378 & -0.088 & 0.043 & 0.289\\
    \addlinespace
    Party strength: Weak (\%) & 0.454 & 0.483 & -0.029 & 0.533 & 1.000\\
    Party strength: strong (\%) & 0.546 & 0.517 & 0.029 & 0.533 & 1.000\\
    \addlinespace
    Party importance (0 - 100) & 67.429 & 63.935 & 3.494 & 0.158 & 0.501\\
    \addlinespace
    Gender: Female (\%) & 0.542 & 0.448 & 0.094 & 0.042 & 0.289\\
    Gender: Male (\%) & 0.441 & 0.526 & -0.085 & 0.066 & 0.289\\
    Gender: Other (\%) & 0.017 & 0.026 & -0.009 & 0.490 & 1.000\\
    \addlinespace
    Age & 42.916 & 41.470 & 1.446 & 0.266 & 0.918\\
    \addlinespace
    Race: White (\%) & 0.777 & 0.757 & 0.021 & 0.596 & 1.000\\
    Race: Black (\%) & 0.134 & 0.139 & -0.005 & 0.883 & 1.000\\
    Race: Hispanic (\%) & 0.080 & 0.135 & -0.055 & 0.056 & 0.289\\
    Race: Asian (\%) & 0.092 & 0.048 & 0.045 & 0.058 & 0.289\\
    Race: Other (\%) & 0.034 & 0.043 & -0.010 & 0.581 & 1.000\\
    \addlinespace
    Education: High school or less (\%) & 0.139 & 0.139 & 0.000 & 0.988 & 1.000\\
    Education: Some college (\%) & 0.361 & 0.383 & -0.021 & 0.635 & 1.000\\
    Education: Bachelor's degree (\%) & 0.319 & 0.322 & -0.002 & 0.956 & 1.000\\
    Education: Master's, PhD, JD, MD (\%) & 0.181 & 0.157 & 0.024 & 0.486 & 1.000\\
    \addlinespace
    Income: less than \$30,000 (\%) & 0.193 & 0.165 & 0.028 & 0.430 & 1.000\\
    Income: \$30,000 - \$80,000 (\%) & 0.412 & 0.422 & -0.010 & 0.827 & 1.000\\
    Income: \$80,000 - \$125,000 (\%) & 0.164 & 0.252 & -0.088 & 0.019 & 0.289\\
    Income: more than \$125,000 (\%) & 0.231 & 0.161 & 0.070 & 0.056 & 0.289\\
    \addlinespace
    SES Ladder: Bottom rungs (\%) & 0.231 & 0.217 & 0.014 & 0.723 & 1.000\\
    SES Ladder: Middle rungs (\%) & 0.685 & 0.722 & -0.037 & 0.384 & 1.000\\
    SES Ladder: Top rungs (\%) & 0.084 & 0.061 & 0.023 & 0.334 & 1.000\\
    \addlinespace
    Recruitment platform: Bovitz (\%) & 0.576 & 0.578 & -0.003 & 0.954 & 1.000\\
    Recruitment platform: CloudResearch (\%) & 0.424 & 0.422 & 0.003 & 0.954 & 1.000\\
    \bottomrule
    \end{tabular}
}
\caption{Univariate Covariate Balance: Increased Exposure Experiment}
\label{table:covariate-balance-increase}
\end{table}

\pagebreak

\section{Adjustment for Multiple Comparisons}
\label{sec:p-values-corrections}
We perform False Discovery Rate (FDR) adjustment using the method described in~\cite{anderson2008multiple}, considering each experiment (Reduced and Increased Exposure) separately and grouping the variables as:
\begin{itemize}
    \item $K_1$ primary outcomes: Not adjusted,
    \item $K_2$ secondary outcomes: Sharpened FDR-adjusted p-values with $K_1 + K_2$ outcomes,
    \item $K_3$ tertiary outcomes: Sharpened FDR-adjusted p-values with $K_1 + K_2 + K_3$ outcomes.
\end{itemize}

\noindent
We also perform FRD adjustment when examining heterogeneous treatment effects:
\begin{itemize}
    \item Primary tests of heterogeneity: We used sharpened FDR-adjusted p-values with $L_x * K_x$ hypothesis tests, where $L_x$ is the number of primary moderators for the corresponding group, and $K_x$ is the number of outcomes in the corresponding group (i.e., $K_1$, $K_2$, or $K_3$ above).
    \item Secondary tests of heterogeneity: We specify the secondary moderators that we considered for each group of outcomes (primary, secondary, tertiary). We used sharpened FDR-adjusted p-values with $(L_x + L_y) * K_x$ hypothesis tests, where $L_x$ is the number of primary and $L_y$ is the number of secondary moderators for the corresponding group, and $K_x$ is the number of outcomes in the corresponding group (i.e., $K_1$, $K_2$, or $K_3$ above).
    \item Auxiliary tests: Not adjusted, generally these are variables for which statistical significance should not matter.
\end{itemize}

\section{Political Attitude Shifts}
\label{sec:political-attitude-shifts}
\begin{figure}[t]
\centering
\includegraphics[width=\textwidth]{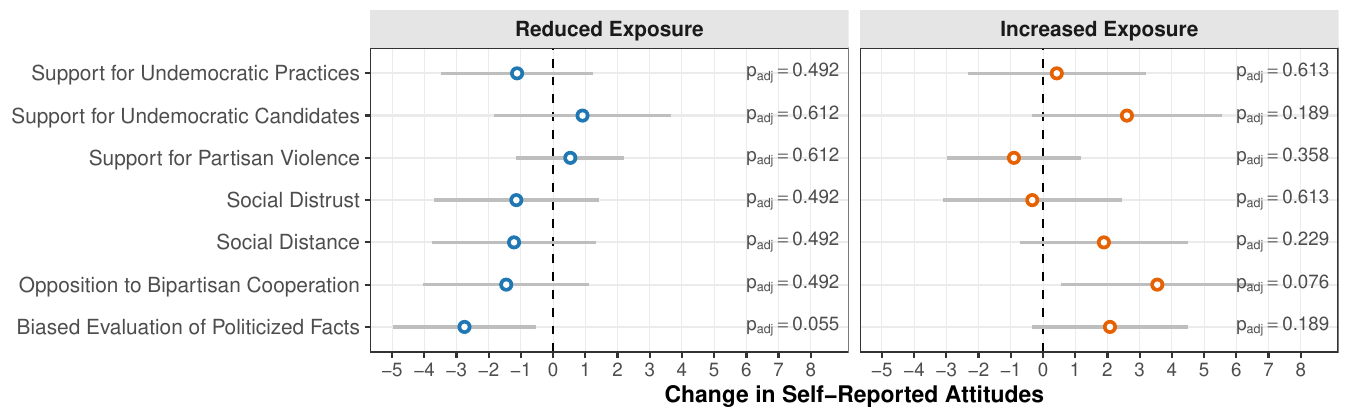}
\caption{Treatment effects on political attitudes estimated by regressing the participants' post-experiment survey responses on the treatment indicator, their pre-experiment survey response to the same question, and an indicator of the platform they were recruited on. The p-values are adjusted for multiple hypothesis testing as preregistered.}
\label{fig:si_sdc}
\end{figure}

As preregistered, we report the effect of the intervention on the seven remaining attitudes we used to identify the AAPA posts. 
We find that Reduced Exposure to AAPA content reduces the \textit{Biased Evaluation of Politicized Facts} while increasing the exposure increases the \textit{Opposition to Bipartisan Cooperation} (\Figref{fig:si_sdc}). However, none of these effects are significant after adjusting for multiple hypothesis testing as preregistered~\Appref{sec:p-values-corrections}. We note that our experiments were designed to detect effects on affective polarization~\Appref{sec:experimental-design}, and are likely not as well-powered to detect effects on these outcomes.

Consistent with \cite{voelkel2024megastudy}, our work also shows a larger treatment effect on partisan animosity compared to other factors such as support for undemocratic practices and support for partisan violence. These findings also align with recent studies which have observed that affective polarization is not robustly linked to antidemocratic tendencies~\cite{landry2024partisan}.

\begin{figure}[p]
\centering
\includegraphics[width=\textwidth]{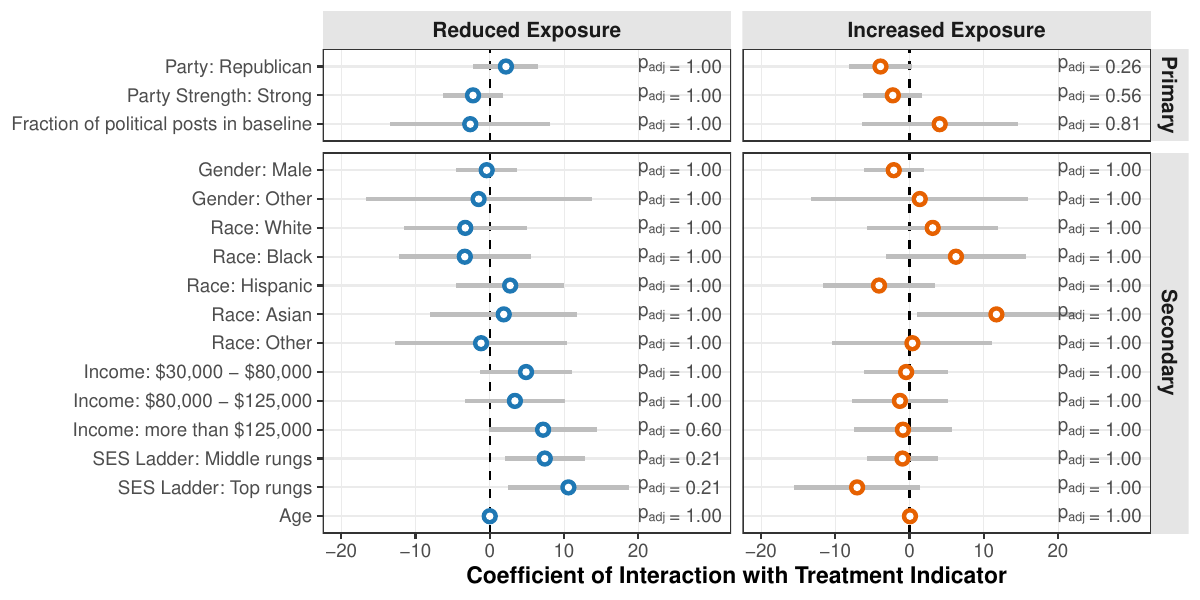}
\caption{Heterogeneous treatment effect estimates of reducing and increasing exposure to AAPA content on affective polarization based on the in-feed measurements. The p-values are adjusted for multiple hypothesis testing as preregistered.}
\label{fig:si_hte_affective_polarization_infeed}
\end{figure}

\begin{figure}[p]
\centering
\includegraphics[width=\textwidth]{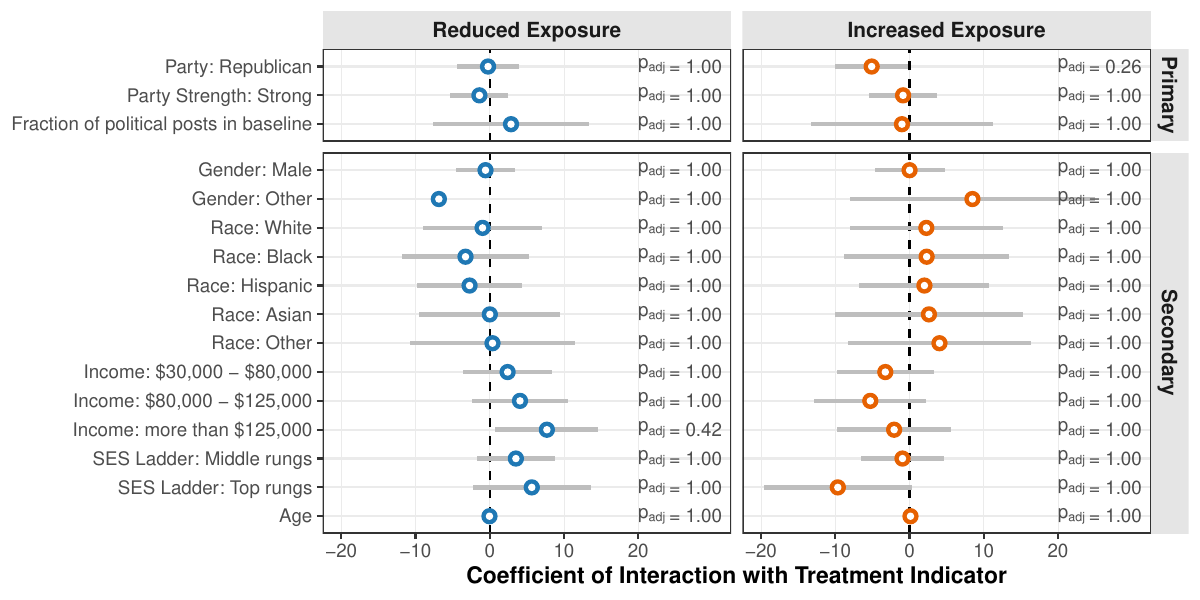}
\caption{Heterogeneous treatment effect estimates of reducing and increasing exposure to AAPA content on affective polarization based on the post-experiment measurements. The p-values are adjusted for multiple hypothesis testing as preregistered.}
\label{fig:si_hte_affective_polarization_prepost}
\end{figure}

\section{Heterogeneous Treatment Effects}
\label{sup:heterogeneous}
\Figref{fig:si_hte_affective_polarization_infeed} and \Figref{fig:si_hte_affective_polarization_prepost} show the heterogeneous treatment effects on affective polarization in the two experiments.  We use the same regression specifications as in the main analyses, but interact the treatment indicator with the moderator and run one regression per moderator. As in the covariate balance and attrition tests, we coarsened the education, income, and SES ladder covariates to avoid small categories. We consider only a subset of covariates and split them into primary and secondary moderators as preregistered. After adjusting the p-values for multiple hypothesis testing, none of the covariates reach statistical significance.

\section{Engagement}
\label{sup:engagement}

We examine three aspects of engagement: (i) activity frequency and duration, i.e., number of sessions and time spent on the platform, (ii) engagement volume, i.e., absolute number of posts viewed, liked, and reposted by the participants, and (iii) engagement rate, i.e., rate at which the participants liked or reposted the posts they viewed.

\textbf{Activity frequency and duration.}
We start by analyzing the number of sessions and the amount of time the participants spent on the platform. To track their passive activity on the platform, we configured our browser extension to send a request to our server every minute while the participant's browser tab with the X page was active. We delineate the end of a session and the beginning of a new one by identifying gaps of over one hour between consecutive requests. To calculate the time a participant spent on the platform per session, we summed the number of requests the browser sent during the session, excluding periods when the participant browsed other websites. To prevent overcounting unusually long sessions—likely caused by participants leaving their browsers open without actively using the platform—we capped all sessions longer than two hours at two hours. Finally, to compute the total time spent per user, we sum the time spent across all of their sessions. 

To estimate the effects of our interventions, we adjust for participants’ baseline activity and recruitment platform, similar to our main analysis (Tables~\ref{table:engagement_sessions} and ~\ref{table:engagement_timespent}). We report results from linear models but note that using linear models with log-transformed outcomes, Poisson models, or quasi-Poisson models yields similar results. We also note that changing the gap between sessions from one hour to 45 or 90 minutes, or adjusting the maximum session length from two hours to 1.5 or 2.5 hours, did not substantively affect the results reported below.

In the Reduced Exposure experiment, we find that participants in the treatment group had a similar number of sessions per day (estimate: -0.06; 95\% CI: [-0.16, 0.04]), but spent -4.71 minutes less per day (95\% CI: [-8.22, -1.20]) on the platform compared to those in the control group. In the Increased Exposure experiment, participants in the treatment group had both a similar number of sessions per day (estimate: -0.04; 95\% CI: [-0.16, 0.08]) and spent a similar amount of time (estimate: -1.76; 95\% CI: [-6.17, 2.66]) as those in the control group. 

\begin{figure}[t]
\centering
\includegraphics[width=\textwidth]{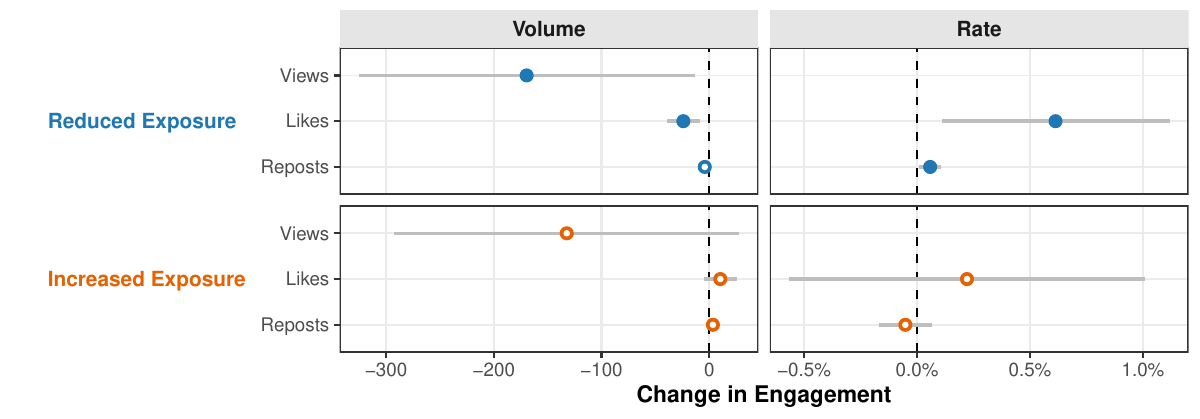}
\caption{Treatment effects on participants' volume and rate of engagement. The points show average marginal effects, and the error bars represent 95\% confidence intervals. Effects on engagement volume are estimated using quasi-Poisson regression models, while effects on engagement rates are estimated using Beta regression.}
\label{fig:engagement-main-figure}
\end{figure}

\textbf{Engagement volume. }
Next, we analyze the frequency of the participants' engagement with posts on the platform. As is common in social media data, we find that the distributions of our measures of absolute engagement are highly skewed: most participants view, like, and repost a small number of posts and a small number of participants view, like, and repost many posts. To appropriately model this skewness in the data, we fit quasi-Poisson models using a similar specification as our main analyses, i.e., adjusting for the participant's volume of the same type of engagement during the baseline period and the platform they were recruited on. We chose a quasi-Poisson regression as it effectively modeled the overdispersion in the data and was less sensitive to outliers than a negative binomial model. While our discussion focuses on the results obtained with quasi-Poisson models, we report the regression tables of the equivalent analyses with a linear model (Tables~\ref{table:engagement_n_view_1_sec_R}--\ref{table:engagement_n_retweet_I}). To ease the interpretation of the effects estimated by the quasi-Poisson models, we report average marginal effects. We count a post as viewed if it was in the viewport of the participant’s browser for at least one second.

In the Reduced Exposure experiment, we find that participants in the treated group viewed -169.52 fewer posts (95\% CI: [-325.48, -13.56]) over the 7-day intervention period (Figure~\ref{fig:engagement-main-figure}, left panel). They also liked -24.06 fewer posts (95\% CI: [-39.53, -8.59]) compared to those in the control condition. Participants in the treatment condition, however,  reposted a similar number of posts (estimate: -4.18; 95\% CI: [-9.89, 1.53]) as those in the control condition. In the Increased Exposure experiment, we do not find any statistically significant differences in the number of views (estimate: -132.3; 95\% CI: [-292.5, 27.94]), likes (estimate: 10.25; 95\% CI: [-5.23, 25.74]), or reposts (estimate: 3.38; 95\% CI: [-0.06, 6.82]) between the participants in the treatment and control conditions.

\textbf{Engagement rate. }
Finally, we examine the rate at which participants engaged with the posts they viewed. Unlike the engagement volume analysis, which counts views of at least one second, here we consider all views, since participants could engage even with posts that briefly entered their viewport. We use Beta regression models to appropriately account for the fact that the outcomes are rates. To accommodate rates of zero or one, we transform the outcomes using the commonly used transformation by Smithson and Verkuilen~\cite{smithson2006better}. In addition to the results from the Beta regression models, we also report regression tables from an equivalent analysis using linear models (Tables~\ref{table:engagement_f_likes_R}--\ref{table:engagement_f_retweets_I}). As in the engagement volume analysis, we use a model specification similar to our main analyses (for both the mean and precision parameters) and report average marginal effects.

In the Reduced Exposure experiment, we observe that although participants in the treatment group viewed and liked fewer posts overall, they engage with posts they viewed at higher rates than those in the control condition: 0.61\% higher like rate (95\% CI: [0.11\%, 1.12\%]) and 0.058\% higher repost rate (95\% CI: [0.009\%, 0.11\%]), as shown in Figure~\ref{fig:engagement-main-figure} (right panel). In the Increased Exposure experiment, we do not find any significant effects on the like (estimate: 0.22\%; 95\% CI: [-0.57\%, 1.01\%]) and repost (estimate: -0.052; 95\% CI: [-0.17\%, 0.066\%]) engagement rates.

Finally, as we discuss in Section~\ref{sec:experimental-design}, our experiments were designed to detect meaningful effects on affective polarization and are not necessarily powered to detect small effects on engagement. To contextualize the results, Table~\ref{table:engagement_MDEs} reports the observed and post-hoc Minimum Detectable Effects (MDE) for each of the engagement metrics discussed above.

\begin{table}
\centering
\small
{
    \renewcommand{\arraystretch}{0.95}
    
\begin{tabular}{l c c}
\toprule
 & \multicolumn{2}{c}{Number of Sessions (Daily)} \\
\cmidrule(lr){2-3}
 & Reduced Exposure & Increased Exposure \\
\midrule
(Intercept)                          & $0.43^{*}$       & $0.51^{*}$       \\
                                     & $ [ 0.32; 0.54]$ & $ [ 0.37; 0.64]$ \\
Condition: Treatment                 & $-0.06$          & $-0.04$          \\
                                     & $ [-0.16; 0.04]$ & $ [-0.16; 0.08]$ \\
Baseline: Number of Sessions (Daily) & $0.68^{*}$       & $0.65^{*}$       \\
                                     & $ [ 0.63; 0.72]$ & $ [ 0.59; 0.70]$ \\
Recruitment Platform: CloudResearch  & $0.19^{*}$       & $0.22^{*}$       \\
                                     & $ [ 0.08; 0.29]$ & $ [ 0.10; 0.34]$ \\
\midrule
Num. obs.                            & $622$            & $468$            \\
\bottomrule
\multicolumn{3}{l}{\scriptsize{$^*$ Null hypothesis value outside the confidence interval.}}
\end{tabular}

}
\caption{Engagement: Number of Sessions. Regression tables of the analysis of the number of sessions per day using linear models.}
\label{table:engagement_sessions}
\end{table}

\begin{table}
\centering
\small
{
    \renewcommand{\arraystretch}{0.95}
    
\begin{tabular}{l c c}
\toprule
 & \multicolumn{2}{c}{Time Spent (Daily in Minutes)} \\
\cmidrule(lr){2-3}
 & Reduced Exposure & Increased Exposure \\
\midrule
(Intercept)                             & $8.80^{*}$        & $8.89^{*}$        \\
                                        & $ [ 5.55; 12.06]$ & $ [ 4.77; 13.01]$ \\
Condition: Treatment                    & $-4.71^{*}$       & $-1.76$           \\
                                        & $ [-8.22; -1.20]$ & $ [-6.17;  2.66]$ \\
Baseline: Time Spent (Daily in Minutes) & $0.77^{*}$        & $0.79^{*}$        \\
                                        & $ [ 0.73;  0.81]$ & $ [ 0.73;  0.84]$ \\
Recruitment Platform: CloudResearch     & $4.03^{*}$        & $3.78$            \\
                                        & $ [ 0.39;  7.66]$ & $ [-0.69;  8.25]$ \\
\midrule
Num. obs.                               & $622$             & $468$             \\
\bottomrule
\multicolumn{3}{l}{\scriptsize{$^*$ Null hypothesis value outside the confidence interval.}}
\end{tabular}

}
\caption{Engagement: Time Spent. Regression tables of the analysis of the time spent in minutes per day using linear models.}
\label{table:engagement_timespent}
\end{table}

\begin{table}
\centering
\small
{
    \renewcommand{\arraystretch}{0.95}
    
\begin{tabular}{l c c}
\toprule
 & \multicolumn{2}{c}{Reduced Exposure, Number of Views} \\
\cmidrule(lr){2-3}
 & Linear & Quasi-Poisson \\
\midrule
(Intercept)                         & $326.62^{*}$         & $6.71^{*}$        \\
                                    & $ [ 209.39; 443.85]$ & $ [ 6.59;  6.84]$ \\
Condition: Treatment                & $-213.21^{*}$        & $-0.15^{*}$       \\
                                    & $ [-342.39; -84.02]$ & $ [-0.30; -0.01]$ \\
Baseline: Number of Views           & $1.96^{*}$           & $0.00^{*}$        \\
                                    & $ [   1.86;   2.07]$ & $ [ 0.00;  0.00]$ \\
Recruitment Platform: CloudResearch & $85.98$              & $0.14^{*}$        \\
                                    & $ [ -47.65; 219.61]$ & $ [ 0.00;  0.29]$ \\
\midrule
Num. obs.                           & $622$                & $622$             \\
\bottomrule
\multicolumn{3}{l}{\scriptsize{$^*$ Null hypothesis value outside the confidence interval.}}
\end{tabular}

}
\caption{Engagement: Reduced Exposure Experiment, Number of Views. Regression results for the number of posts viewed during the 7-day intervention period, using linear and quasi-Poisson~models.}
\label{table:engagement_n_view_1_sec_R}
\end{table}

\begin{table}
\centering
\small
{
    \renewcommand{\arraystretch}{0.95}
    
\begin{tabular}{l c c}
\toprule
 & \multicolumn{2}{c}{Increased Exposure, Number of Views} \\
\cmidrule(lr){2-3}
 & Linear & Quasi-Poisson \\
\midrule
(Intercept)                         & $311.00^{*}$         & $6.57^{*}$       \\
                                    & $ [ 181.04; 440.97]$ & $ [ 6.44; 6.71]$ \\
Condition: Treatment                & $-54.96$             & $-0.13$          \\
                                    & $ [-199.62;  89.69]$ & $ [-0.28; 0.03]$ \\
Baseline: Number of Views           & $1.83^{*}$           & $0.00^{*}$       \\
                                    & $ [   1.70;   1.96]$ & $ [ 0.00; 0.00]$ \\
Recruitment Platform: CloudResearch & $119.21$             & $0.19^{*}$       \\
                                    & $ [ -27.24; 265.66]$ & $ [ 0.04; 0.35]$ \\
\midrule
Num. obs.                           & $468$                & $468$            \\
\bottomrule
\multicolumn{3}{l}{\scriptsize{$^*$ Null hypothesis value outside the confidence interval.}}
\end{tabular}

}
\caption{Engagement: Increased Exposure Experiment, Number of Views. Regression results for the number of posts viewed during the 7-day intervention period, using linear and quasi-Poisson~models.}
\label{table:engagement_n_view_1_sec_I}
\end{table}

\begin{table}
\centering
\small
{
    \renewcommand{\arraystretch}{0.95}
    
\begin{tabular}{l c c}
\toprule
 & \multicolumn{2}{c}{Reduced Exposure, Number of Likes} \\
\cmidrule(lr){2-3}
 & Linear & Quasi-Poisson \\
\midrule
(Intercept)                         & $16.41^{*}$        & $3.72^{*}$        \\
                                    & $ [  4.18; 28.63]$ & $ [ 3.46;  3.98]$ \\
Condition: Treatment                & $-19.57^{*}$       & $-0.47^{*}$       \\
                                    & $ [-34.06; -5.08]$ & $ [-0.78; -0.16]$ \\
Baseline: Number of Likes           & $1.93^{*}$         & $0.00^{*}$        \\
                                    & $ [  1.85;  2.01]$ & $ [ 0.00;  0.00]$ \\
Recruitment Platform: CloudResearch & $9.90$             & $0.25$            \\
                                    & $ [ -5.09; 24.88]$ & $ [-0.05;  0.56]$ \\
\midrule
Num. obs.                           & $622$              & $622$             \\
\bottomrule
\multicolumn{3}{l}{\scriptsize{$^*$ Null hypothesis value outside the confidence interval.}}
\end{tabular}

}
\caption{Engagement: Reduced Exposure Experiment, Number of Likes. Regression results for the number of posts liked during the 7-day intervention period, using linear and quasi-Poisson models.}
\label{table:engagement_n_favorite_R}
\end{table}

\begin{table}
\centering
\small
{
    \renewcommand{\arraystretch}{0.95}
    
\begin{tabular}{l c c}
\toprule
 & \multicolumn{2}{c}{Increased Exposure, Number of Likes} \\
\cmidrule(lr){2-3}
 & Linear & Quasi-Poisson \\
\midrule
(Intercept)                         & $2.44$             & $3.08^{*}$       \\
                                    & $ [ -8.17; 13.05]$ & $ [ 2.76; 3.39]$ \\
Condition: Treatment                & $-0.73$            & $0.22$           \\
                                    & $ [-13.20; 11.74]$ & $ [-0.11; 0.55]$ \\
Baseline: Number of Likes           & $2.43^{*}$         & $0.01^{*}$       \\
                                    & $ [  2.29;  2.56]$ & $ [ 0.01; 0.01]$ \\
Recruitment Platform: CloudResearch & $11.85$            & $0.46^{*}$       \\
                                    & $ [ -0.76; 24.47]$ & $ [ 0.14; 0.78]$ \\
\midrule
Num. obs.                           & $468$              & $468$            \\
\bottomrule
\multicolumn{3}{l}{\scriptsize{$^*$ Null hypothesis value outside the confidence interval.}}
\end{tabular}

}
\caption{Engagement: Increased Exposure Experiment, Number of Likes. Regression results for the number of posts liked during the 7-day intervention period, using linear and quasi-Poisson models.}
\label{table:engagement_n_favorite_I}
\end{table}

\begin{table}
\centering
\small
{
    \renewcommand{\arraystretch}{0.95}
    
\begin{tabular}{l c c}
\toprule
 & \multicolumn{2}{c}{Reduced Exposure, Number of Reposts} \\
\cmidrule(lr){2-3}
 & Linear & Quasi-Poisson \\
\midrule
(Intercept)                         & $4.29^{*}$       & $2.13^{*}$       \\
                                    & $ [ 1.38; 7.21]$ & $ [ 1.60; 2.66]$ \\
Condition: Treatment                & $-2.45$          & $-0.51$          \\
                                    & $ [-5.96; 1.07]$ & $ [-1.28; 0.25]$ \\
Baseline: Number of Reposts         & $2.07^{*}$       & $0.01^{*}$       \\
                                    & $ [ 2.02; 2.11]$ & $ [ 0.00; 0.01]$ \\
Recruitment Platform: CloudResearch & $-1.81$          & $-0.71$          \\
                                    & $ [-5.45; 1.82]$ & $ [-1.59; 0.18]$ \\
\midrule
Num. obs.                           & $622$            & $622$            \\
\bottomrule
\multicolumn{3}{l}{\scriptsize{$^*$ Null hypothesis value outside the confidence interval.}}
\end{tabular}

}
\caption{Engagement: Reduced Exposure Experiment, Number of Reposts. Regression results for the number of reposts during the 7-day intervention period, using linear and quasi-Poisson models.}
\label{table:engagement_n_retweet_R}
\end{table}

\begin{table}
\centering
\small
{
    \renewcommand{\arraystretch}{0.95}
    
\begin{tabular}{l c c}
\toprule
 & \multicolumn{2}{c}{Increased Exposure, Number of Reposts} \\
\cmidrule(lr){2-3}
 & Linear & Quasi-Poisson \\
\midrule
(Intercept)                         & $0.49$           & $0.98^{*}$       \\
                                    & $ [-3.27; 4.24]$ & $ [ 0.45; 1.52]$ \\
Condition: Treatment                & $0.92$           & $0.46$           \\
                                    & $ [-3.58; 5.41]$ & $ [-0.02; 0.93]$ \\
Baseline: Number of Reposts         & $2.27^{*}$       & $0.02^{*}$       \\
                                    & $ [ 2.14; 2.39]$ & $ [ 0.02; 0.03]$ \\
Recruitment Platform: CloudResearch & $0.99$           & $0.40$           \\
                                    & $ [-3.56; 5.55]$ & $ [-0.19; 0.99]$ \\
\midrule
Num. obs.                           & $468$            & $468$            \\
\bottomrule
\multicolumn{3}{l}{\scriptsize{$^*$ Null hypothesis value outside the confidence interval.}}
\end{tabular}

}
\caption{Engagement: Increased Exposure Experiment, Number of Reposts. Regression results for the number of reposts during the 7-day intervention period, using linear and quasi-Poisson models.}
\label{table:engagement_n_retweet_I}
\end{table}

\begin{table}
\centering
\small

\begin{tabular}{l c c}
\toprule
 & \multicolumn{2}{c}{Reduced Exposure, Like Rate} \\
\cmidrule(lr){2-3}
 & Linear & Beta \\
\midrule
(Intercept)                                    & $0.00$           & $-4.67^{*}$       \\
                                               & $ [-0.00; 0.01]$ & $ [-4.83; -4.50]$ \\
Condition: Treatment                           & $0.00$           & $0.22^{*}$        \\
                                               & $ [-0.00; 0.01]$ & $ [ 0.04;  0.40]$ \\
Baseline: Like Rate                            & $0.86^{*}$       & $14.66^{*}$       \\
                                               & $ [ 0.83; 0.89]$ & $ [13.84; 15.48]$ \\
Recruitment Platform: CloudResearch            & $0.00$           & $0.06$            \\
                                               & $ [-0.01; 0.01]$ & $ [-0.13;  0.25]$ \\
Precision, (Intercept)                         &                  & $4.08^{*}$        \\
                                               &                  & $ [ 3.85;  4.30]$ \\
Precision, Condition: Treatment                &                  & $-0.32^{*}$       \\
                                               &                  & $ [-0.57; -0.07]$ \\
Precision, Baseline: Like Rate                 &                  & $-6.91^{*}$       \\
                                               &                  & $ [-7.93; -5.89]$ \\
Precision, Recruitment Platform: CloudResearch &                  & $-0.09$           \\
                                               &                  & $ [-0.34;  0.17]$ \\
\midrule
Num. obs.                                      & $622$            & $622$             \\
\bottomrule
\multicolumn{3}{l}{\scriptsize{$^*$ Null hypothesis value outside the confidence interval.}}
\end{tabular}

\caption{Engagement: Reduced Exposure Experiment, Like Rate. Regression results for the like rate, i.e., number of posts liked over number of posts viewed, during the 7-day intervention period, using linear and Beta models.}
\label{table:engagement_f_likes_R}
\end{table}

\begin{table}
\centering
\small

\begin{tabular}{l c c}
\toprule
 & \multicolumn{2}{c}{Increased Exposure, Like Rate} \\
\cmidrule(lr){2-3}
 & Linear & Beta \\
\midrule
(Intercept)                                    & $0.01$           & $-4.18^{*}$       \\
                                               & $ [-0.00; 0.02]$ & $ [-4.38; -3.99]$ \\
Condition: Treatment                           & $-0.00$          & $0.06$            \\
                                               & $ [-0.01; 0.01]$ & $ [-0.16;  0.29]$ \\
Baseline: Like Rate                            & $0.82^{*}$       & $11.38^{*}$       \\
                                               & $ [ 0.76; 0.87]$ & $ [10.44; 12.33]$ \\
Recruitment Platform: CloudResearch            & $0.00$           & $0.23$            \\
                                               & $ [-0.01; 0.01]$ & $ [-0.00;  0.46]$ \\
Precision, (Intercept)                         &                  & $3.59^{*}$        \\
                                               &                  & $ [ 3.33;  3.84]$ \\
Precision, Condition: Treatment                &                  & $-0.17$           \\
                                               &                  & $ [-0.47;  0.12]$ \\
Precision, Baseline: Like Rate                 &                  & $-5.92^{*}$       \\
                                               &                  & $ [-7.05; -4.79]$ \\
Precision, Recruitment Platform: CloudResearch &                  & $-0.36^{*}$       \\
                                               &                  & $ [-0.66; -0.06]$ \\
\midrule
Num. obs.                                      & $468$            & $468$             \\
\bottomrule
\multicolumn{3}{l}{\scriptsize{$^*$ Null hypothesis value outside the confidence interval.}}
\end{tabular}

\caption{Engagement: Increased Exposure Experiment, Like Rate. Regression results for the like rate, i.e., number of posts liked over number of posts viewed, during the 7-day intervention period, using linear and Beta models.}
\label{table:engagement_f_likes_I}
\end{table}

\begin{table}
\centering
\small

\begin{tabular}{l c c}
\toprule
 & \multicolumn{2}{c}{Reduced Exposure, Repost Rate} \\
\cmidrule(lr){2-3}
 & Linear & Beta \\
\midrule
(Intercept)                                    & $0.00$           & $-6.24^{*}$         \\
                                               & $ [-0.00; 0.00]$ & $ [ -6.38;  -6.11]$ \\
Condition: Treatment                           & $0.00$           & $0.19^{*}$          \\
                                               & $ [-0.00; 0.00]$ & $ [  0.03;   0.35]$ \\
Baseline: Repost Rate                          & $0.98^{*}$       & $44.61^{*}$         \\
                                               & $ [ 0.95; 1.00]$ & $ [ 40.38;  48.83]$ \\
Recruitment Platform: CloudResearch            & $-0.00$          & $-0.11$             \\
                                               & $ [-0.00; 0.00]$ & $ [ -0.27;   0.05]$ \\
Precision, (Intercept)                         &                  & $6.22^{*}$          \\
                                               &                  & $ [  6.01;   6.43]$ \\
Precision, Condition: Treatment                &                  & $-0.30^{*}$         \\
                                               &                  & $ [ -0.55;  -0.05]$ \\
Precision, Baseline: Repost Rate               &                  & $-42.55^{*}$        \\
                                               &                  & $ [-46.96; -38.14]$ \\
Precision, Recruitment Platform: CloudResearch &                  & $0.31^{*}$          \\
                                               &                  & $ [  0.06;   0.57]$ \\
\midrule
Num. obs.                                      & $622$            & $622$               \\
\bottomrule
\multicolumn{3}{l}{\scriptsize{$^*$ Null hypothesis value outside the confidence interval.}}
\end{tabular}

\caption{Engagement: Reduced Exposure Experiment, Repost Rate. Regression results for the repost rate, i.e., number of posts reposted over number of posts viewed, during the 7-day intervention period, using linear and Beta models.}
\label{table:engagement_f_retweets_R}
\end{table}

\begin{table}
\centering
\small

\begin{tabular}{l c c}
\toprule
 & \multicolumn{2}{c}{Increased Exposure, Repost Rate} \\
\cmidrule(lr){2-3}
 & Linear & Beta \\
\midrule
(Intercept)                                    & $0.00$           & $-5.92^{*}$         \\
                                               & $ [-0.00; 0.00]$ & $ [ -6.07;  -5.78]$ \\
Condition: Treatment                           & $0.00$           & $-0.08$             \\
                                               & $ [-0.00; 0.00]$ & $ [ -0.28;   0.11]$ \\
Baseline: Repost Rate                          & $0.89^{*}$       & $35.43^{*}$         \\
                                               & $ [ 0.81; 0.96]$ & $ [ 32.04;  38.82]$ \\
Recruitment Platform: CloudResearch            & $0.00$           & $0.78^{*}$          \\
                                               & $ [-0.00; 0.01]$ & $ [  0.56;   1.00]$ \\
Precision, (Intercept)                         &                  & $6.04^{*}$          \\
                                               &                  & $ [  5.80;   6.28]$ \\
Precision, Condition: Treatment                &                  & $0.06$              \\
                                               &                  & $ [ -0.23;   0.35]$ \\
Precision, Baseline: Repost Rate               &                  & $-25.04^{*}$        \\
                                               &                  & $ [-29.03; -21.05]$ \\
Precision, Recruitment Platform: CloudResearch &                  & $-1.58^{*}$         \\
                                               &                  & $ [ -1.89;  -1.28]$ \\
\midrule
Num. obs.                                      & $468$            & $468$               \\
\bottomrule
\multicolumn{3}{l}{\scriptsize{$^*$ Null hypothesis value outside the confidence interval.}}
\end{tabular}

\caption{Engagement: Increased Exposure Experiment, Repost Rate. Regression results for the repost rate, i.e., number of posts reposted over number of posts viewed, during the 7-day intervention period, using linear and Beta models.}
\label{table:engagement_f_retweets_I}
\end{table}

\begin{table}
\centering
\small
\begin{tabular}{lcccc}
\toprule
 & \multicolumn{2}{c}{Reduced Exposure} & \multicolumn{2}{c}{Increased Exposure} \\
\cmidrule(l{2pt}r{2pt}){2-3} \cmidrule(l{2pt}r{2pt}){4-5}
Metric                        & Observed Effect & MDE      & Observed Effect & MDE \\
\midrule
Number of Sessions (Daily)    & -0.06           & 0.14     & -0.04           & 0.17   \\
Time spent (Daily in Minutes) & -4.71           & 5.02     & -1.76           & 6.31   \\
\midrule
Number of Views (Total)       & -170.00         & 249.41   & -132.00         & 254.99 \\
Number of Likes (Total)       & -24.10          & 29.84    & 10.30           & 27.46  \\
Number of Reposts (Total)     & -4.18           & 17.67    & 3.38            & 7.33   \\
\midrule
Like Rate                     & 0.61\%          & 0.79\%   & 0.22\%          & 1.25\% \\
Repost Rate                   & 0.06\%          & 0.08\%   & -0.05\%         & 0.19\% \\
\bottomrule
\end{tabular}
\caption{Engagement: Observed and post-hoc Minimum Detectable Effects (MDE) in the two experiments (power = 0.8, $\alpha = 0.05$). For the engagement volume (number of views, likes, and reports) and engagement rate (like and repost rate) metrics, we first computed the MDE coefficients from the non-linear models used to estimate the treatment effects, and then calculated the corresponding average marginal effects, which are reported here.}
\label{table:engagement_MDEs}
\end{table}

\section{Population Average Treatment Effects}
\label{sec:reweighting}
To examine the generalizability of our treatment effect estimates, we reweighted our sample to reflect the X user population in terms of party affiliation, education, and race. To determine the proportion of X users across these dimensions, we relied on the 2024 National Public Opinion Reference Survey conducted by Pew Research. The survey was fielded between February 1 and June 10, 2024, just before the start of our experiment (July 7, 2024) and after the rebranding of Twitter to X.
Since we did not preregister this analysis, we minimize the number of subjective modeling decisions by considering the same participant characteristics and following a similar reweighting procedure as the Facebook and Instagram Election studies~\cite{nyhan2023like, guess2023social}. We reweight our sample based on the following characteristics: (a) Party affiliation: Democrat vs. Republican, considering Independents as partisans depending on which party they lean towards; (b) Education: less than a college degree vs. college degree or more; and (c) Race: white vs. non-white.

We used raking to generate the weights that calibrate to the population estimates of these participant characteristics. We did not trim the weights since there were no extreme values. We consider the participants who completed the post-experiment survey and at least one affective polarization or emotions in-feed survey during the intervention period (N=1,112). Compared to the X population, our sample included fewer Republicans (Reduced Exposure sample: 31.6\%, Increased Exposure sample: 33\%, X population: 48.4\%), fewer participants with less than a college degree (Reduced Exposure sample: 43.7\%, Increased Exposure sample: 50.9\%, X population: 65.6\%), and a similar proportion of white participants (Reduced Exposure sample: 75.7\%, Increased Exposure sample: 76.4\%, X population: 74.4\%). We compute the effective sample size~\cite{kish1965survey} of the reweighted sample to estimate the proportional increase in variance due to reweighting. We find that our reweighted sample corresponds to a 23.3\% smaller sample size in the Reduced Exposure experiment and a 17.3\% smaller sample size in the Increased Exposure experiment.

\begin{figure}[t!]
\centering
\includegraphics[width=0.95\textwidth]{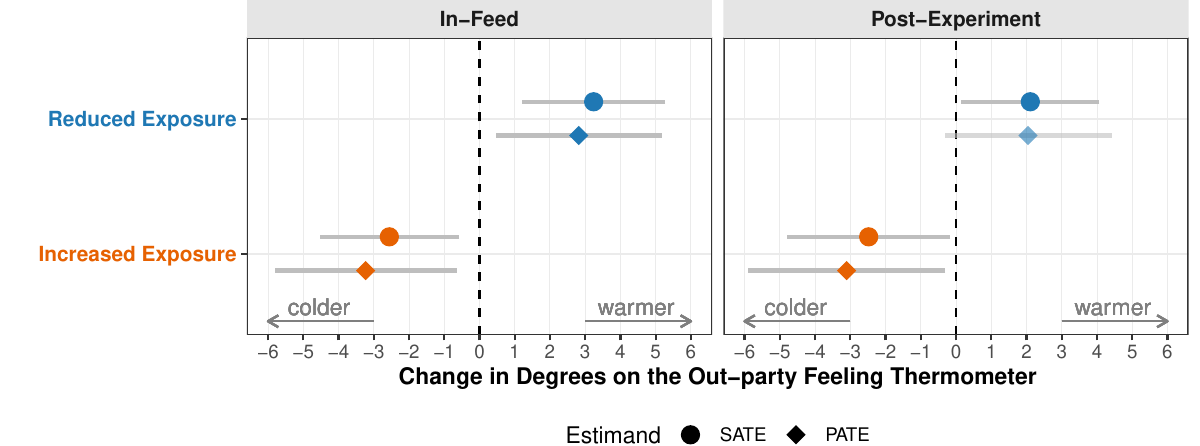}
\caption{Affective polarization: Comparison between the Population Average Treatment Effect estimates (PATE)---reweighting the sample to be representative of the X population in terms of party affiliation, education, and race---and the Sample Average Treatment Effects (SATE) estimates reported in the main text (Figure~\ref{fig:results_polarization}). The error bars represent 95\% confidence intervals. Point estimates for which the 95\% confidence intervals include zero are displayed with lighter colors.}
\label{fig:affective-pol-pate}
\end{figure}

\begin{figure}[h!]
\centering
\includegraphics[width=0.95\textwidth]{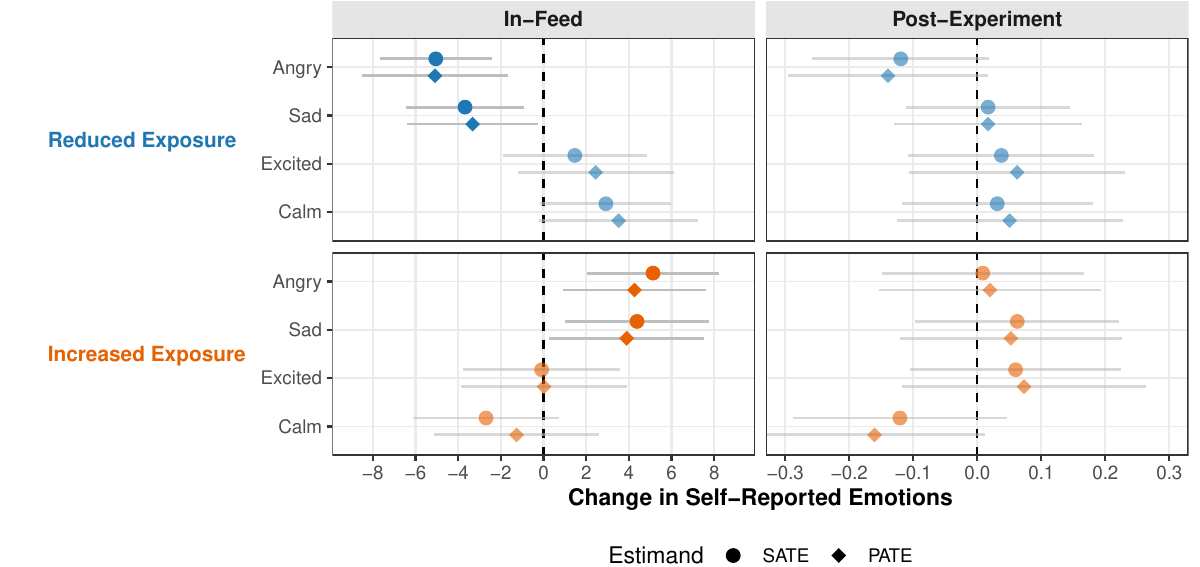}
\caption{Emotions: Comparison between the Population Average Treatment Effect estimates (PATE)---reweighting the sample to be representative of the X population in terms of party affiliation, education, and race---and the Sample Average Treatment Effects (SATE) estimates reported in the main text (Figure~\ref{fig:results_emotions}). The error bars represent 95\% confidence intervals. Point estimates for which the 95\% confidence intervals include zero are displayed with lighter colors.}
\label{fig:emotions-pol-pate}
\end{figure}

To incorporate the raking weights in our analyses, we used the \verb|svyglm| function from the \verb|survey| R package~\cite{rsurvey2024, lumley2010}. Since we are unaware of any well-established method for incorporating raking weights into mixed-effect models, we first averaged each participant's responses to the in-feed surveys and then ran weighted regression using \verb|svyglm|. We note that when we apply the same averaging approach to the unweighted data and run an ordinary linear regression, the resulting point estimates and confidence intervals are similar to those obtained with the mixed-effects models.

We find that the Population Average Treatment Effect (PATE) estimates closely align with the Sample Average Treatment Effect (SATE) estimates (Figures \ref{fig:affective-pol-pate} and \ref{fig:emotions-pol-pate}). The significant treatment effects observed for affective polarization and emotional outcomes remain significant after reweighting. The only exception is the treatment effect of reduced exposure to AAPA post-experiment: although the PATE point estimate is similar to the SATE, its confidence interval now includes zero, reflecting a loss of precision from reweighting. In the Increased Exposure experiment, the PATE estimate for affective polarization in-feed is slightly larger than the SATE, while in the Reduced Exposure experiment, it is slightly smaller. The PATE estimates of the treatment effects on emotions are similar to the SATE estimates.

\section{Supplementary Results}
\label{sup:results}

\subsection{Affective Polarization: Regression Tables}
\label{sup:regression_tables}

The preregistered regression analysis for the in-feed effect is reported in \Tabref{table:affective_polarization_infeed_regressions}, and the regression analyses for the post-experiment effect are reported in \Tabref{table:affective_polarization_prepost_regressions}.

\Tabref{table:polarization_infeed_all} and \Tabref{table:polarization_prepost_all} show the equivalent regressions but include the entire population (N=1,256), regardless of whether the participants completed only the post-experiment survey or only at least one in-feed survey. For both experiments, the treatment effect remains statistically~significant. 

We note that in this sample there is also no evidence of covariate imbalance (reduce, post-experiment: $p=0.86$; reduce, in-feed: $p=0.59$; increase, post-experiment: $p=0.77$; increase, in-feed: $p = 0.31$). The attrition rates were as follows---reduce, post-experiment, treatment: 27.7\% vs. control: 26.1\%; reduce, in-feed, treatment: 33.1\% vs. control: 36.4\%; increase, post-experiment, treatment: 19.4\% vs. control: 22.0\%; increase, in-feed, treatment: 26.2\% vs. control: 24.2\%. We also find no evidence of differential attrition rates (reduce, post-experiment: $p=0.55$; reduce, in-feed: $p=0.28$; increase, post-experiment: $p=0.4$; increase, in-feed: $p=0.56$), or differential attrition patterns (reduce, post-experiment: $p = 0.49$; reduce, in-feed: $p = 0.69$; increase, post-experiment: $p = 0.78$; increase, in-feed: $p = 0.98$). The testing procedures are described in Section~\ref{sec:experimental-design}.

\begin{table}
\centering
\small
{
    \renewcommand{\arraystretch}{0.9}
    
\begin{tabular}{l c c}
\toprule
 & \multicolumn{2}{c}{In-Feed} \\
\cmidrule(lr){2-3}
 & Reduced Exposure & Increased Exposure \\
\midrule
(Intercept)                                 & $0.90$           & $3.06^{*}$        \\
                                            & $ [-1.08; 2.87]$ & $ [ 1.08;  5.05]$ \\
Condition: Treatment                        & $3.24^{*}$       & $-2.56^{*}$       \\
                                            & $ [ 1.21; 5.27]$ & $ [-4.53; -0.59]$ \\
Baseline: Mean Outparty Feeling Thermometer & $0.87^{*}$       & $0.84^{*}$        \\
                                            & $ [ 0.83; 0.92]$ & $ [ 0.79;  0.88]$ \\
Recruitment Platform: CloudResearch         & $-0.58$          & $-1.58$           \\
                                            & $ [-2.69; 1.54]$ & $ [-3.58;  0.43]$ \\
\midrule
Num. obs.                                   & $6116$           & $4778$            \\
Num. groups.                       & $622$            & $468$             \\
\bottomrule
\multicolumn{3}{l}{\scriptsize{$^*$ Null hypothesis value outside the confidence interval.}}
\end{tabular}

}
\caption{Affective Polarization: In-Feed. Regression tables of the analysis of the affective polarization outcome for the in-feed survey responses.}
\label{table:affective_polarization_infeed_regressions}
\end{table}

\begin{table}
\centering
\small
{
    \renewcommand{\arraystretch}{0.9}
    
\begin{tabular}{l c c}
\toprule
 & \multicolumn{2}{c}{Post-Experiment} \\
\cmidrule(lr){2-3}
 & Reduced Exposure & Increased Exposure \\
\midrule
(Intercept)                              & $1.26$           & $3.89^{*}$        \\
                                         & $ [-0.76; 3.27]$ & $ [ 1.47;  6.31]$ \\
Condition: Treatment                     & $2.11^{*}$       & $-2.48^{*}$       \\
                                         & $ [ 0.15; 4.06]$ & $ [-4.79; -0.17]$ \\
Pre-survey: Outparty Feeling Thermometer & $0.81^{*}$       & $0.83^{*}$        \\
                                         & $ [ 0.77; 0.85]$ & $ [ 0.77;  0.88]$ \\
Recruitment Platform: CloudResearch      & $-0.76$          & $-2.72^{*}$       \\
                                         & $ [-2.79; 1.28]$ & $ [-5.06; -0.37]$ \\
\midrule
Num. obs.                                & $622$            & $468$             \\
\bottomrule
\multicolumn{3}{l}{\scriptsize{$^*$ Null hypothesis value outside the confidence interval.}}
\end{tabular}

}
\caption{Affective Polarization: Post-Experiment. Regression tables of the analysis of the affective polarization outcome for the post-experiment survey responses.}
\label{table:affective_polarization_prepost_regressions}
\end{table}

\begin{table}
\centering
\small
{
    \renewcommand{\arraystretch}{0.9}
    
\begin{tabular}{l c c}
\toprule
 & \multicolumn{2}{c}{In-Feed} \\
\cmidrule(lr){2-3}
 & Reduced Exposure & Increased Exposure \\
\midrule
(Intercept)                                 & $0.81$           & $2.42^{*}$        \\
                                            & $ [-1.13; 2.75]$ & $ [ 0.44;  4.39]$ \\
Condition: Treatment                        & $3.30^{*}$       & $-2.45^{*}$       \\
                                            & $ [ 1.30; 5.30]$ & $ [-4.42; -0.48]$ \\
Baseline: Mean Outparty Feeling Thermometer & $0.88^{*}$       & $0.85^{*}$        \\
                                            & $ [ 0.83; 0.92]$ & $ [ 0.81;  0.89]$ \\
Recruitment Platform: CloudResearch         & $-0.58$          & $-0.82$           \\
                                            & $ [-2.68; 1.51]$ & $ [-2.84;  1.19]$ \\
\midrule
Num. obs.                                   & $6273$           & $4871$            \\
Num. groups.                       & $649$            & $499$             \\
\bottomrule
\multicolumn{3}{l}{\scriptsize{$^*$ Null hypothesis value outside the confidence interval.}}
\end{tabular}

}
\caption{Full population -- Affective Polarization: In-Feed. Regression tables of the analysis of the affective polarization outcome for the in-feed survey responses.}
\label{table:polarization_infeed_all}
\end{table}

\begin{table}
\centering
\small
{
    \renewcommand{\arraystretch}{0.9}
    
\begin{tabular}{l c c}
\toprule
 & \multicolumn{2}{c}{Post-Experiment} \\
\cmidrule(lr){2-3}
 & Reduced Exposure & Increased Exposure \\
\midrule
(Intercept)                              & $1.72$           & $3.63^{*}$        \\
                                         & $ [-0.14; 3.59]$ & $ [ 1.34;  5.91]$ \\
Condition: Treatment                     & $1.87^{*}$       & $-2.24^{*}$       \\
                                         & $ [ 0.05; 3.69]$ & $ [-4.40; -0.08]$ \\
Pre-survey: Outparty Feeling Thermometer & $0.81^{*}$       & $0.84^{*}$        \\
                                         & $ [ 0.77; 0.85]$ & $ [ 0.80;  0.89]$ \\
Recruitment Platform: CloudResearch      & $-0.79$          & $-3.07^{*}$       \\
                                         & $ [-2.72; 1.14]$ & $ [-5.30; -0.84]$ \\
\midrule
Num. obs.                                & $727$            & $529$             \\
\bottomrule
\multicolumn{3}{l}{\scriptsize{$^*$ Null hypothesis value outside the confidence interval.}}
\end{tabular}

}
\caption{Full population -- Affective Polarization: Post-Experiment. Regression tables of the analysis of the affective polarization outcome for the post-experiment survey responses.}
\label{table:polarization_prepost_all}
\end{table}

\subsection{Emotions: Regression Tables}
\label{sup:emotions_regression_tables}

The preregistered regression analysis for the in-feed effect in the reduced exposure is reported in \Tabref{table:emotions_infeed_reduce} and increased exposure in \Tabref{table:emotions_infeed_increase}.
The post-experiment effect regression results for the reduced exposure experiment are reported in \Tabref{table:emotions_prepost_reduce} and for the increased exposure in \Tabref{table:emotions_prepost_increase}.

\Tabref{table:emotions_infeed_reduce_all}, \Tabref{table:emotions_infeed_increase_all}, \Tabref{table:emotions_prepost_reduce_all}, and \Tabref{table:emotions_prepost_increase_all} report the equivalent regressions but include the entire population (N=1,256), regardless of whether the participants completed only the post-experiment survey or only at least one in-feed survey. The results are qualitatively similar, the variation in negative emotions is statistically significant in both experiments: increased in the increased exposure experiment and reduced in the reduced exposure experiment. 

\begin{table}
\centering
\scriptsize
\resizebox{\textwidth}{!}{%
    
\begin{tabular}{l c c c c}
\toprule
 & \multicolumn{4}{c}{Reduced Exposure, In-Feed} \\
\cmidrule(lr){2-5}
 & Angry & Sad & Excited & Calm \\
\midrule
(Intercept)                         & $9.26^{*}$        & $7.77^{*}$        & $12.53^{*}$       & $17.67^{*}$       \\
                                    & $ [ 6.67; 11.85]$ & $ [ 5.04; 10.49]$ & $ [ 9.03; 16.04]$ & $ [12.78; 22.57]$ \\
Condition: Treatment                & $-5.05^{*}$       & $-3.68^{*}$       & $1.47$            & $2.92$            \\
                                    & $ [-7.68; -2.42]$ & $ [-6.42; -0.94]$ & $ [-1.91;  4.84]$ & $ [-0.15;  5.99]$ \\
Baseline: Mean Answer               & $0.54^{*}$        & $0.68^{*}$        & $0.75^{*}$        & $0.68^{*}$        \\
                                    & $ [ 0.47;  0.62]$ & $ [ 0.60;  0.75]$ & $ [ 0.68;  0.82]$ & $ [ 0.61;  0.75]$ \\
Recruitment Platform: CloudResearch & $0.56$            & $2.42$            & $-5.05^{*}$       & $-3.09$           \\
                                    & $ [-2.16;  3.27]$ & $ [-0.41;  5.24]$ & $ [-8.54; -1.55]$ & $ [-6.25;  0.08]$ \\
\midrule
Num. obs.                           & $3057$            & $3007$            & $2990$            & $3098$            \\
Num. groups.               & $574$             & $566$             & $572$             & $567$             \\
\bottomrule
\multicolumn{5}{l}{\scriptsize{$^*$ Null hypothesis value outside the confidence interval.}}
\end{tabular}

}
\caption{Emotions: Reduced Exposure Experiment, In-Feed. Regression tables of the analysis of the emotion outcomes for the in-feed survey responses in the reduced exposure experiment.}
\label{table:emotions_infeed_reduce}
\end{table}

\begin{table}
\centering
\scriptsize
\resizebox{\textwidth}{!}{%
    
\begin{tabular}{l c c c c}
\toprule
 & \multicolumn{4}{c}{Increased Exposure, In-Feed} \\
\cmidrule(lr){2-5}
 & Angry & Sad & Excited & Calm \\
\midrule
(Intercept)                         & $5.98^{*}$       & $6.43^{*}$      & $14.91^{*}$       & $20.87^{*}$       \\
                                    & $ [ 3.05; 8.90]$ & $ [3.07; 9.79]$ & $ [10.94; 18.89]$ & $ [15.33; 26.40]$ \\
Condition: Treatment                & $5.13^{*}$       & $4.38^{*}$      & $-0.09$           & $-2.69$           \\
                                    & $ [ 2.05; 8.22]$ & $ [1.00; 7.76]$ & $ [-3.75;  3.57]$ & $ [-6.09;  0.70]$ \\
Baseline: Mean Answer               & $0.56^{*}$       & $0.55^{*}$      & $0.68^{*}$        & $0.66^{*}$        \\
                                    & $ [ 0.47; 0.64]$ & $ [0.46; 0.64]$ & $ [ 0.60;  0.76]$ & $ [ 0.59;  0.74]$ \\
Recruitment Platform: CloudResearch & $1.46$           & $5.10^{*}$      & $-6.31^{*}$       & $-3.93^{*}$       \\
                                    & $ [-1.65; 4.57]$ & $ [1.70; 8.50]$ & $ [-9.99; -2.62]$ & $ [-7.35; -0.51]$ \\
\midrule
Num. obs.                           & $2451$           & $2380$          & $2434$            & $2409$            \\
Num. groups.               & $443$            & $442$           & $441$             & $453$             \\
\bottomrule
\multicolumn{5}{l}{\scriptsize{$^*$ Null hypothesis value outside the confidence interval.}}
\end{tabular}

}
\caption{Emotions: Increased Exposure Experiment, In-Feed. Regression tables of the analysis of the emotion outcomes for the in-feed survey responses in the increased exposure experiment.}
\label{table:emotions_infeed_increase}
\end{table}

\begin{table}
\centering
\scriptsize
\resizebox{\textwidth}{!}{%
    
\begin{tabular}{l c c c c}
\toprule
 & \multicolumn{4}{c}{Reduced Exposure, Post-Experiment} \\
\cmidrule(lr){2-5}
 & Angry & Sad & Excited & Calm \\
\midrule
(Intercept)                         & $1.20^{*}$       & $1.06^{*}$       & $1.28^{*}$       & $2.12^{*}$       \\
                                    & $ [ 1.01; 1.39]$ & $ [ 0.88; 1.24]$ & $ [ 1.08; 1.48]$ & $ [ 1.89; 2.36]$ \\
Condition: Treatment                & $-0.12$          & $0.02$           & $0.04$           & $0.03$           \\
                                    & $ [-0.26; 0.02]$ & $ [-0.11; 0.14]$ & $ [-0.11; 0.18]$ & $ [-0.12; 0.18]$ \\
Pre-survey Answer                   & $0.46^{*}$       & $0.45^{*}$       & $0.51^{*}$       & $0.40^{*}$       \\
                                    & $ [ 0.39; 0.53]$ & $ [ 0.39; 0.52]$ & $ [ 0.44; 0.57]$ & $ [ 0.34; 0.47]$ \\
Recruitment Platform: CloudResearch & $0.06$           & $0.05$           & $-0.07$          & $-0.04$          \\
                                    & $ [-0.08; 0.21]$ & $ [-0.08; 0.18]$ & $ [-0.22; 0.08]$ & $ [-0.20; 0.11]$ \\
\midrule
Num. obs.                           & $617$            & $617$            & $617$            & $617$            \\
\bottomrule
\multicolumn{5}{l}{\scriptsize{$^*$ Null hypothesis value outside the confidence interval.}}
\end{tabular}

}
\caption{Emotions: Reduced Exposure Experiment, Post-Experiment. Regression tables of the analysis of the emotion outcomes for the post-experiment survey responses.}
\label{table:emotions_prepost_reduce}
\end{table}

\begin{table}
\centering
\scriptsize
\resizebox{\textwidth}{!}{%
    
\begin{tabular}{l c c c c}
\toprule
 & \multicolumn{4}{c}{Increased Exposure, Post-Experiment} \\
\cmidrule(lr){2-5}
 & Angry & Sad & Excited & Calm \\
\midrule
(Intercept)                         & $0.88^{*}$       & $0.95^{*}$       & $1.28^{*}$        & $2.14^{*}$       \\
                                    & $ [ 0.67; 1.10]$ & $ [ 0.72; 1.17]$ & $ [ 1.05;  1.51]$ & $ [ 1.86; 2.42]$ \\
Condition: Treatment                & $0.01$           & $0.06$           & $0.06$            & $-0.12$          \\
                                    & $ [-0.15; 0.17]$ & $ [-0.10; 0.22]$ & $ [-0.10;  0.22]$ & $ [-0.29; 0.05]$ \\
Pre-survey Answer                   & $0.59^{*}$       & $0.54^{*}$       & $0.53^{*}$        & $0.41^{*}$       \\
                                    & $ [ 0.52; 0.67]$ & $ [ 0.46; 0.62]$ & $ [ 0.45;  0.61]$ & $ [ 0.33; 0.48]$ \\
Recruitment Platform: CloudResearch & $0.08$           & $0.17^{*}$       & $-0.25^{*}$       & $-0.12$          \\
                                    & $ [-0.08; 0.24]$ & $ [ 0.01; 0.34]$ & $ [-0.42; -0.08]$ & $ [-0.28; 0.05]$ \\
\midrule
Num. obs.                           & $472$            & $472$            & $472$             & $472$            \\
\bottomrule
\multicolumn{5}{l}{\scriptsize{$^*$ Null hypothesis value outside the confidence interval.}}
\end{tabular}

}
\caption{Emotions: Increased Exposure Experiment, Post-Experiment. Regression tables of the analysis of the emotion outcomes for the post-experiment survey responses.}
\label{table:emotions_prepost_increase}
\end{table}

\begin{table}
\centering
\scriptsize
\resizebox{\textwidth}{!}{%
    
\begin{tabular}{l c c c c}
\toprule
 & \multicolumn{4}{c}{Reduced Exposure, In-Feed} \\
\cmidrule(lr){2-5}
 & Angry & Sad & Excited & Calm \\
\midrule
(Intercept)                         & $9.14^{*}$        & $7.44^{*}$        & $12.11^{*}$       & $17.23^{*}$       \\
                                    & $ [ 6.59; 11.69]$ & $ [ 4.79; 10.09]$ & $ [ 8.65; 15.56]$ & $ [12.36; 22.11]$ \\
Condition: Treatment                & $-4.65^{*}$       & $-3.28^{*}$       & $1.39$            & $2.56$            \\
                                    & $ [-7.24; -2.06]$ & $ [-5.96; -0.60]$ & $ [-1.94;  4.73]$ & $ [-0.47;  5.60]$ \\
Baseline: Mean Answer               & $0.54^{*}$        & $0.68^{*}$        & $0.74^{*}$        & $0.69^{*}$        \\
                                    & $ [ 0.47;  0.61]$ & $ [ 0.61;  0.76]$ & $ [ 0.67;  0.82]$ & $ [ 0.63;  0.76]$ \\
Recruitment Platform: CloudResearch & $0.62$            & $2.52$            & $-4.23^{*}$       & $-3.10$           \\
                                    & $ [-2.07;  3.32]$ & $ [-0.26;  5.30]$ & $ [-7.71; -0.76]$ & $ [-6.24;  0.04]$ \\
\midrule
Num. obs.                           & $3120$            & $3081$            & $3062$            & $3162$            \\
Num. groups.               & $595$             & $587$             & $596$             & $587$             \\
\bottomrule
\multicolumn{5}{l}{\scriptsize{$^*$ Null hypothesis value outside the confidence interval.}}
\end{tabular}

}
\caption{Full population -- Emotions: Reduced Exposure Experiment, In-Feed. Regression tables of the analysis of the emotion outcomes for the in-feed survey responses.}
\label{table:emotions_infeed_reduce_all}
\end{table}

\begin{table}
\centering
\scriptsize
\resizebox{\textwidth}{!}{%
    
\begin{tabular}{l c c c c}
\toprule
 & \multicolumn{4}{c}{Increased Exposure, In-Feed} \\
\cmidrule(lr){2-5}
 & Angry & Sad & Excited & Calm \\
\midrule
(Intercept)                         & $6.03^{*}$       & $6.21^{*}$      & $14.45^{*}$       & $20.12^{*}$       \\
                                    & $ [ 3.16; 8.91]$ & $ [2.95; 9.47]$ & $ [10.53; 18.37]$ & $ [14.72; 25.52]$ \\
Condition: Treatment                & $4.89^{*}$       & $4.61^{*}$      & $0.23$            & $-2.76$           \\
                                    & $ [ 1.86; 7.91]$ & $ [1.30; 7.92]$ & $ [-3.37;  3.83]$ & $ [-6.08;  0.57]$ \\
Baseline: Mean Answer               & $0.55^{*}$       & $0.57^{*}$      & $0.68^{*}$        & $0.67^{*}$        \\
                                    & $ [ 0.47; 0.63]$ & $ [0.49; 0.66]$ & $ [ 0.60;  0.76]$ & $ [ 0.60;  0.75]$ \\
Recruitment Platform: CloudResearch & $1.95$           & $5.04^{*}$      & $-6.02^{*}$       & $-3.70^{*}$       \\
                                    & $ [-1.10; 5.00]$ & $ [1.70; 8.37]$ & $ [-9.65; -2.40]$ & $ [-7.05; -0.35]$ \\
\midrule
Num. obs.                           & $2501$           & $2427$          & $2480$            & $2458$            \\
Num. groups.               & $463$            & $465$           & $458$             & $473$             \\
\bottomrule
\multicolumn{5}{l}{\scriptsize{$^*$ Null hypothesis value outside the confidence interval.}}
\end{tabular}

}
\caption{Full population -- Emotions: Increased Exposure Experiment, In-Feed. Regression tables of the analysis of the emotion outcomes for the in-feed survey responses.}
\label{table:emotions_infeed_increase_all}
\end{table}

\begin{table}
\centering
\scriptsize
\resizebox{\textwidth}{!}{%
    
\begin{tabular}{l c c c c}
\toprule
 & \multicolumn{4}{c}{Reduced Exposure, Post-Experiment} \\
\cmidrule(lr){2-5}
 & Angry & Sad & Excited & Calm \\
\midrule
(Intercept)                         & $1.14^{*}$       & $1.08^{*}$       & $1.19^{*}$       & $2.09^{*}$       \\
                                    & $ [ 0.97; 1.32]$ & $ [ 0.91; 1.24]$ & $ [ 1.01; 1.37]$ & $ [ 1.88; 2.31]$ \\
Condition: Treatment                & $-0.08$          & $-0.00$          & $0.04$           & $0.04$           \\
                                    & $ [-0.21; 0.04]$ & $ [-0.12; 0.12]$ & $ [-0.09; 0.18]$ & $ [-0.10; 0.17]$ \\
Pre-survey Answer                   & $0.45^{*}$       & $0.44^{*}$       & $0.54^{*}$       & $0.41^{*}$       \\
                                    & $ [ 0.39; 0.52]$ & $ [ 0.38; 0.51]$ & $ [ 0.48; 0.60]$ & $ [ 0.35; 0.47]$ \\
Recruitment Platform: CloudResearch & $0.11$           & $0.09$           & $-0.08$          & $-0.04$          \\
                                    & $ [-0.03; 0.24]$ & $ [-0.04; 0.21]$ & $ [-0.22; 0.06]$ & $ [-0.18; 0.11]$ \\
\midrule
Num. obs.                           & $727$            & $727$            & $727$            & $727$            \\
\bottomrule
\multicolumn{5}{l}{\scriptsize{$^*$ Null hypothesis value outside the confidence interval.}}
\end{tabular}

}
\caption{Full population -- Emotions: Reduced Exposure Experiment, Post-Experiment. Regression tables of the analysis of the emotion outcomes for the post-experiment survey responses.}
\label{table:emotions_prepost_reduce_all}
\end{table}

\begin{table}
\centering
\scriptsize
\resizebox{\textwidth}{!}{%
    
\begin{tabular}{l c c c c}
\toprule
 & \multicolumn{4}{c}{Increased Exposure, Post-Experiment} \\
\cmidrule(lr){2-5}
 & Angry & Sad & Excited & Calm \\
\midrule
(Intercept)                         & $0.90^{*}$       & $0.87^{*}$       & $1.19^{*}$        & $2.05^{*}$       \\
                                    & $ [ 0.70; 1.11]$ & $ [ 0.66; 1.09]$ & $ [ 0.97;  1.41]$ & $ [ 1.78; 2.31]$ \\
Condition: Treatment                & $-0.02$          & $0.08$           & $0.09$            & $-0.07$          \\
                                    & $ [-0.17; 0.13]$ & $ [-0.07; 0.23]$ & $ [-0.06;  0.25]$ & $ [-0.23; 0.09]$ \\
Pre-survey Answer                   & $0.58^{*}$       & $0.55^{*}$       & $0.56^{*}$        & $0.43^{*}$       \\
                                    & $ [ 0.50; 0.65]$ & $ [ 0.47; 0.63]$ & $ [ 0.49;  0.63]$ & $ [ 0.36; 0.50]$ \\
Recruitment Platform: CloudResearch & $0.13$           & $0.22^{*}$       & $-0.27^{*}$       & $-0.12$          \\
                                    & $ [-0.03; 0.28]$ & $ [ 0.06; 0.37]$ & $ [-0.43; -0.11]$ & $ [-0.28; 0.05]$ \\
\midrule
Num. obs.                           & $528$            & $528$            & $528$             & $528$            \\
\bottomrule
\multicolumn{5}{l}{\scriptsize{$^*$ Null hypothesis value outside the confidence interval.}}
\end{tabular}

}
\caption{Full population -- Emotions: Increased Exposure Experiment, Post-Experiment. Regression tables of the analysis of the emotion outcomes for the post-experiment survey responses.}
\label{table:emotions_prepost_increase_all}
\end{table}

\begin{figure}[p]
\centering
\includegraphics[width=0.85\textwidth]{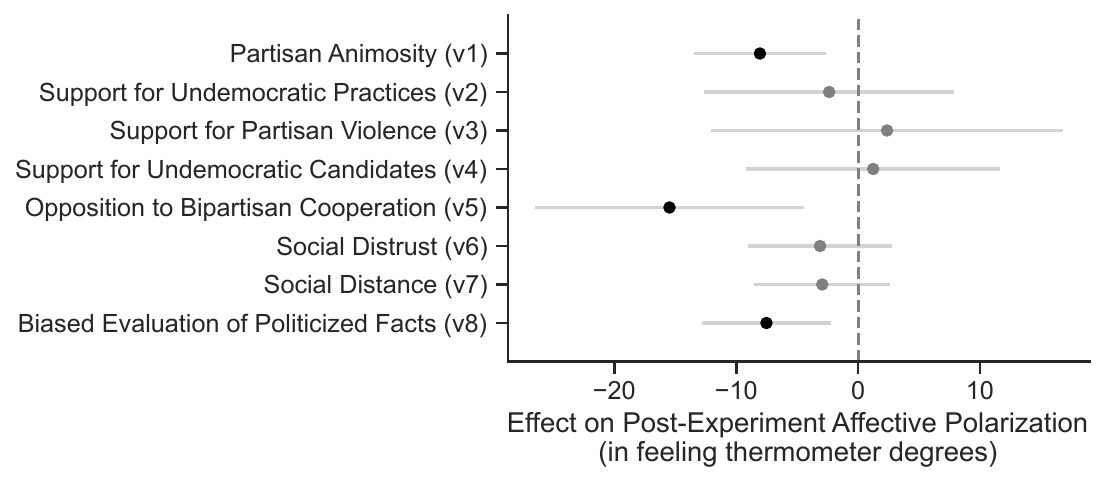}
\caption{Relationship between the effect on affective polarization (post-experiment) and exposure to posts with each of the eight AAPA factors. Each value represents the estimated effect from a separate regression model. The dependent variable is the political feeling response in the post-experiment survey, controlling for both the pre-experiment survey response and the fraction of political posts viewed over a ten-day period for at least one second, which expressed support for a particular factor (i.e., $\text{\textit{affective\_polarization}}_{\text{\textit{post}}} \sim \text{\textit{affective\_polarization}}_{\text{\textit{pre}}} + \text{\textit{fraction\_}}v_n$, where $v_n$ is a specific factor such as partisan animosity or social distrust). The coefficient represents the estimated change (on a 0-100 scale) in feelings toward people from the opposite party when 100\% of the viewed political posts expressed support for the corresponding factor ($v_n$). All participants who completed the post-experiment survey are included in the analysis (N = 1,256).}\label{fig:factors_contribution}
\end{figure}

\subsection{AAPA Factors Analysis}
\label{sup:infeed_factors}
To investigate how individual AAPA factors influence in-feed survey responses, we first analyzed the relationship between participants' reported out-party feelings and the aggregate AAPA score of the post immediately preceding each in-feed survey. Using a mixed-effects regression model with participants as a random effect, we found a significant negative association (coefficient = –0.124; 95\% CI: [–0.236, –0.012]; P = 0.029). This suggests that posts containing more AAPA factors were associated with colder feelings toward the opposing party.

We then extended this analysis to test the effect of individual AAPA factors, included as binary indicators in the same mixed-effects framework. However, none of the individual factors reached statistical significance, likely due to limited statistical power and collinearity among the predictors.

To further examine the impact of these factors, we investigated the correlation between the participants' response to the post-experiment out-party feeling thermometer and the overall exposure to posts with a certain factor. As shown in \Figref{fig:factors_contribution}, certain types of AAPA factors, such as expressions of partisan animosity, opposition to bipartisan collaboration, and biased evaluations of political facts, do predict changes in participants' post-experiment attitudes. This figure reports the variation in post-experiment scores as a function of the fraction of content consumed that reflects each specific factor. For example, the regression coefficients suggest that if a participant's feed contained exclusively (100\%) posts expressing opposition to bipartisan collaboration, their attitude would shift by approximately 15 degrees.
It is essential to note that other types of content, such as support for political violence, contribute to the final score and may influence changes in attitude. However, their effects may not reach statistical significance, possibly because they are rare (appearing in only 5.3\% of political posts and 1.62\% of the posts seen by participants in their feeds), and we are not powered to detect effects for these rarer factors when analyzed individually.

\subsection{Qualitative Feedback on the Impact of the Experiment}
\label{sup:impact}

At the end of the study, in the post-experiment survey, we asked participants: ``Did the Chrome extension affect your experience on Twitter/X in any way?" This question was open-ended, and 1,218 participants (96.9\%) provided a non-empty response. Most answers were short and indicated no impact, such as ``No" or ``Not that I could tell." We individually coded each response with the support of AI (GPT-4), providing the original survey question for context and asking whether the participant reported any impact on their experience on Twitter/X.

The majority of participants (overall: 74.2\%; increase, control: 73.7\%; increase, treatment: 72.9\%; reduce, control: 76.4\%; reduce, treatment: 73.4\%) explicitly stated that they did not notice any impact on their experience. We manually reviewed a random sample of 100 responses and found complete agreement with GPT-4's annotations.
Among the remaining responses, we manually coded 10.3\% as reporting a ``positive experience," often for reasons unrelated to the extension, such as: ``Twitter seemed better organized," ``It made me feel informed," and ``It helped me explore the social network better." Additionally, 4.1\% reported observing more political content; however, this observation was uniformly distributed across conditions, likely reflecting an overall increase in political content during the election period. Finally, 2.8\% of participants mentioned noticing a slowdown in the website's performance. The remaining answers mention details unrelated to the assigned condition, such as ``It made me use it on my laptop more, which was a bit annoying" or ``Having to use the For You page I noticed a lot of tweets about AI."

To investigate this further, we adopted a quantitative approach by embedding participant responses using a pre-trained BERT-based model (\textit{all-MiniLM-L6-v2}). We then predicted the assigned experiment using logistic regression with L2 regularization, employing 5-fold cross-validation. The classification achieved an F1 score of 53.9\% ($\pm$ 4.7\%) on a balanced dataset, performance statistically indistinguishable from random guessing. Similarly, we predicted the experimental condition (treatment or control) for each experiment. For the reduced experiment, the F1 score was 47.4\% ($\pm$ 6.8\%), and for the increased experiment, it was 50.5\% ($\pm$ 4.7\%); performance no better than random guessing.

Together, these results suggest that the experimental condition did not significantly impact participants' experience, as they did not perceive or report notable differences.

\subsection{Decay of Treatment Effects}
\label{sec:time_gap}
To investigate whether there is any short-term decay in the treatment effects of the interventions reducing and increasing exposure to AAPA content, we examined (1) how the treatment effects varied over the course of the experiment and (2) whether the time gap between the end of the intervention and participants’ completion of the post-experiment survey was correlated with the size of the treatment effects.

\begin{figure}[t]
\centering
\includegraphics[width=0.9\textwidth]{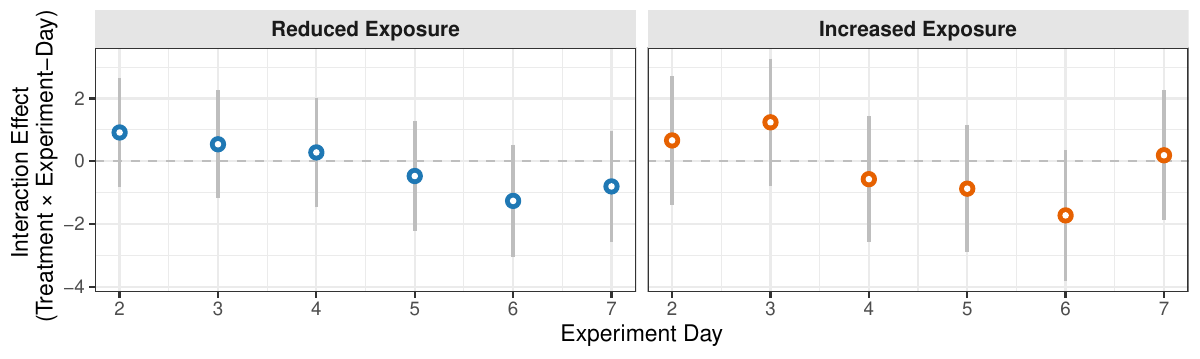}
\caption{Variation in treatment effects on affective polarization for each day of the intervention, relative to the first day, as measured by the in-feed surveys. Points show the coefficients of the interactions between treatment assignment and experiment day, and error bars represent the 95\%~CIs.}
\label{fig:si_in_feed_experiment_day}
\end{figure}

To examine how the treatment effects varied as the experiment progressed, we reanalyzed the in-feed responses, extending our model specification to include an indicator variable for the day of the experiment (relative to the start of the intervention for each participant) and its interaction with the experimental condition indicator. Figure~\ref{fig:si_in_feed_experiment_day} shows the point estimates and 95\% confidence intervals of the interaction coefficients. We find that none of the interactions between the experimental condition and day of the experiment are statistically significant in either the Reduced or Increased Exposure experiments. These results suggest that the treatment effects do not significantly vary over time and that there is no evidence of decay effects within the week of the intervention.

We also tested for potential decay effects based on the timing of participants' last exposure to treatment. We can perform this analysis because of natural variation in the time delta between (a) when our browser extension automatically turned off the reranking at the end of the final day of the experiment and (b) when participants filled out our post-survey. Participants could take the survey immediately after the reranking turned off (after midnight on the last day of the experiment), but most of them completed the survey in the following days. In the interim the participants could still browse their (unmodified) X feeds, but were shown a large banner with a link to the post-experiment survey. We extended our model specification for analyzing the post-experiment effects to include the total hours between when a participant last viewed a post (for at least one second) during the experiment and when they began the post-survey. The median delay gap was 25.5 hours (mean~=~42.3). Table~\ref{table:polarization_prepost_time_gap} shows the results of the regression analysis. We find that the time difference between the last exposure and the start of the post-survey is not a significant predictor of changes in affective polarization and does not alter the significance of the treatment effect.

Together, these results suggest that the treatment effects do not decay in the short term.

\begin{table}
\centering
\small
\begin{tabular}{l c c}
\toprule
 & \multicolumn{2}{c}{Post-Experiment} \\
\cmidrule(lr){2-3}
 & Reduced Exposure & Increased Exposure \\
\midrule
(Intercept)                         & $0.41$           & $4.59^{*}$         \\
                                    & $[-1.88; 2.70]$  & $[1.70; 7.49]$     \\
Condition: Treatment                & $2.10^{*}$       & $-2.44^{*}$        \\
                                    & $[0.14; 4.05]$   & $[-4.76; -0.13]$   \\
Pre-survey Answer                  & $0.81^{*}$       & $0.83^{*}$         \\
                   & $[0.77; 0.85]$   & $[0.77; 0.88]$     \\
Recruitment Platform: CloudResearch & $-0.54$          & $-3.04^{*}$        \\
                                    & $[-2.60; 1.52]$  & $[-5.50; -0.57]$   \\
Hours Difference                           & $0.02$           & $-0.02$            \\
                                    & $[-0.01; 0.05]$  & $[-0.05; 0.02]$    \\
\midrule
Num. obs.                           & $622$            & $468$              \\
\bottomrule
\multicolumn{3}{l}{\scriptsize{$^*$ Null hypothesis value outside the confidence interval.}}
\end{tabular}

\caption{Affective Polarization: Post-Experiment. Regression tables of the analysis of the
affective polarization outcome for the post-experiment survey responses, including the number of hours from the last post view (at least 1 second) and the answer to the post-experiment survey.}
\label{table:polarization_prepost_time_gap}
\end{table}

\begin{figure}[t]
\centering
\includegraphics[width=\textwidth]{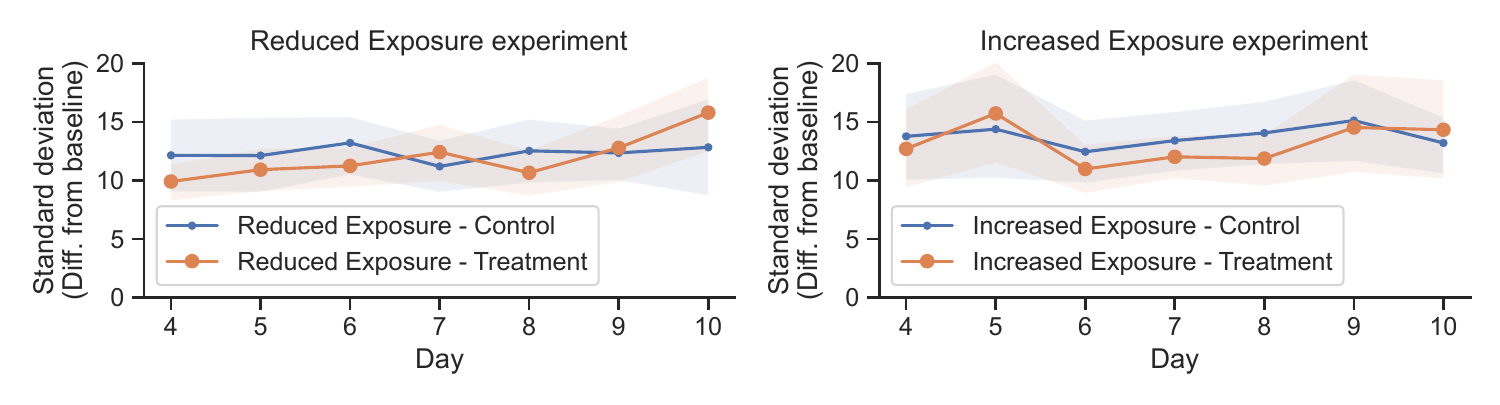}
\caption{Standard deviation of daily participant responses (difference from baseline) by day, experiment, and condition. Shaded areas represent 95\% bootstrapped confidence intervals.}
\label{fig:in_feed_std_by_day}
\end{figure}

\subsection{Survey Fatigue and Response Lock-in}
A potential risk in our experimental design is that participants repeatedly asked to report their feelings toward political outgroups may experience survey fatigue and settle on a fixed response over time, responding habitually rather than reflectively.
This may cause responses to converge to a preferred value, regardless of the actual stimuli. To examine this possibility, we computed, for each day, the standard deviation of the difference between each participant's in-feed response and their own baseline. This metric captures the dispersion of daily responses relative to each participant's initial reporting level, aggregated across users.
If participants were anchoring to a fixed response, we would expect this marginal standard deviation to decline over time. However, as shown in Figure~\ref{fig:in_feed_std_by_day}, we observe no consistent downward trend in either experiment or condition. This suggests that participants continued to respond with a similar degree of variation, providing no evidence of convergence.

\subsection{Controlling for Major Political Events}
The experiment was conducted during a period of heightened political activity preceding the 2024 US election. Several major events during this time may have influenced participants' feelings toward the opposing party, potentially affecting levels of affective polarization. These events include:
\begin{itemize}
    \item The assassination attempt on Donald Trump (July 13),
    \item The Republican National Convention (July 15--18),
    \item Joe Biden dropping out of the presidential race (July 21),
    \item The presumptive nomination of Kamala Harris (July 22),
    \item Kamala Harris's official nomination as the Democratic candidate (August 5).
\end{itemize}

To assess whether these events were driving the observed effects, we conducted an additional analysis using binary indicator variables. Each indicator was set to 1 if a survey response occurred on the day of a given event or the day immediately following it. The exception was the Republican National Convention, for which the indicator covered the entire three-day span.
These event indicators were then interacted with the treatment variable in our regression models. The results, summarized in \Tabref{table:polarization_infeed_events}, show that our core findings remain robust: the treatment variable remains statistically significant even after accounting for these events.

In the Reduced Exposure experiment, none of the event indicators had a statistically significant interaction with the treatment, indicating that the observed effects are not attributable to these external events.
In the Increased Exposure experiment, the main treatment effect remains statistically significant. While some event indicators---such as the assassination attempt on Trump and the presumptive nomination of Harris---reach the conventional 5\% significance threshold, these effects do not survive Bonferroni correction for multiple comparisons (five hypotheses). Furthermore, their interaction terms with the treatment condition are not significant after correction.

Overall, these results suggest that the main effects of our intervention on affective polarization persist even when accounting for the influence of major political events during the study period.

\begin{table}
\centering
\scriptsize
\resizebox{!}{0.46\textheight}{
    
\begin{tabular}{l c c}
\toprule
 & \multicolumn{2}{c}{In-Feed} \\
\cmidrule(lr){2-3}
 & Reduced Exposure & Increased Exposure \\
\midrule
(Intercept)                                 & $-0.14$           & $3.12^{*}$        \\
                                            & $[-3.01;\ 2.73]$  & $[1.11;\ 5.14]$   \\
Recruitment Platform: CloudResearch         & $-1.32$           & $-1.05$           \\
                                            & $[-4.15;\ 1.51]$  & $[-3.09;\ 0.99]$  \\
Treatment                                   & $3.88^{*}$        & $-2.65^{*}$       \\
                                            & $[1.14;\ 6.61]$   & $[-4.68;\ -0.62]$ \\
Baseline: Outparty Feeling                  & $0.92^{*}$        & $0.84^{*}$        \\
                                            & $[0.85;\ 1.00]$   & $[0.80;\ 0.88]$   \\
Assassination Attempt                       & $-1.72$           & $-3.77^{*}$       \\
                                            & $[-4.62;\ 1.17]$  & $[-6.63;\ -0.91]$ \\
Assassination Attempt $\times$ Treatment    & $1.64$            & $4.50^{*}$        \\
                                            & $[-2.67;\ 5.95]$  & $[0.41;\ 8.58]$   \\
Republican Convention                       & $-0.37$           & $-0.45$           \\
                                            & $[-2.35;\ 1.61]$  & $[-2.49;\ 1.59]$  \\
Republican Convention $\times$ Treatment    & $-0.20$           & $-1.87$           \\
                                            & $[-3.08;\ 2.68]$  & $[-4.93;\ 1.20]$  \\
Biden Dropout                               & $0.01$            & $-0.48$           \\
                                            & $[-1.33;\ 1.35]$  & $[-1.83;\ 0.87]$  \\
Biden Dropout $\times$ Treatment            & $0.36$            & $1.82$            \\
                                            & $[-1.53;\ 2.26]$  & $[-0.16;\ 3.81]$  \\
Harris Presumptive Nomination               & $0.41$            & $1.43^{*}$        \\
                                            & $[-0.94;\ 1.76]$  & $[0.09;\ 2.78]$   \\
Harris Presumptive $\times$ Treatment       & $0.99$            & $-2.19^{*}$       \\
                                            & $[-0.87;\ 2.84]$  & $[-4.12;\ -0.26]$ \\
Harris Official Nomination                  & $-0.22$           & $-2.96^{*}$       \\
                                            & $[-1.71;\ 1.27]$  & $[-4.67;\ -1.24]$ \\
Harris Official Nomination $\times$ Treatment          & $0.92$            & $2.81^{*}$        \\
                                            & $[-1.24;\ 3.09]$  & $[0.28;\ 5.34]$   \\
\midrule
Num. obs.                                   & $6,116$           & $4,778$           \\
Num. groups                                 & $622$             & $468$             \\
\bottomrule
\multicolumn{3}{l}{\scriptsize{$^*$ Null hypothesis value outside the confidence interval.}} \\
\end{tabular}

}
\caption{Affective Polarization: In-Feed. Regression tables of the analysis of the affective polarization outcome for the in-feed survey responses, controlling for major political events.}
\label{table:polarization_infeed_events}
\end{table}

\clearpage

\end{document}